\documentclass[twocolumn,apj,tighten]{openjournal}
\pdfoutput=1
\usepackage{graphicx}
\usepackage{amsfonts,amsmath}
\usepackage{bm}
\usepackage[section]{placeins}
\usepackage[breaklinks,colorlinks,urlcolor=blue,citecolor=blue,linkcolor=blue]{hyperref}
\usepackage{chngcntr}

\makeatletter
\renewcommand{\appendix}{
  \par
  \setcounter{section}{0}
  \setcounter{subsection}{0}
  \renewcommand\thesection{\Alph{section}}%
  \counterwithin{figure}{section}
  \counterwithin{table}{section}
  \counterwithin{equation}{section}
  \renewcommand\thefigure{\thesection\arabic{figure}}%
  \renewcommand\thetable{\thesection\arabic{table}}%
  \renewcommand\theequation{\thesection\arabic{equation}}%
}
\makeatother
\def\mone{{\rm M}_1}
\def\mtwo{{\rm M}_2}
\def\mfirst{{\rm M}_{\rm first}}
\def\msecond{{\rm M}_{\rm second}}

\def\aone{{\rm a}_1}
\def\atwo{{\rm a}_2}

\def\xeff{\chi_{\rm eff}}

\def\zmax{z_{\rm max}}

\def\tdelay{t_{\rm delay}}

\def\etabbh{\eta_{\rm BBH}}
\def\etabns{\eta_{\rm BNS}}
\def\etansbh{\eta_{\rm NSBH}}

\def\porb{P_{\rm orb}}
\def\vorb{V_{\rm orb}}
\def\pspin{P_{\rm spin}}

\def\abh{a_{\rm BH,sec}}
\def\abhzero{a_{\rm BH0}}
\def\abhfirst{a_{\rm BH,fir}}
\def\abhone{a_{\rm BH1}}
\def\abhtwo{a_{\rm BH2}}
\def\abhmean{\langle a_{\rm BH,sec} \rangle}
\def\frev{f_{\rm rev}}
\def\conea{c_1^\alpha}
\def\ctwoa{c_2^\alpha}
\def\cthreea{c_3^\alpha}
\def\coneb{c_1^\beta}
\def\ctwob{c_2^\beta}
\def\cthreeb{c_3^\beta}
\def\mwr{M_{\rm WR}}
\def\qcrit{q_{\rm crit}}
\def\qcrzero{q_{\rm crit0}}
\def\tilt{\nu_{\rm sec}}
\def\tiltfirst{\nu_{\rm fir}}
\def\thls{\theta_{\rm sec}}
\def\thlsfirst{\theta_{\rm fir}}
\def\fdyn{f_{\rm dyn}}
\def\lzerocap{\hat{L_0}}
\def\lonecap{\hat{L_1}}
\def\ltwocap{\hat{L_2}}
\def\lcap{\hat{L}}
\def\aonecap{\hat{a_1}}
\def\atwocap{\hat{a_2}}
\def\ivec{\hat{i}}
\def\jvec{\hat{j}}
\def\kvec{\hat{k}}
\def\ombmax{\omega_{\rm b,max}}
\newcommand{\Rs}{\ensuremath{{\rm~R}_{\odot}}}
\newcommand{\Ms}{\ensuremath{{\rm~M}_{\odot}}}

\newcommand{\eg}{{\it e.g.}}

\newcommand{\ie}{{\it i.e.}}

\newcommand{\beq}{\begin{equation}}
\newcommand{\eeq}{\end{equation}}
\newcommand{\mtot}{\ensuremath{M_{\rm tot}}}
\newcommand{\mtotst}{\ensuremath{M_{\rm tot,\ast}}}

\newcommand{\peryr}{\ensuremath{{\rm~yr}^{-1}}}

\newcommand{\nbseven}{{\tt NBODY7}}

\newcommand{\nbpp}{{\tt NBODY6++GPU}}

\newcommand{\bse}{{\tt BSE}}

\newcommand{\ssebse}{{\tt SSE/BSE}}

\newcommand{\mocca}{{\tt MOCCA}}
\newcommand{\cmc}{{\tt CMC}}
\newcommand{\startrack}{{\tt StarTrack}}

\newcommand{\mco}{\ensuremath{M_{\rm CO}}}

\newcommand{\ace}{\ensuremath{\alpha_{\rm CE}}}

\makeatletter
\long\def\@makecaption#1#2{%
 \noindent\begin{minipage}{0.9999\linewidth}
   \if\csname ftype@\@captype\endcsname 2
   \vskip 2ex\noindent \centering\@table@type@size{\@eapj@cap@font  #1}\par
    #2\par\medskip
   \else
   \vspace*{\abovecaptionskip}\noindent\footnotesize #1 #2\par\vskip \belowcaptionskip
   \fi
 \end{minipage}\par
}
\makeatother

\begin{document}

\title[Effective spin distribution in $\bse$]
{On the effective spin - mass ratio relation of binary black hole mergers that evolved in isolation}

\author{Sambaran Banerjee\altaffilmark{1}, Aleksandra Olejak\altaffilmark{2}}
\affil{$^1$Helmholtz-Instituts f\"ur Strahlen- und Kernphysik (HISKP),
Nussallee 14-16, D-53115 Bonn, Germany}
\affil{$^2$Max Planck Institut für Astrophysik (MPA),
Karl-Schwarzschild-Straße 1, 85748 Garching bei M\"unchen, Germany}

\begin{abstract}
The ground-based measurement of gravitational waves (GW) from merging binary black holes (BBH) allows independent determination of spins of stellar-remnant black holes (BH), thereby offering insights into their formation mechanisms. The observed population of BBH mergers exhibits two intriguing peculiarities related to BH spins, namely, a positively biased distribution of the effective spin parameter, $\xeff$, and an apparent anti-correlation between merger mass ratio, $q$, and $\xeff$. Here we investigate the potential mechanisms for such observed properties, in BBH mergers via isolated binary evolution. We synthesise BBH mergers with the fast binary evolution code BSE. The role of various physical assumptions is explored, including tidal spin-up, compact remnant mass, and mass transfer physics. We compare the properties of BBHs that form through stable mass transfer (SMT) and common envelope evolution (CE). We find that both the asymmetry in the $\xeff$ distribution and the $\xeff-q$ anti-correlation can be natural outcomes of isolated-binary BBH formation. The anti-correlation is especially pronounced for SMT-channel BBH mergers that experience a mass-ratio reversal, i.e., those where the second-born BH is the more massive one. The anti-correlation arises from the dependence of orbital shrinking during mass transfer and the Roche lobe size on the system's mass ratio. This characteristic $\xeff-q$ trend diminishes with increasing metallicity and when the isolated-binary BBH merger population is mixed with a significant contribution of dynamically formed BBH mergers or the newly formed BH’s spin is misaligned relative to the parent star’s spin. Our results demonstrate that isolated massive binary evolution via the SMT sub-channel can reproduce trends in the observable BBH merger population, with the characteristic signatures in mass, mass ratio, and spin distributions.
\end{abstract}


\maketitle

\section{Introduction}\label{intro}

The transient gravitational-wave events, as detected by the LIGO-Virgo-KAGRA (hereafter LVK)
gravitational-wave (hereafter GW) detector network \citep[][]{Asai_2015,Acernese_2015,KAGRA_2020},
continue to cause widespread speculations about their origin and formation scenarios.
So far, the LVK collaboration has published, in their GW transient catalogue (hereafter GWTC)
\footnote{\url{https://gwosc.org/eventapi/html/GWTC/}},
more than 200 detected candidates of compact binary merger events until the end of their fourth observing run, `O4'. The up-to-date GWTC includes all event candidates
from LVK's first, second (`O1', `O2'; \citealt{Abbott_GWTC1}),
third (`O3'; \citealt{Abbott_GWTC2,Abbott_GWTC2.1,Abbott_GWTC3}) and the first part of `O4'  (`O4a' \citealt{GWTC4a_cat}).
The vast majority of  the detections are classified as binary black hole mergers (hereafter BBH).
The rest comprise neutron star-neutron star (hereafter NS), NS-BH mergers, and events possibly involving compact objects within the so-called lower mass gap between $2\Ms-5\Ms$ \citep[e.g.,][]{LowerMassGap2024, Unequal_masss_2020}.

A key aspect of GW detection from a double compact object close to its merger
is the ability to infer about the spins of the merging members, as encoded in the
GW signal.
The observable quantity, effective spin parameter \citep{Ajith_2011}, $\xeff$, is a measure of the spin-orbit alignment of the merging system and is defined as
\begin{equation}
	\xeff = \frac{(\mone\vec\aone + \mtwo\vec\atwo)\cdot\lcap}{\mone+\mtwo}
	= \frac{(\mone\aone\aonecap\cdot\lcap + \mtwo\atwo\atwocap\cdot\lcap)}{\mone+\mtwo}
\label{eq:xeffvec}
\end{equation}
or
\begin{equation}
\xeff = \frac{\mone\aone\cos\theta_1 + \mtwo\atwo\cos\theta_2}{\mone+\mtwo}
	= \frac{\aone\cos\theta_1 +  q\atwo\cos\theta_2}{1 + q}.
\label{eq:xeffdef}
\end{equation}
Here, the merging masses $\mone$, $\mtwo$, with mass ratio
$q\equiv\mtwo/\mone$ ($q\leq1$)\footnote{In this paper, we follow the convention that
the more (less) massive merging member is the primary (secondary)
that is denoted by the subscript 1 (2).}, have, respectively,
dimensionless spin vectors or Kerr vectors $\vec\aone=\aonecap\aone$, $\vec\atwo=\atwocap\atwo$.
They make angles
$\theta_1$, $\theta_2$, respectively, with the direction of the orbital angular momentum, $\vec L$,
defined by the unit vector $\lcap\equiv{\vec L}/L$
\footnote{In this paper, we follow the notation convention that a vector $\vec A$ has magnitude
$A$ and the unit vector in its direction is $\hat{A}$.}.
During a GR inspiral, $\xeff$ remains nearly invariant \citep{Yu_2020,Gerosa_2021b}.
Therefore, one can, to a good approximation, apply $\theta_1$, $\theta_2$
at the formation of the double compact binary to infer the LVK-measured $\xeff$
at a later time, in the final inspiral phase.

%
The observed BBH-merger GW events hint at peculiarities in their $\xeff$ distribution. Until GWTC-3, the overall $\xeff$
distribution was positively skewed and peaked at $\xeff\approx0.07$ \citep[][their Fig.~16]{Abbott_GWTC3}, which implies that aligned mergers ($\xeff>0$)
are preferred to anti-aligned mergers. In the more recent GWTC-4 from LVK's O4a, the preference towards $\xeff>0$ remains.
Another reported feature is an apparent anti-correlation
on the $\xeff-q$ plane --- higher spinning mergers tend to have more unequally massive components \citep{Callister_2021,Abbott_GWTC3,Adamcewicz2023,Heinzel2024}.
While this anti-correlation was merely hinted in GWTC-3, it could not be better confirmed in GWTC-4: in fact, GWTC-4 may instead suggest a 
$\xeff-q$ correlation \citep{GWTC4a_pop}.
To date, the existence and origins of this peculiariy in the observed BBH-merger data remain open. Therefore, it is,
nevertheless, important to investigate possible origins of $\xeff-q$ anti-correlation in BBH-merger formation channels.

A pure dynamical pairing of merging BBHs would produce an isotropic spin distribution for both of the merging components. That would result in a $\xeff$ distribution symmetric around zero \citep[\eg,][]{Rodriguez_2018,ArcaSedda_2021b}, even after taking into account spin precession of the merging binary \citep[\eg,][]{Yu_2020}.
A proposed origin of the positive-biased asymmetry in $\xeff$ distribution (also referred to as `symmetry breaking') is a contribution from BBH mergers formed via isolated binary evolution \citep{Belczynski_2020,Bavera_2020,Olejak_2021,Broekgaarden2022,Perigois_2023,Banerjee_2023}. 

In this work, we aim to explore BHs' spins and their alignments in BBH mergers originating from
isolated massive binary evolution, based on evolutionary population synthesis of massive binary stars. 
For this purpose, we utilize an updated version of the rapid binary evolution code $\bse$ \citep{Hurley_2002,Banerjee_2020}
that incorporates BH spins based on stellar angular momentum transport models and binary tidal interaction. We explore
the regimes of BBH formation via both stable mass transfer
(hereafter SMT; \eg, \citealt{Marchant_2021,Gallegos_2021, Olejak_2021a,vanSon_2022,Briel2023,Dorozsmai_2024,Picco_2024}) and common envelope
(hereafter CE) evolution \citep{Belczynski_2016,Eldridge_2016,Stevenson_2017,Kruckow_2018,Spera_2019,Mapelli_2019,Belczynski_2020,Broekgaarden2022b,Boesky2024,Romagnolo2024}. In particular, we show that isolated binary evolution with a fraction of mass-ratio reversed BBH mergers can produce a positive-asymmetric $\xeff$ distribution and a $\xeff-q$ anti-correlation. Moreover, the SMT subchannel simultaneously produces the $\sim 10 M_{\odot}$ peak in the distribution of primary BH reported for the detected population of GW mergers \citep{Abbott_GWTC3_prop}. See the companion paper by \citet{Sen_2025} for further
discussions, also in the context of the Galactic population and possible X-ray binary signatures of such systems.

The possible anti-correlation between $\xeff - q$ trend for BBH mergers formed from isolated binaries, for which the progenitors undergo a mass ratio reversal, has been also explored by \citet{Broekgaarden2022}; see also the work of \citet{Korb_2024} in this context. The origin of the reported $\xeff - q$ anti-correlation from a combination of isolated binary sub-channels (SMT and CE) has recently been proposed and explored in the work by \citet{Olejak_2024}. Alternatively, the contribution of hierarchical mergers produced in active galactic nuclei environment has been suggested as an another possible explanation for the observed $\xeff - q$ anti-correlation, see, e.g., \citet{Santini2023} and \citet{Delfavero2024}.

In this study, we perform isolated binary population syntheses by considering various physical assumptions on binary interaction physics (\eg, regarding mass transfer stability and efficiency) and a wide range of metallicity to study the spin characteristics of merging BBHs. We also explore different recipes for spin-up of the second-born BH, implemented within
the $\bse$ code, which enables a self-consistent treatment of BH spin-up including the effects of stellar wind. Furthermore, we study how different fractions of dynamical-origin BBH mergers and BH spin tossing affect the $\xeff-q$ anti-correlation.

This paper is organized as follows. In Sec.~\ref{method}, we describe the
updates to the $\bse$ code and the isolated binary population syntheses.  
In Sec.~\ref{result}, we present our results, focussing on mass, mass ratio,
and spin of the BBH merger populations as obtained from the model isolated binary
populations. In Sec.~\ref{discuss}, we compare our results with other recent
studies and assess our various assumptions.
In Sec.~\ref{summary}, we summarize our results. In Appendix~\ref{spevol}, we present individual
examples of binary evolution, focussing on tidal spin synchronisation of Wolf-Rayet members.
In Appendix~\ref{dists_more}, we present results for several additional cases. 
In Appendix~\ref{bhspin_app}, we elaborate further 
on the new code implementations. In Appendix~\ref{popsynth_app}, we elaborate further
on practical strategies for evolving a large number of isolated binaries.
In Appendix~\ref{angles}, we derive the composite spin-orbit tilt angles that we use
in this work.
 
\section{Methods}\label{method}

\subsection{The updated $\bse$ code}\label{newbse}

In this study, we develop further on the version of $\bse$ described in \citet{Banerjee_2020}. This previous
version is a direct derivative of the original $\ssebse$ code by \citet{Hurley_2000,Hurley_2002}. The public version
of this branch of $\bse$ incorporates the so-called `rapid' and `delayed' remnant mass models of \citet{Fryer_2012} and 
prescriptions for pulsation pair instability supernova (hereafter PPSN) and pair instability supernova (hereafter PSN),
following those of \citet{Belczynski_2016}. Apart from the standard, momentum-conserving natal kick \citep{Belczynski_2008}
of compact remnants (NSs and BHs). 
the `bi-stability jump' in the wind of OB stars \citep{Vink_2001} is also implemented.
The physics ingredients of binary interaction and orbital evolution
remains the same as in \citet{Hurley_2002}. A private update of this branch also incorporates natal spins
of BHs based on stellar-evolutionary models, namely, the `Geneva', `MESA', and `FM19' BH-spin models as
described in \citet{Banerjee_2020c}; see also \citet{Kamlah_2021}.

In the present work, we perform further developments related to BH spins in $\bse$. We introduce, inside the $\bse$
code, a mechanism for obtaining an elevated BH spin when the BH's progenitor Wolf-Rayet (hereafter WR) star
is a He-MS or He-giant star that remains a very fast rotator until
the time of the BH formation. Typically, irrespective of the rotation rate of the pre-core-collapse star,
differential rotation within the star gives rise to what is known as the Tayler-Spruit 
(hereafter TS) dynamo \citep{Spruit_2002},
that efficiently transports angular momentum from the star's core to its envelope, resulting in a low-spinning
remnant BH. Detailed stellar evolutionary calculations suggest that the magnitude of the resulting BH's dimensionless
spin vector (or Kerr parameter), $\abhzero$, can be as low as
$0.05\lesssim\abhzero\lesssim0.15$ \citep{Belczynski_2020} or even vanishing \citep{Fuller_2019a}.

In the present update of $\bse$, we apply the `MESA' prescription of \citet[][hereafter Be20]{Belczynski_2020} for the TS-dynamo spin,
$\abhzero$.
This recipe summarises the outcomes of rotating MESA stellar models that include TS dynamo (magnetic angular momentum transport):
\begin{equation}
	\abhzero = \begin{cases}
		a_1\mco + b_1 & \text{$\mco\leq m_0$}\\
		a_2\mco + b_2 & \text{$\mco > m_0$}
	\end{cases}
\label{eq:abh0}
\end{equation}
Here, $\mco$ is the carbon-oxygen core mass of the BH-progenitor star in solar masses.
As done in recent studies \citep{Banerjee_2020c,Kamlah_2021}, we use the values of the constants
$a_1$, $b_1$, $a_2$, $b_2$, and $m_0$ as stated in Be20, after mapping them into
piecewise metallicity ranges as in \citet{Morawski_2018}. The value of $\abhzero$, obtained
in this way, lies between $0.05-0.15$ for all metallicities.

However, if the pre-SN WR star rotates very fast, with spin period $\pspin\lesssim1$ day, then the resulting
BH can have a spin that is higher than the spin it would have via TS-dynamo. This is also suggested by
detailed stellar- and binary-evolutionary calculations \citep{Detmers_2008,Kushnir_2016,Qin_2018,Belczynski_2020,Bavera_2020b,Fuller_2022}.
Such a high spin
of a BH's immediate progenitor star can potentially occur in massive binary evolution via the formation
of a tight WR-BH binary, of orbital period $\porb\lesssim1$ day. In such a binary, the WR star,
which is a naked He-MS or He-giant star, potentially spins up to $\pspin \sim \porb$
being subjected to the companion BH's tidal torque, as a part of the binary's tidal synchronization process.
Binary population synthesis (hereafter BPS) studies suggest that such tight WR-BH binary formation can be a
common pathway towards the occurrence of BBH mergers from isolated
massive binary evolution \citep{Belczynski_2020,Bavera_2020b,Korb_2024}. The intermediate
WR-BH binary can form via either CE ejection or SMT between a first-born BH and a core
helium burning (hereafter CHeB) donor star; see, \eg,
\citet{Olejak_2021}. That way, the second-born BH from the spun-up WR star can possess a \emph{natal} spin that
is higher than that due to TS-dynamo, resulting in a higher-$\xeff$ BBH merger.
In the following, we shall refer to such a BH as a `spun-up BH'.

\subsubsection{Spins of tidally spun-up BH}\label{bhspin}

In the present update of $\bse$, we implement two recently proposed analytical
prescriptions for the spin of the second-born, tidally spun-up BH that
express the BH's dimensionless spin magnitude, $\abh$\footnote{In this paper,
we shall use the subscript `fir' or `first' (`sec' or `second') to denote the attributes of a
first-born (second-born) remnant.}, as a function of the WR-BH orbital period, $\porb$.
One is the $\abh-\porb$ relation of Be20 (their Eqn.~15),
which is based on MESA \citep{Paxton_2011,Paxton_2015} 
models of fast and uniformly rotating single He-stars. This recipe is given by:
\begin{equation}
	\abh = \begin{cases}
		1.0       {\rm ~~~~~}                    \text{$\porb<0.1$ day,} & \\
		                            &    \\
		e^{{-0.1(\porb^\prime/{P_0}-1)}^{1.1}} + &  \\
		{\rm ~~} 0.125 {\rm ~~~~~} \text{$0.1{\rm~day}\leq\porb\leq1.3{\rm~day,}$} & \\
		                            &    \\
		\abhzero        {\rm ~~~~~}               \text{$\porb>1.3$ day.}
	       \end{cases}
\label{eq:abh_be20}
\end{equation}
Here, $\porb^\prime$ is the WR-BH orbital period in seconds, $P_0=4000.0$s, and $\abhzero$ is the BH-spin
solely due to TS-dynamo (see above). The Be20 prescription inherently assumes
that a WR-BH binary of $\porb\leq1.3$ day is fully synchronised, with the WR star rotating as
a rigid body with $\pspin=\porb$. 

The second implemented prescription, namely, that of
\citet[][hereafter Ba21]{Bavera_2021}, are based on the grid of WR-BH binaries computed with the MESA code by \citet{Bavera_2020b}.
The latter work has sampled the WR-BH binary masses and orbital parameters from the BPS outcomes of \citet{Neijssel_2019}.
The MESA-modelled WR members of the binaries in \citet{Bavera_2020b} are differentially rotating and
a complete tidal synchronisation is not enforced. The Ba21 prescription is given by:
\\\\
\begin{equation}
	\abh = \begin{cases}
		\alpha\log_{10}^2(\porb/[{\rm day}]) + \\
		{\rm ~~~~~~}\beta\log_{10}(\porb/[{\rm day}]) & \text{$\porb\leq1$ day,}\\

                                                  &         \\
                \abhzero  & \text{$\porb>1$ day,}
	\end{cases}
\label{eq:abh_ba21}
\end{equation}
where
\begin{equation}
\begin{aligned}
\alpha = f(\mwr,\conea,\ctwoa,\cthreea),\\
\beta = f(\mwr,\coneb,\ctwob,\cthreeb)
\end{aligned}
\label{eq:def1}
\end{equation}
with
\begin{equation}
f(\mwr,c_1,c_2,c_3) = \frac{-c_1}{c_2 + \exp(-c_3\mwr/[\Ms])}.
\label{eq:def2}
\end{equation}
Here: $\mwr$ is the instantaneous mass of the WR star in solar masses.
The best-fit values of the constants are $(\conea,\ctwoa,\cthreea) = (0.059305,0.035552,0.270245)$ and
$(\coneb,\ctwob,\cthreeb) = (0.026960,0.011001,0.420739)$ for $\porb$ and $\mwr$
at He-depletion and $(\conea,\ctwoa,\cthreea) = (0.051237,0.029928,0.282998)$ and
$(\coneb,\ctwob,\cthreeb) = (0.027090,0.010905,0.422213)$ for $\porb$ and $\mwr$
at C-depletion. Notably, the two sets of constants result in nearly identical functional
dependence from Eqn.~\ref{eq:def2}.

We implement the above recipes for $\abh$ within the $\bse$ code, as described in Appendix~\ref{bhspin_app}.
This enables taking into account, within the limitations and approximations of $\bse$'s binary-interaction
physics, the effect of the evolution of $\mwr$, $\porb$, and $\pspin$ due to mass loss and nuclear
evolution. For example, it is possible that a WR-BH binary, during its lifetime until the second BH formation,
widens due to rapid wind mass loss from the WR star, desynchronizing and spinning down the star
(due to angular momentum being carried away by the wind), so as to form a less spun-up or non-spun-up BH.
In our version of $\bse$, this can be tracked explicitly for all WR-BH binaries (in general,
for all kinds of binaries).

For the purposes of continuity and numerical stability and to avoid Kerr parameter having
unphysical values, we apply bounds as follows:
\begin{equation}
\begin{aligned}
	& \abhzero = \max \lbrace \abhzero,0 \rbrace,\\
	& \abhzero = \min \lbrace 1, \abhzero \rbrace,\\
	& \abh = \max \lbrace \abh, \abhzero \rbrace,\\  
	& \abh = \min \lbrace 1, \abh \rbrace.
\end{aligned}
\label{eq:lims}
\end{equation}

\begin{table*}[ht]
\centering
\caption{Model assumptions for the various model isolated binary sets computed in this study. The columns from left to right
are, respectively, the model's reference serial number, reference name, BH natal spin ($\abhzero$) choice,  BH spin-up model choice,
$\qcrit$ choice (high $\Rightarrow \qcrit=8$), CE treatment for a HG donor (pessimistic $\Rightarrow$ CE disallowed;
optimistic $\Rightarrow$ CE allowed, as defaulted in BSE), and remnant mass model choice. For each set of choices,
sets of $2\times10^6$ binaries are evolved at the metallicities 0.0001, 0.0002, 0.001, 0.005, 0.01, 0.02. See text for
more details.}
\label{tab:models}
\begin{tabular}{lllllll}
\hline
No. & Model & BH natal spin & BH spin-up & $q_{\rm crit}$ & HG CE & Remnant mass \\
\hline
1 & Be20\_SMT\_delayed & MESA & Be20 & high & pessimistic & delayed \\[3pt]
2 & Be20\_CE\_delayed & MESA & Be20 & BSE default & pessimistic & delayed \\[3pt]
3 & Ba21\_SMT\_rapid & MESA & Ba21 & high & pessimistic & rapid \\[3pt]
4 & Be20\_SMT\_rapid & MESA & Be20 & high & pessimistic & rapid \\[3pt]
5 & Ba21\_default & MESA & Ba21 & BSE default & optimistic & delayed \\[3pt]
6 & Ba21\_SMT\_delayed & MESA & Ba21 & high & pessimistic & delayed \\[3pt]
7 & Ba21\_CE\_delayed & MESA & Ba21 & BSE default & pessimistic & delayed \\[3pt]
8 & Ba21\_SMT\_rapid\_M300 & MESA & Ba21 & high & pessimistic & rapid \\[3pt]
9 & Ba21\_default\_aBH0 & $\abhzero=0$ & Ba21 & BSE default & optimistic & delayed \\[3pt]
10 & Ba21\_CE\_rapid & MESA & Ba21 & BSE default & pessimistic & rapid \\[3pt]
11 & Ba21\_default\_rapid & MESA & Ba21 & BSE default & optimistic & rapid \\
\hline
\end{tabular}
\end{table*}

In particular, the third condition implies that formation of a `spun-down BH' is prevented. Although we adopt
finite but small $\abhzero$ (see above),  the recipe-derived $\abh$ can be smaller than $\abhzero$ for marginal spin-up cases.
However, the underlying MESA computations, on which the above $\abh$ prescriptions are based, never consider
a reduction of BH spin via tidal synchronisation. Therefore, we prevent the formation of spun-down BHs, to avoid occurrences that
are purely due to the properties of the specific fitting formulae. This restriction applies only to a small range of
WR-BH systems with relatively wide $\porb$ that lead to small $\abh$, and is unlikely to influence the current results.

Note that the $\abh-\porb$ relation is applied at both BHs' formation, irrespective of the
BH being first or second born. However, in our evolutionary scenarios leading to a BBH merger,
$\porb$ is typically well above 1 day at the formation of the first born BH,
resulting in a BH of spin $\abhzero$. An exception is a contact
binary between evolved stars with mass ratio $\approx1$. In $\bse$, such a binary is arranged
to enter a double-cored CE phase. If the post-CE WR-WR binary, following a successful
envelope ejection, is tight enough,
both BHs (that would form nearly simultaneously) will be formed spun up, with the
first-born BH's spin $\abhfirst\approx\abh$. Fig~\ref{fig:dcecase}
demonstrates this scenario.

The in-code implementation of spun-up BH formation allows for explicitly testing whether
the WR-star has indeed undergone a significant tidal spin-up, which is key to BH formation with elevated
natal spin. Especially, the Be20 formulation assumes a complete tidal synchronization
of the WR-BH binary and a rigid-body rotation of the WR companion until the second BH formation,
for $\porb=\pspin\leq1.3$ day.
This means that Eqn.~\ref{eq:abh_be20} is as well valid with $\porb$
in its right-hand side replaced by $\pspin$.
We preliminarily apply the Be20 relation in this way to take into account partial
tidal spin-orbit synchronization of the WR-BH binary as it forms and evolves -- in particular, if $\pspin>1.3$ day
despite $\porb<1.3$ day, the WR-star yields a second-born BH of spin $\abhzero$. In the following, we shall refer to
this variant of the Be20 recipe as `Be20 with spin-up check' (hereafter Be20 SC).
See Appendix~\ref{spevol} for further details and some typical examples of tidal spin-orbit synchronization 
in WR-BH systems, in $\bse$.

The Ba21 prescription, on the other hand, is a population-level formula and does
not require spin-orbit synchronization \citep{Bavera_2020b}. 
Therefore, while applying the Ba21 prescription using Eqn.~\ref{eq:abh_ba21},
we do not perform any spin-up check and simply use the value of $\porb$ at the
second BH formation. That way, we track the entire WR-BH phase, when using the Ba21 option. (We do add a spin-up check option for
the Ba21 prescription too, likewise that for Be20, but this option for Ba21
is never used for the results presented in this paper; see the linked $\bse$ code for the details.)

We also evaluate $\abh$ as has originally been done in Be20. For this, we simply use $\porb$ at the formation
of the WR-BH binary and apply Eqn.~\ref{eq:abh_be20}, during the post-processings (see below).

No innermost stable circular orbit (hereafter ISCO) physics \citep{Mondal_2020} has yet been implemented in $\bse$,
which regulates the wind or Roche lobe overflow (hereafter RLO) accretion onto the first-born BH.
However, it has been shown that the overall mass and spin gain
of the first-born BH due to accretion is generally negligible to small, especially
when the accretion rate onto the BH is constrained by the Eddington limit (Be20).
Therefore, in this work, we neglect the first-born
BH's evolution and assume natal mass and spin values for both BHs.

\subsubsection{Binary mass transfer physics}\label{qcr}

With $\bse$'s default mass transfer stability criteria based on donor-to-companion critical mass ratios $\qcrit$ \citep{Hurley_2002}, the vast majority ($>90$\%) of the BBH mergers forms via CE ejection, see e.g. \citep{Olejak_2020,Banerjee_2021b}. Motivated by recent studies revisiting the stability of mass transfer in BBH progenitors, see, e.g., \citet{Pavlovskii_2017} and \citet{Gallegos_2021}, here we test models with much more conservative criteria for CE development than the original $\bse$ one. In particular, we facilitate stable mass transfer contribution by simply overriding the default recipe of $\qcrit$ in $\bse$ and setting a constant, high value for this parameter: 
$\qcrit=\qcrzero=8.0$. This increases the SMT-origin BBH mergers fraction to $>90$\%. This $\qcrit$ is mainly motivated by the maximum (donor-to-accretor) mass ratio for stable mass transfer obtained in the simulations by \citet{Pavlovskii_2017} for isotropic re-emission angular-momentum loss. This value is also motivated by the typical threshold where a binary system may become Darwin unstable, \ie, a binary's tidal synchronisation extracts angular momentum from the orbit in a runaway manner, potentially leading to a stellar merger \citep{Darwin1879, Zahn1977, Hut1980}. The focus on the SMT channel can be further motivated by its lower predicted low-redshift BBH merger rate compared to the CE channel, which is in better agreement with the merger rate inferred from LVK observations \citep{Sgalletta_2025}.

To parallelise with other recent studies/codes \citep{Belczynski_2008,Giacobbo_2018,Riley_2022,Iorio_2023,Chattaraj_2026},
we distinguish between the two models by either allowing or disallowing CE survival of a system with an HG-star donor.
By default, $\bse$ assumes mass transfer and accretion rate to be limited by thermal and dynamical timescales \citep{Hurley_2002}. Here, we additionally introduce an overall fudge factor, $f_a$, that determines the accretion efficiency during a star-star mass transfer phase. This parameter allows to choose a fixed value for mass transfer efficiency and set it according to some observational constraints; see, e.g., \citet{Vinciguerra_2020}. The default accretion in $\bse$ is restored with $f_a=1$. 
We describe these implementations in more detail in Appendix~\ref{bhspin_app}; see also the linked updated $\bse$ code for further information.

\subsection{Initial conditions for binary population}\label{popsynth}

We perform BPS with the above-described, enhanced $\bse$ code. The initial binary
distribution is taken according to \citet{Sana_2012}:
orbital periods within $0.15\leq\log_{10}(P/{\rm day})\leq3.5$ and following the distribution
$f(\log_{10}P) \propto (\log_{10}P)^{-0.55}$ and eccentricities following
$f(e) \propto e^{-0.45}$. The initial binary components are all taken to be
zero age main sequence (hereafter ZAMS) stars, that are distributed according to the Salpeter
mass function, $f(m) \propto m^{-2.35}$, over the range
$5.0\Ms \leq m \leq 150.0\Ms$ and are paired with each other randomly.

For a given set of global model choices (remnant mass, BH-spin, natal kick, and other
$\bse$ input parameters), we evolve six $2\times10^6$-binary sets (generated with different
random seeds) for the metallicities $Z=0.0001$, 0.0002, 0.001, 0.005, 0.01, and 0.02,
with one set per metallicity. That way, a population of $1.2\times10^7$ binaries,
ranging over $0.0001\leq Z \leq0.02$, are evolved per set of physical assumptions.
Table~\ref{tab:models} lists the various model choices with which the binary sets are
evolved. See Appendix~\ref{popsynth_app} for further details.

Next, the compact object mergers within a Hubble time, from the binary-evolutionary
population, are compiled and co-evolved with a metallicity-dependent cosmic star formation history (hereafter SFH)
and a cosmological model to obtain a present-day-observable (\ie, observable at redshift $z\approx0$)
merger population, using the BPS methodology described in \citet{Banerjee_2021b}.
In this study, we apply
the star formation rate-redshift relation of \citet{Madau_2017}, the `moderate-$Z$' metallicity-redshift
dependence of \citet{Chruslinska_2019}, the $\Lambda$CDM cosmological framework with parameters
from the latest Planck results \citep{Planck_2018}, and a detector visibility horizon
of $\zmax=1$. For both the underlying, computed binary-evolutionary populations and the
present-day-observable merger population, we identify, for each individual merger event,
the formation sub-channel, \ie, CE ejection or SMT.

While calculating the evolution of each binary, the modification of the orbital parameters and the
orbital tilt angle at each remnant formation, due to SN natal kick and/or mass loss, are evaluated as formulated in
Appendix A1 of \citet{Hurley_2002} and implemented in the public version of
$\bse$ (inside the subroutine {\tt KICK}). The tilt, $\nu$, of the orbital plane due to a single SN
with a natal kick of magnitude $v_k$ is given by (\citealt{Hurley_2002}; see their Eqn.~A13)
\begin{equation}
\lcap\cdot\hat{L^\prime} = \cos\nu =
	\frac{\vorb\sin\beta - v_k\cos\omega\cos\phi}
	{\sqrt{v_k^2\sin^2\phi+(\vorb\sin\beta - v_k\cos\omega\cos\phi)^2}}.
\label{eq:nu}
\end{equation}
Here $\lcap$ and $\hat{L^\prime}$ are the directions of the pre- and post-SN orbital angular momenta,
$\vec\vorb$ is the relative orbital velocity that makes an angle $\beta$ with the orbital separation vector
$\vec r$, and $\phi$ and $\omega$ are the declination and azimuthal angles of the natal kick vector $\vec v_k$.

Based on the SN tilt angles, $\tiltfirst$ and $\tilt$
(both lying within the range $0-\pi$), corresponding to,
respectively, the first- and second-born remnant, the respective spin-orbit tilts are assigned
as follows:
\begin{equation}
\begin{aligned}
      &	\thlsfirst = \cos^{-1}(\hat{\abhfirst}\cdot\ltwocap)  \\
      &	 = \cos^{-1}(\cos\tiltfirst\cos\tilt - \sin\tiltfirst\sin\tilt\cos\Omega),\\
      &                                    \\
      &	\thls = \cos^{-1}(\hat{\abh}\cdot\ltwocap) = \tilt.
\end{aligned}
\label{eq:tilts}
\end{equation}
Here $\ltwocap$ is the direction of the orbital angular momentum of the BBH right after
the second remnant formation, and $\Omega$ is the azimuthal angle that $\ltwocap$ makes with respect to the
plane containing $\vec\abhfirst$ and $\vec\abh$, taken to be uniform over $[0,2\pi]$. See Appendix~\ref{angles} for a derivation of
Eqn.~\ref{eq:tilts} and additional explanations.
The Eqn.~\ref{eq:tilts} is based on three idealistic assumptions. First, the initial, ZAMS-ZAMS
binary is always perfectly spin-orbit aligned. Second, the newborn BH's spin maintains
the direction of the spin of its parent WR star - \ie, there is no spin-tossing during
the BH formation (but see \citealt{Tauris_2022}; Sec.~\ref{toss}).
Third, it is assumed that in tight WR-BH systems that lead to
GR coalescence within Hubble time, the WR star's spin realigns with the WR-BH binary's orbital
angular momentum via tidal interaction. In fact, we find that practically all
WR-BH progenitors of GW mergers in our computed binary populations are symbiotic.
We then evaluate the $\xeff$ of the mergers from Eqn.~\ref{eq:xeffdef},
mapping the first- and the second-born remnant's attributes to the primary and the secondary
remnant (see Sec.~\ref{intro}).

We evolve the binary populations for two distinct, main cases, with varied options for each case.
In one case, we set $\qcrit=8.0$ to obtain GW mergers predominantly ($>90$\%) via SMT. For these runs,
we select solely the SMT-driven BBH mergers in order to study a pure population of SMT-channel BBH mergers.
In the other case, we adopt $\bse$'s default $\qcrit$ recipe to obtain predominantly ($>95$\%) CE-channel mergers.
For the latter case, we select only the CE-driven BBH mergers. For the sake of brevity, we shall, hereafter, refer to these cases
simply as `SMT' and `CE', respectively. Also, to parallelise with several other studies, we disallow CE evolution
for HG donors (see Sec.~\ref{qcr} and references therein), unless stated otherwise\footnote{The choice of disallowing
CE for HG donors is often referred to as the `pessimistic' case \citep{Dominik_2012,Giacobbo_2018,Riley_2022}.}.
Throughout this work, we restrict to the `$\bse$-defaults' by setting the CE efficiency parameter $\ace=1$
and choosing the option of angular momentum loss during a star-star mass transfer to be with respect to the donor \citep{Hurley_2002};
see Sec.~\ref{discuss} for a discussion.
While these binary populations also yield NS-BH and binary neutron star (hereafter BNS) mergers,
we restrict mainly to BBH mergers in this paper.

\section{Results}\label{result}

\begin{figure*}
\centering
\includegraphics[width = 0.95\textwidth, angle=0.0]{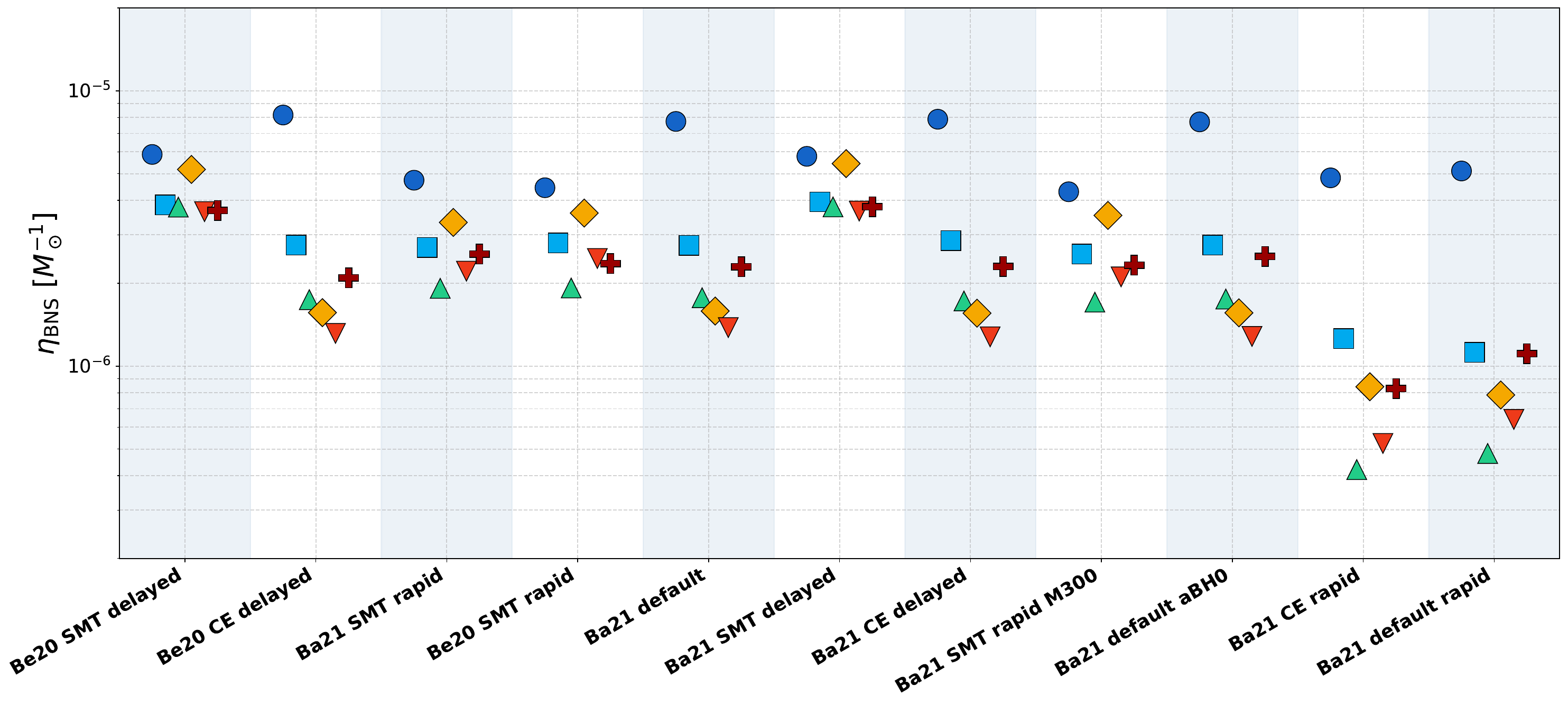}\\
\vspace{-1.55 cm}
\includegraphics[width = 0.95\textwidth, angle=0.0]{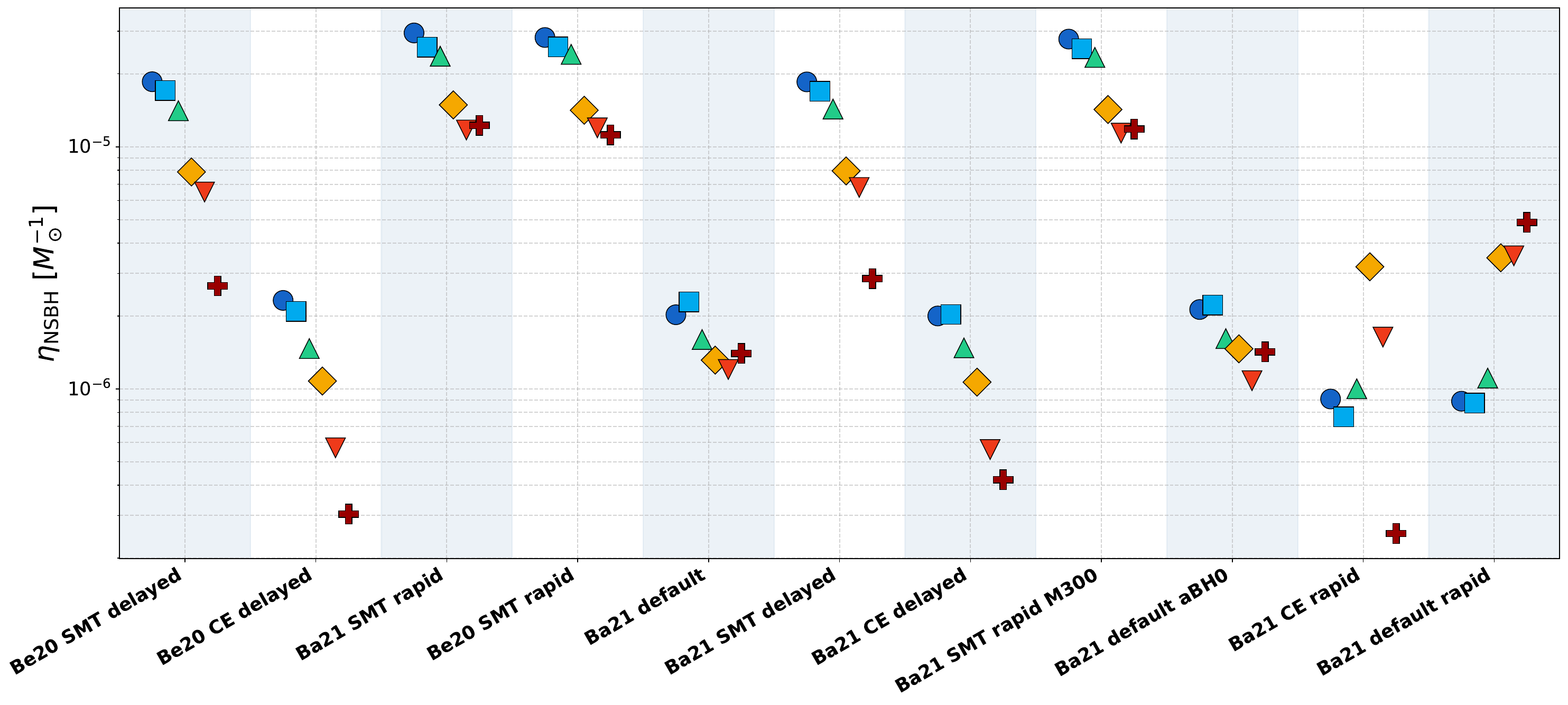}\\
\vspace{-1.55 cm}
\includegraphics[width = 0.95\textwidth, angle=0.0]{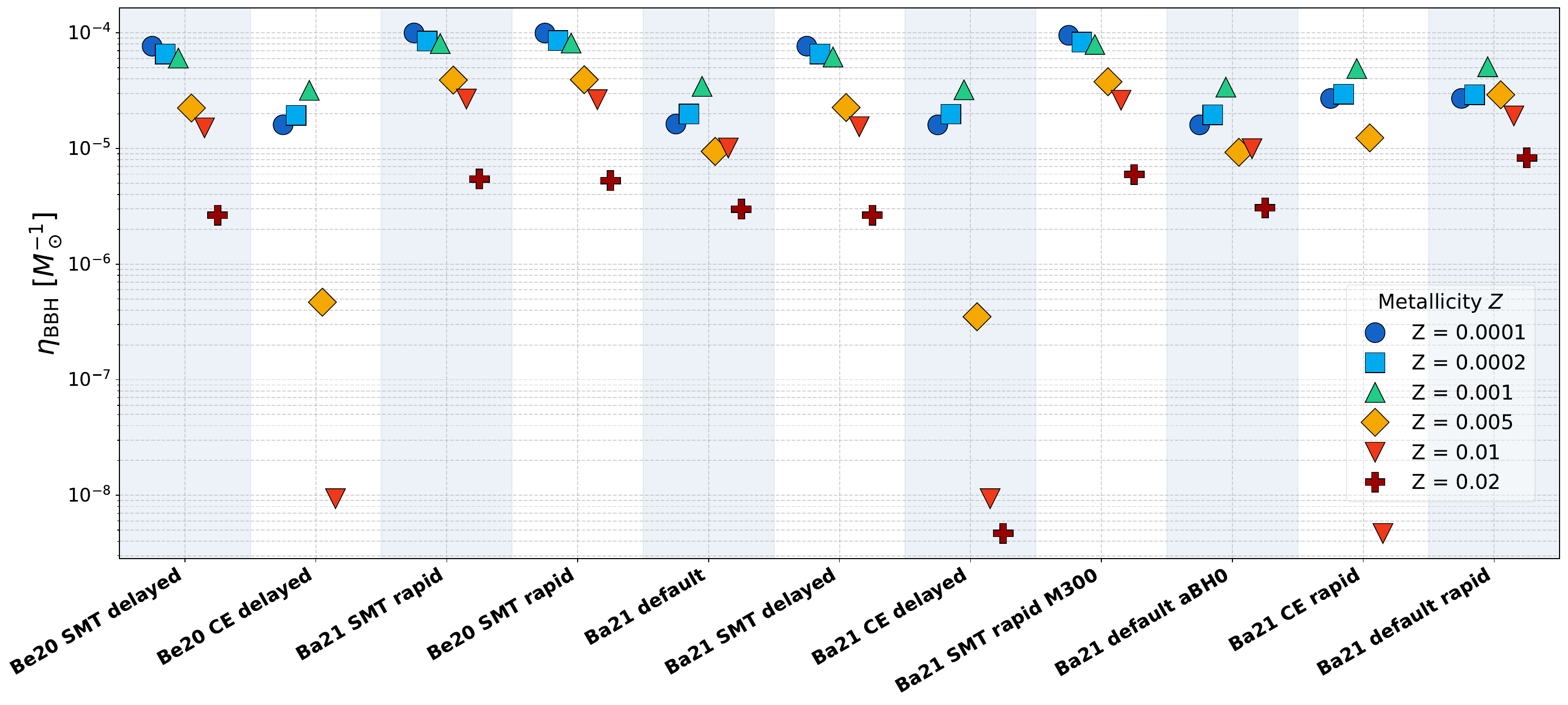}
\caption{Efficiencies of BNS (upper panel), NS-BH (middle), and BBH (lower) mergers, as obtained from
the present model binary population evolutions (Table~\ref{tab:models}).
In each panel, the merger efficiencies at the different
metallicities (legend) are plotted along the Y-axis and they are placed against the corresponding binary-evolution
model indicated along the X-axis.}
\label{fig:eta}
\end{figure*}

\begin{figure*}
\centering
\includegraphics[width = 16.0 cm, angle=0.0]{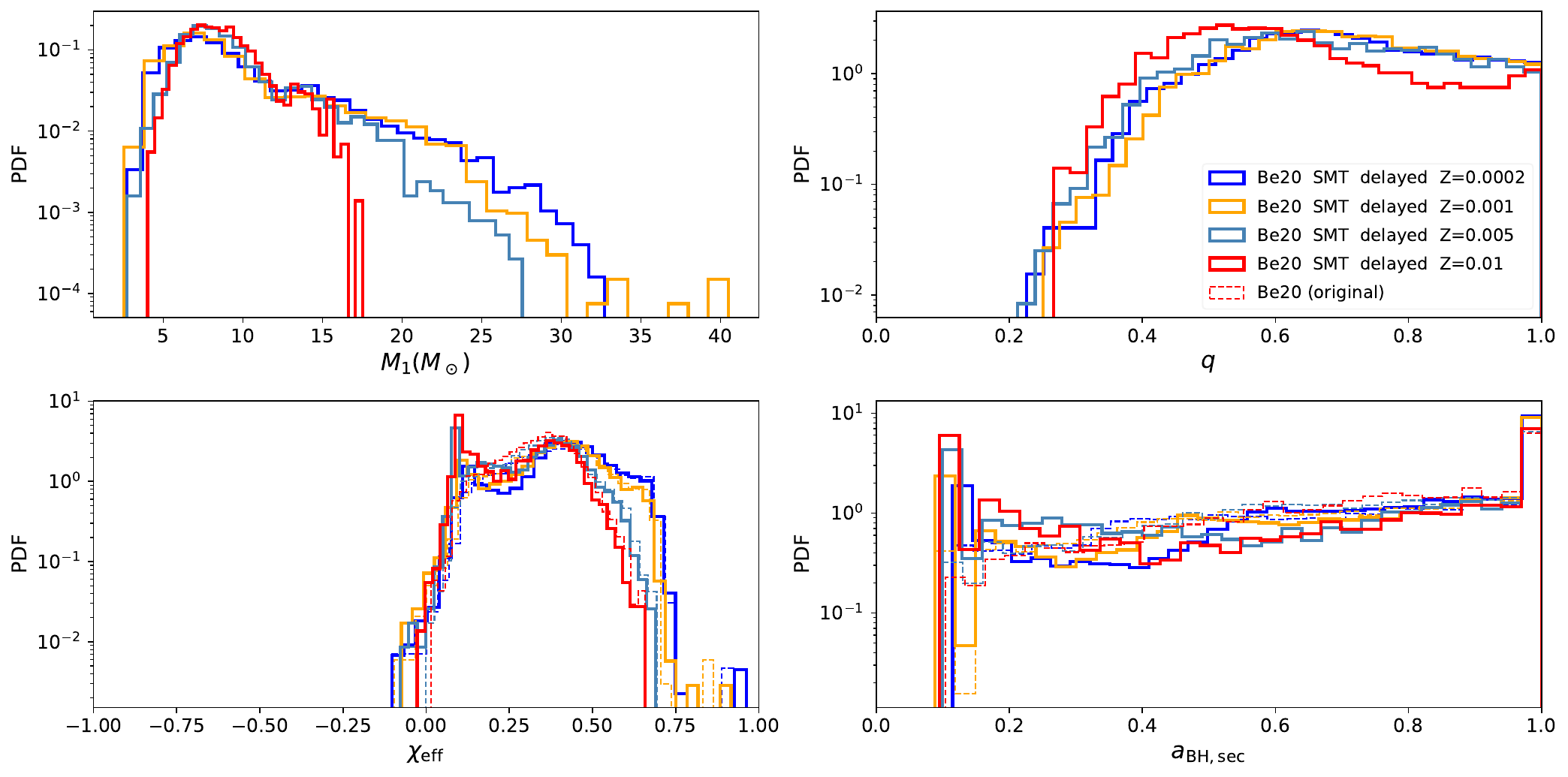}
\includegraphics[width = 16.0 cm, angle=0.0]{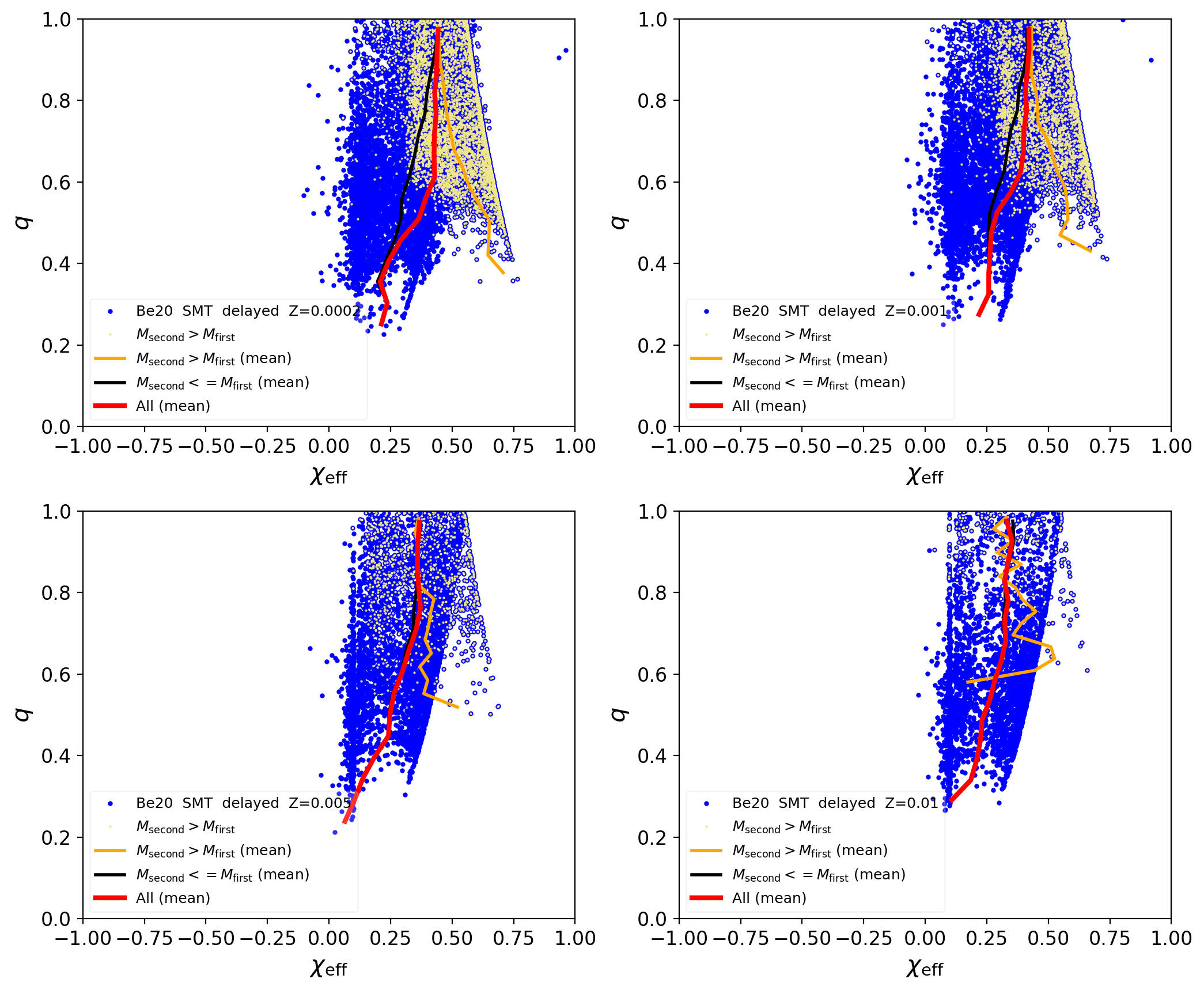}
\vspace{-0.3cm}
\caption{The upper four panels plot the normalized distributions of primary mass, $\mone$,
	mass ratio, $q=\mtwo/\mone$ ($\mtwo\leq\mone$), effective spin parameter, $\xeff$,
	and spin of the spun-up, second-born BH, $\abh$, of only the SMT-channel BBH mergers (see text)
	yielded by the computed isolated binary evolutionary populations with metallicities
	$Z=0.0002$, 0.001, 0.005, and 0.01 (legend). These isolated binary models assume
	the Be20 $\abh-\porb$ prescription and the `delayed'
	remnant mass model. The dashed-lined $\xeff$ and $\abh$ distributions
	are obtained from BH spins determined by applying the original, post-processing approach of Be20.
	The solid-lined $\xeff$ and $\abh$ distributions are obtained from BH spins
	determined at runtime, \ie, without any post-processing, from the tidal spin-up procedure of the $\bse$ code,
	by choosing the option of applying the Be20 $\abh-\porb$ relation after a WR spin-up check
	(see text). The lower four panels
	plot these BBH mergers (blue-filled circles) in the $\xeff-q$ plane
	(the $\xeff$ values being directly from $\bse$)
	for these four metallicities, with one metallicity per panel (legend).
	The subpopulations with the second-born BH, with mass $\msecond$,
	being more massive than the first-born BH, with mass $\mfirst$, are highlighted in the
	$\xeff-q$ scatter plots (khaki-filled circles).
	The loci of mean $\xeff$ as a function of $q$ for the subpopulations
	with $\msecond>\mfirst$ (orange line), $\msecond\leq\mfirst$ (black line), and for the
	overall population (red line) are shown separately on each of the $\xeff-q$ panels.}
\label{fig:dists_wrf2_qc-8_hgf-1}
\end{figure*}

\begin{figure*}[!h]
\centering
\includegraphics[width = 16.7 cm, angle=0.0]{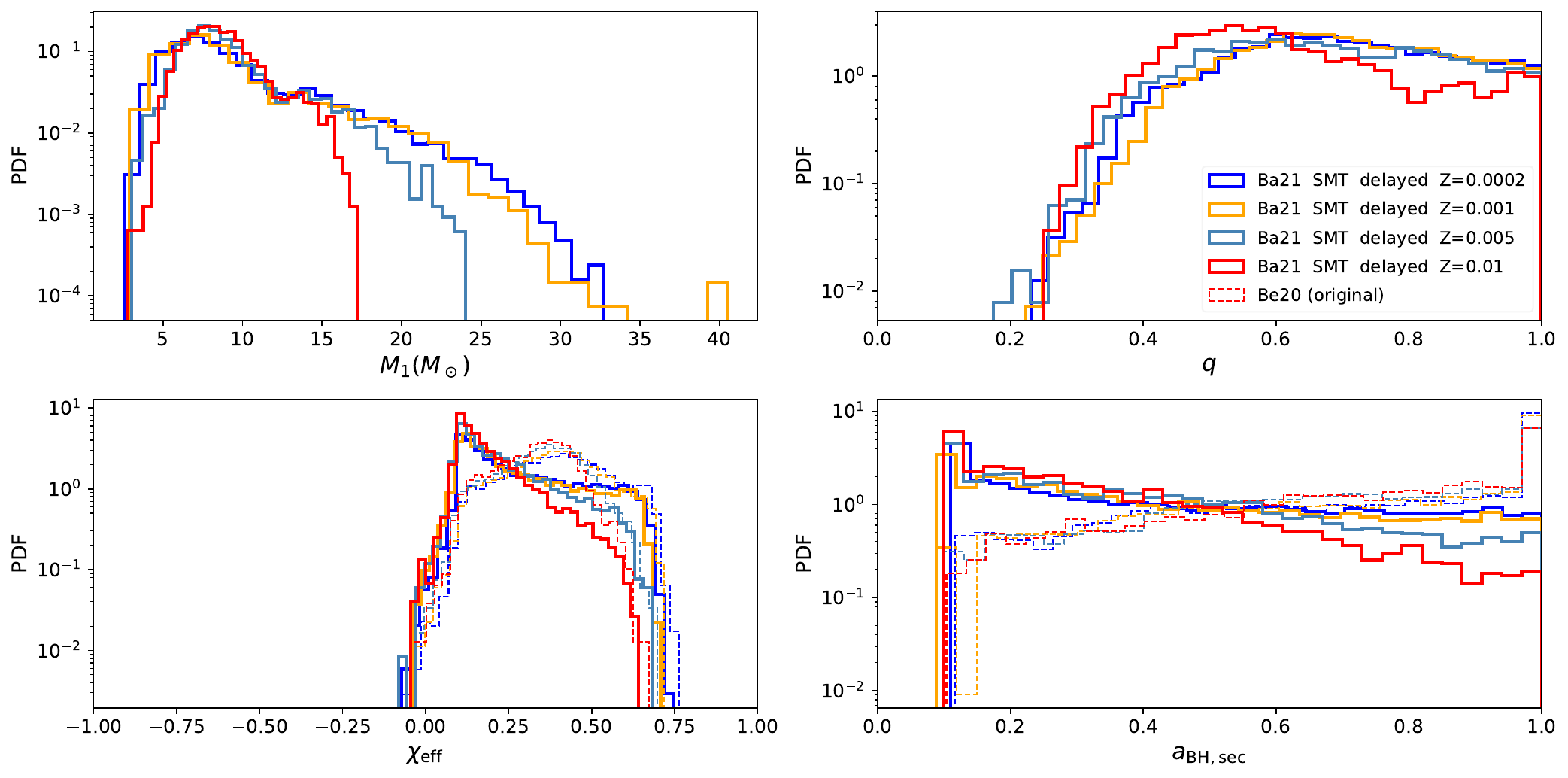}
\includegraphics[width = 16.7 cm, angle=0.0]{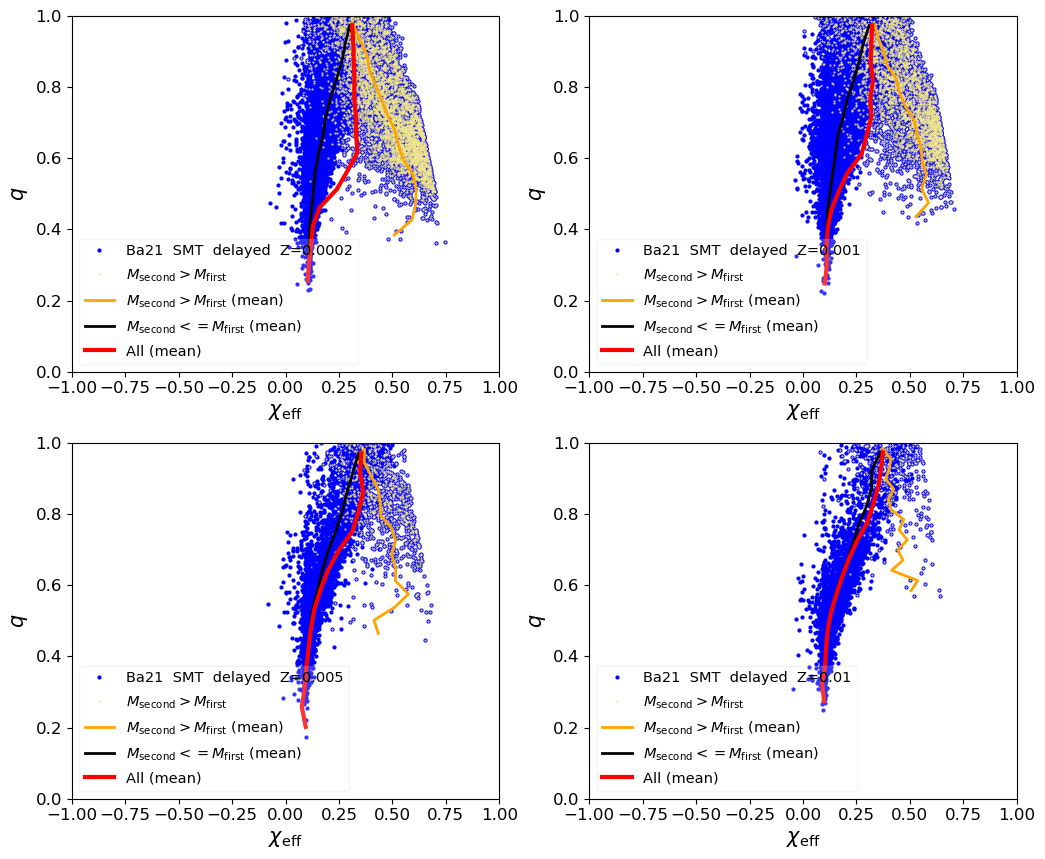}
	\vspace{0.1cm}
	\caption{Same as in Fig.~\ref{fig:dists_wrf2_qc-8_hgf-1}, except that the $\abh-\porb$
	prescription of Ba21 (without spin-up check; see text) is applied when obtaining the
	BH spins from $\bse$ at runtime.}
\label{fig:dists_wrf4_qc-8_hgf-1}
\end{figure*}

\begin{figure*}
\centering
\includegraphics[width = 0.49\textwidth, angle=0.0]{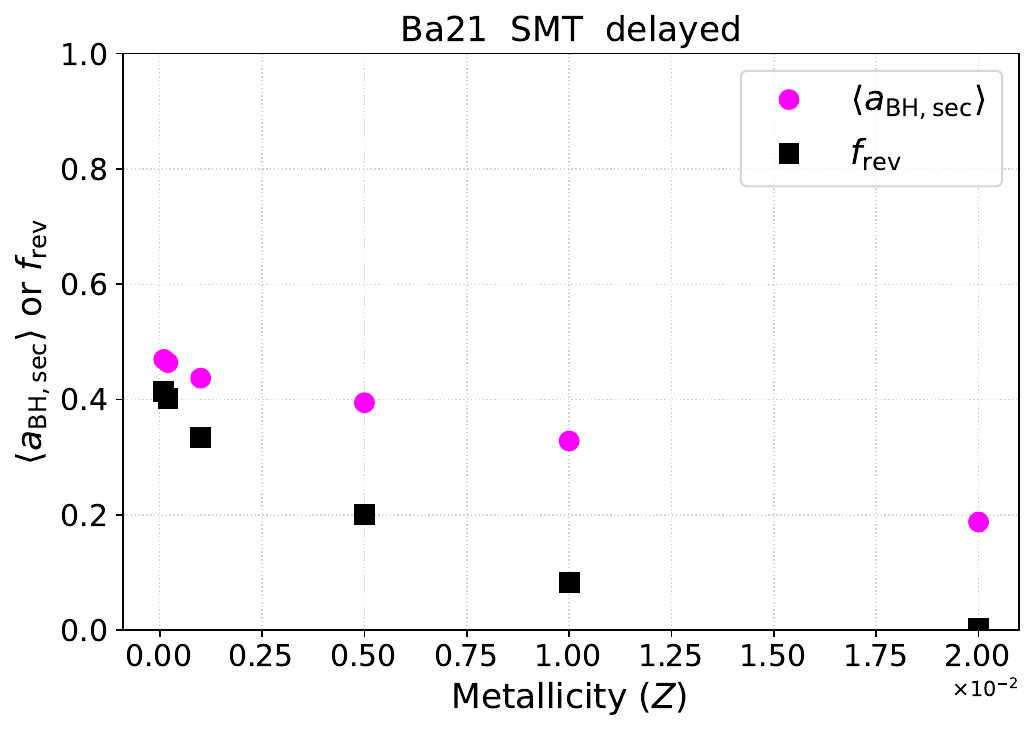}
\includegraphics[width = 0.49\textwidth, angle=0.0]{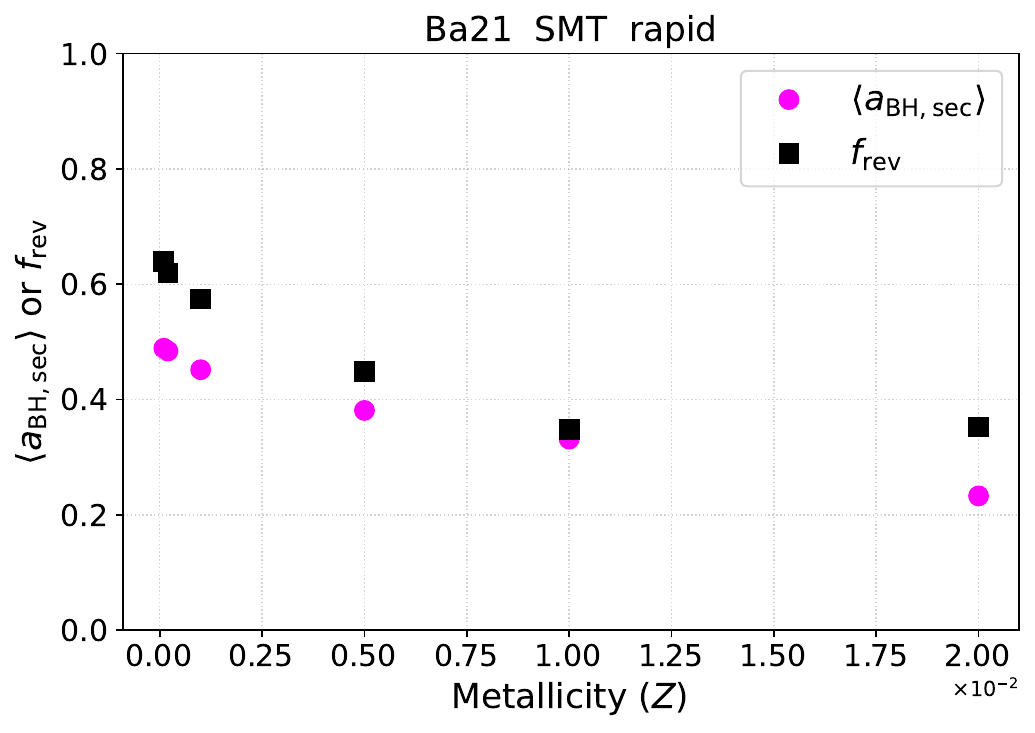}
\includegraphics[width = 0.49\textwidth, angle=0.0]{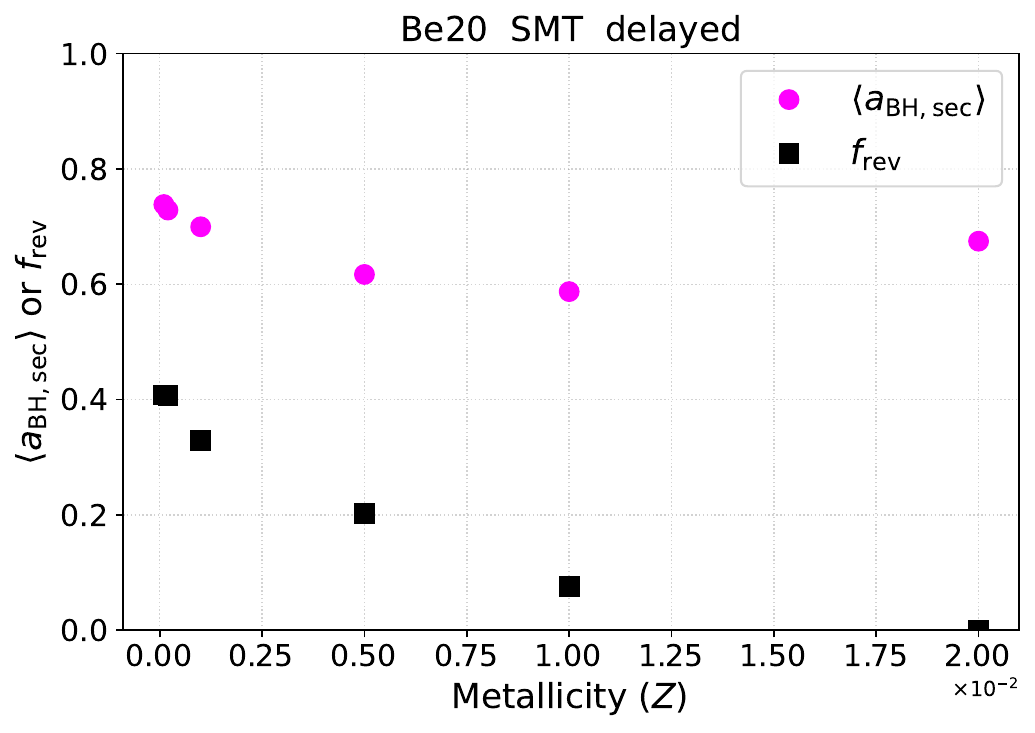}
\includegraphics[width = 0.49\textwidth, angle=0.0]{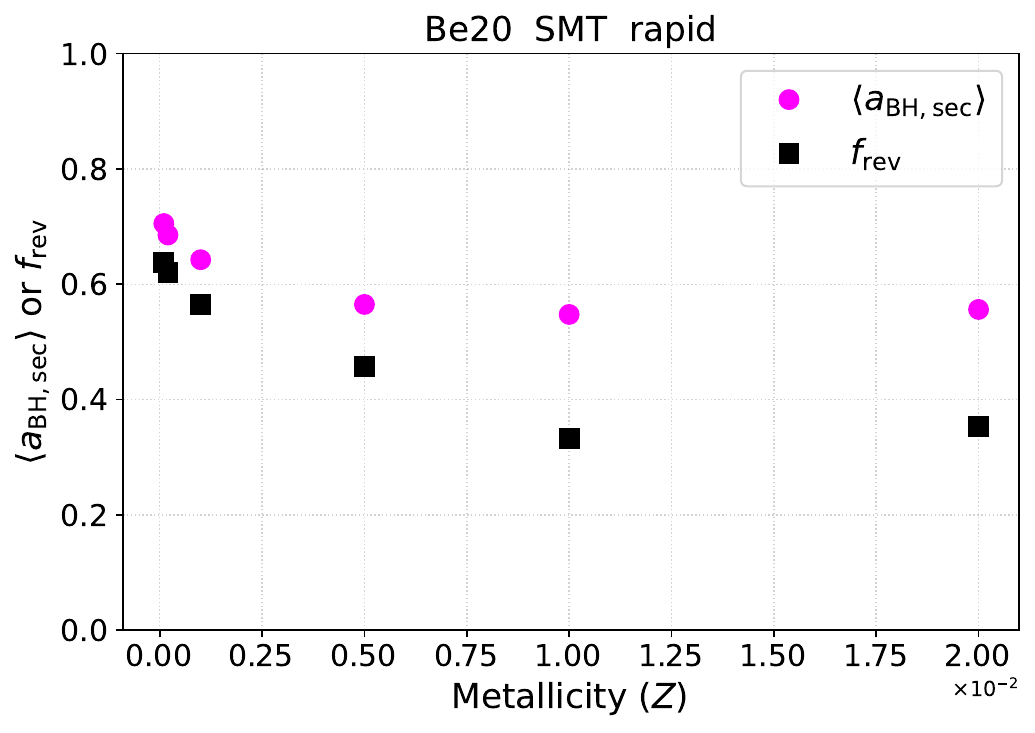}
	\caption{Mean of the spin magnitude (Kerr parameter), $\abhmean$, of the second-born BHs
	and the fraction, $\frev$, of the mass ratio reversed BBH systems (where $\msecond>\mfirst$),	
	for BBH merger events at different metallicities. The four cases considered are mentioned
	in the respective panel's title. 
	}
\label{fig:a2mean_qc-8_hgf-1}
\end{figure*}

In the following, we comparatively explore evolutionary populations of isolated massive binaries incorporating a range of various physical assumptions. In this study, we aim to focus on distribution properties and, hence, do not consider compact-binary
merger rate in detail.
Accordingly, in the following, we shall consider only normalised distributions
(or probability density functions; hereafter PDF). Nevertheless, in Fig.~\ref{fig:eta}, we plot the merger
efficiencies, $\etabns$, $\etansbh$, and $\etabbh$ of, respectively, BNS, NS-BH, and BBH mergers,
as obtained from the present binary-evolutionary models. Here, we define the merger efficiency
of double-compact binaries (hereafter DCB) simply as

\begin{equation}
\eta_{\rm DCB} = \frac{N_{\rm mrg,DCB}}{\mtotst},
\label{eq:etadef}
\end{equation}

where, DCB can be BNS, NS-BH, or BBH, and $N_{\rm mrg,DCB}$ is the total number of DCB mergers produced
from a stellar population of total mass $\mtotst$. For each modelled binary set, we estimate $\mtotst$ by
scaling up the total simulated (binary) mass
assuming a \citet{Kroupa_2001} IMF ranging over $0.08\Ms-150.0\Ms$, except for the sets corresponding to model number 8 (Table~\ref{tab:models}),
where the IMF upper limit is set to $300.0\Ms$. The plotted values correspond to an idealistic, 100\% binary fraction for stars
with ZAMS mass $>5\Ms$. The merger efficiencies are in ballpark agreement with those obtained in recent studies
that utilise different binary evolution models and codes, \eg, \citet{Broekgaarden2022b,Iorio_2023}.

\subsection{Mass distribution of BBH mergers}\label{mqdist}

Fig.~\ref{fig:dists_wrf2_qc-8_hgf-1} (upper four panels) shows the $\mone$, mass ratio $q$, 
$\xeff$, and $\abh$ distributions of BBH mergers for four metallicities and choices of runtime BH spin,
remnant mass, and binary-evolution sub-channel (\ie, either SMT or CE) as indicated in the figure's legend.
This figure (and its following counterparts) presents the entire ensemble of BBH mergers (within a Hubble time)
as obtained for the case and the metallicities specified in the legend. With the delayed remnant mass model,
the $\mone$ distributions reach down to $\approx2.5\Ms$ as this remnant mass model, in contrast to the `rapid' one, allows for the formation of compact objects within the lower mass gap \citep{Fryer_2012}.
In the present remnant scheme in $\bse$,
an NS upper mass limit of $2.5\Ms$ is adopted and all remnants with mass $>2.5\Ms$ (after a
deduction of 10\% due to neutrino mass loss as in \citealt{Belczynski_2008}) are designated as BH.
On the other hand, with the rapid remnant mass model
(\eg, Figs.~\ref{fig:dists_nsf3_wrf4_qc-8_hgf-1},\ref{fig:dists_nsf3_wrf2_qc-8_hgf-1}),
the lower limit of the $\mone$ distribution stays above $5\Ms$ due to the inherent mass gap between $2\Ms-5\Ms$ for this remnant scheme.

Notably, for both delayed and rapid engines, the SMT-channel BBH mergers yield a characteristic
peak in the $\mone$ distribution, at $\approx9\Ms$. This feature for BBH mergers formed via the SMT channel has already been noticed and addressed by, \eg, \citet{vanSon_2022}. This can be seen in all the SMT-only cases
presented in this paper (\eg, Figs.~\ref{fig:dists_wrf4_qc-8_hgf-1} \& \ref{fig:dists_nsf3_wrf4_qc-8_hgf-1}).
A prominent peak at $\approx10\Ms$ (the exact location and shape being dependent on the assumed phenomenological
mass model) in the intrinsic primary mass distribution of the LVK-observed
BBH merger events have been inferred \citep{Tiwari_2021,Abbott_GWTC3_prop}. 
In contrast, the CE-channel mergers comprise more of a plateau in the $\mone$ distribution
whose range depends on the metallicity (Fig.~\ref{fig:dists_wrf4_qcdef_hgfdef} and other CE-only illustrations).
Therefore, the observed $\sim10\Ms$ peak in the LVK primary mass distribution may be revealing
BBH mergers in the Universe originating from SMT in isolated binaries. Notably,
the nature of the primary mass distribution around $10\Ms$ seems to depend rather sensitively on the adopted
stellar- and binary-evolution physics. For example, an opposite trend, \ie, a $\approx10\Ms$ peak
for CE-channel mergers has been found by \citet{Kruckow_2018,Iorio_2023,Briel2023}.

As for the mass ratio distribution, the binary evolution with the delayed remnant model expectedly
produces more asymmetric BBH mergers than with the rapid model
(c.f. Fig.~\ref{fig:dists_wrf4_qc-8_hgf-1} \& Fig.~\ref{fig:dists_nsf3_wrf4_qc-8_hgf-1}).
For the SMT channel, the delayed scheme yields mergers with $q$ reaching down to $\approx0.2$
whereas for the rapid scheme, $q\gtrsim0.4$. The SMT-channel $q$ distributions also exhibit
a broad peak over 0.2 (0.4) $\lesssim q \lesssim 0.8$ for the delayed (rapid) model, which 
is in agreement with recent studies \citep{Belczynski_2020d,Olejak_2024}. These lower limits of $q$ also hold
for CE-channel mergers (Fig.~\ref{fig:dists_wrf4_qcdef_hgfdef}) but the broad peak in the $q$
distribution is generally absent. In general, the SMT-channel BBH mergers from
isolated binary evolution with the delayed remnant scheme can be highly
asymmetric, comparable to the extent of asymmetry in dynamically assembled BBH mergers
from star clusters \citep[\eg][]{Kremer_2020,Rastello_2021,Banerjee_2020c} and
active galactic nucleus (hereafter AGN) disks \citep[\eg][]{McKernan_2020,Vaccaro_2024}.

\begin{figure*}[!h]
\centering
\includegraphics[width = 13.5 cm, angle=0.0]{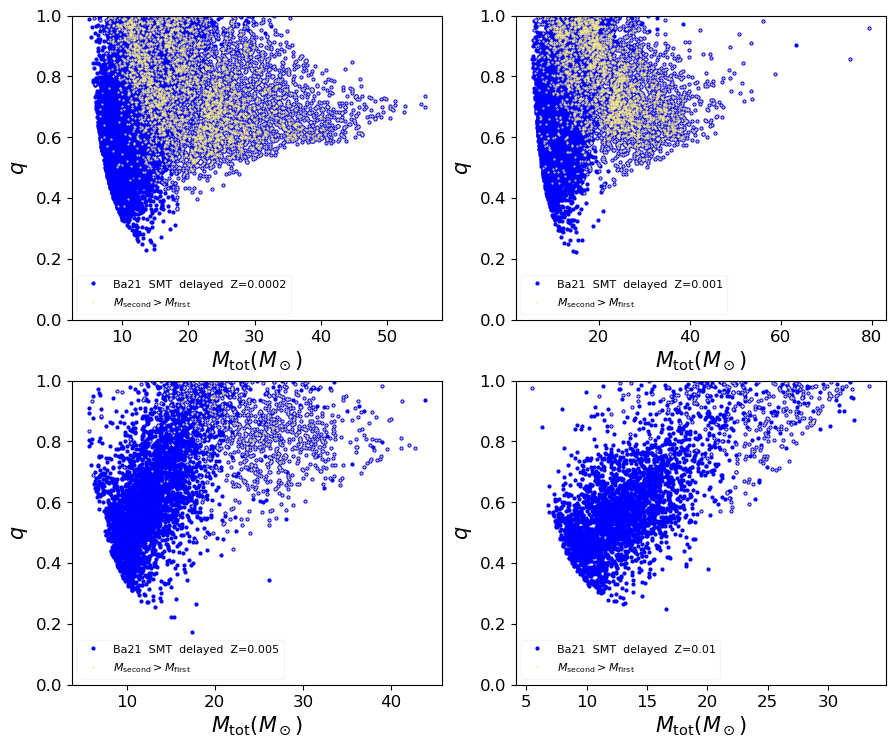}
\includegraphics[width = 13.5 cm, angle=0.0]{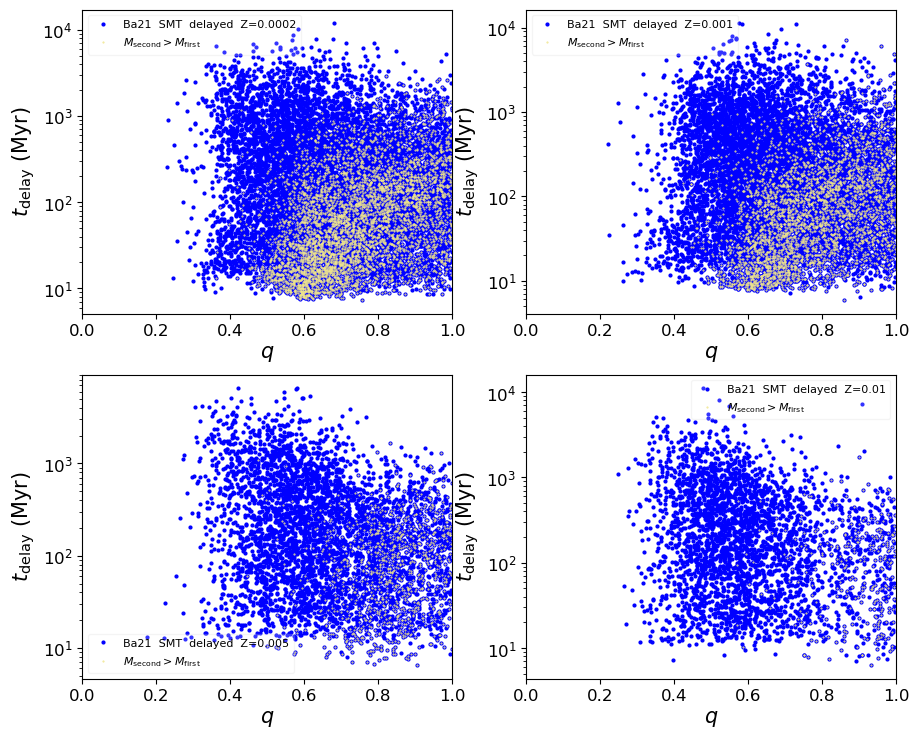}
	\caption{$\mtot-q$ (upper four panels) and $q-\tdelay$ (lower four panels) scatter plots corresponding
	to the BBH-merger events presented in Fig.~\ref{fig:dists_wrf4_qc-8_hgf-1}. Here, $\mtot$ and $\tdelay$ represent
	the total mass and delay time of a BBH merger, respectively. The meanings of the
	symbols are the same as in the other scatter plots. See text for further details.}
\label{fig:dists2_wrf4_qc-8_hgf-1}
\end{figure*}

\begin{figure*}
\centering
\includegraphics[width = 8.5 cm, angle=0.0]{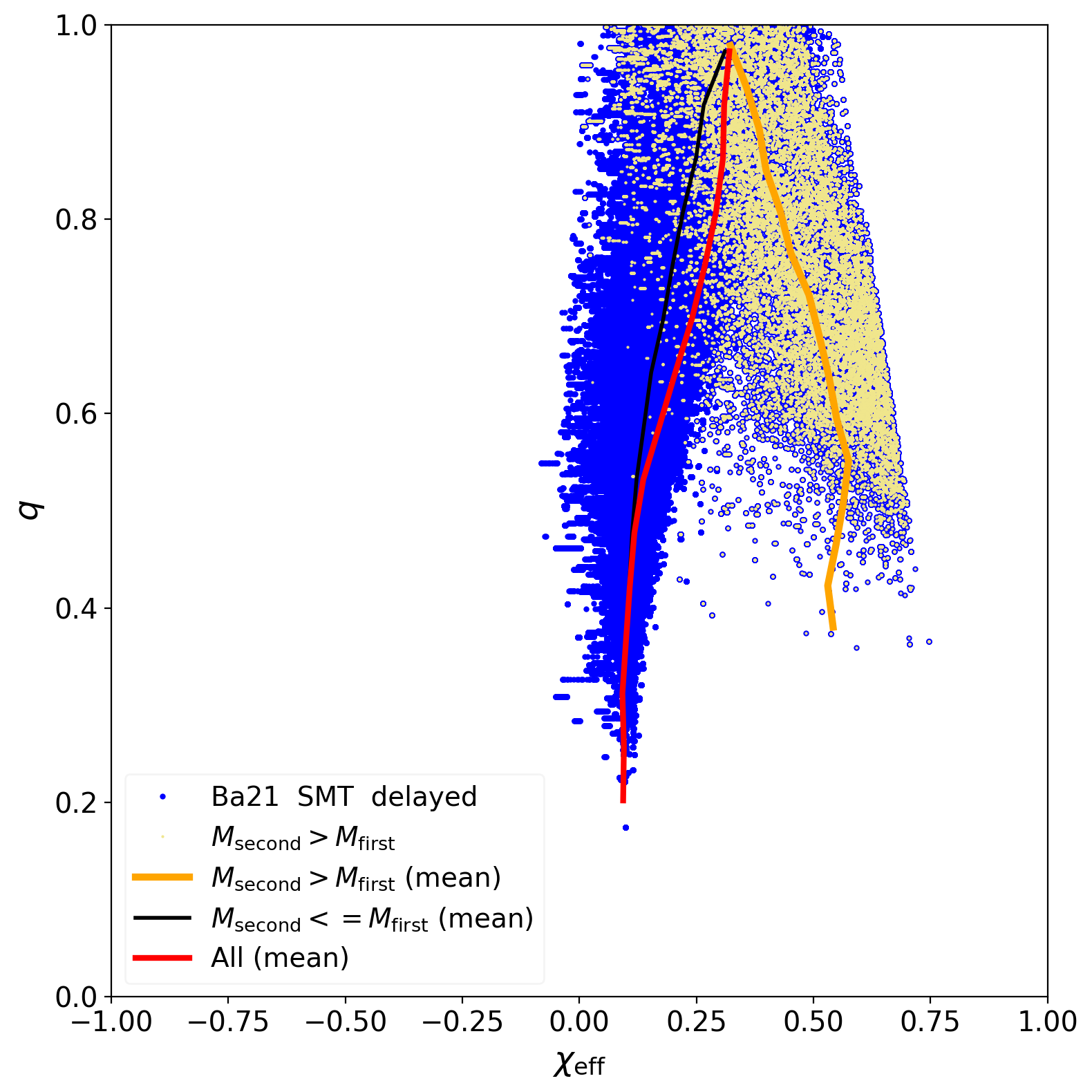}
\includegraphics[width = 8.5 cm, angle=0.0]{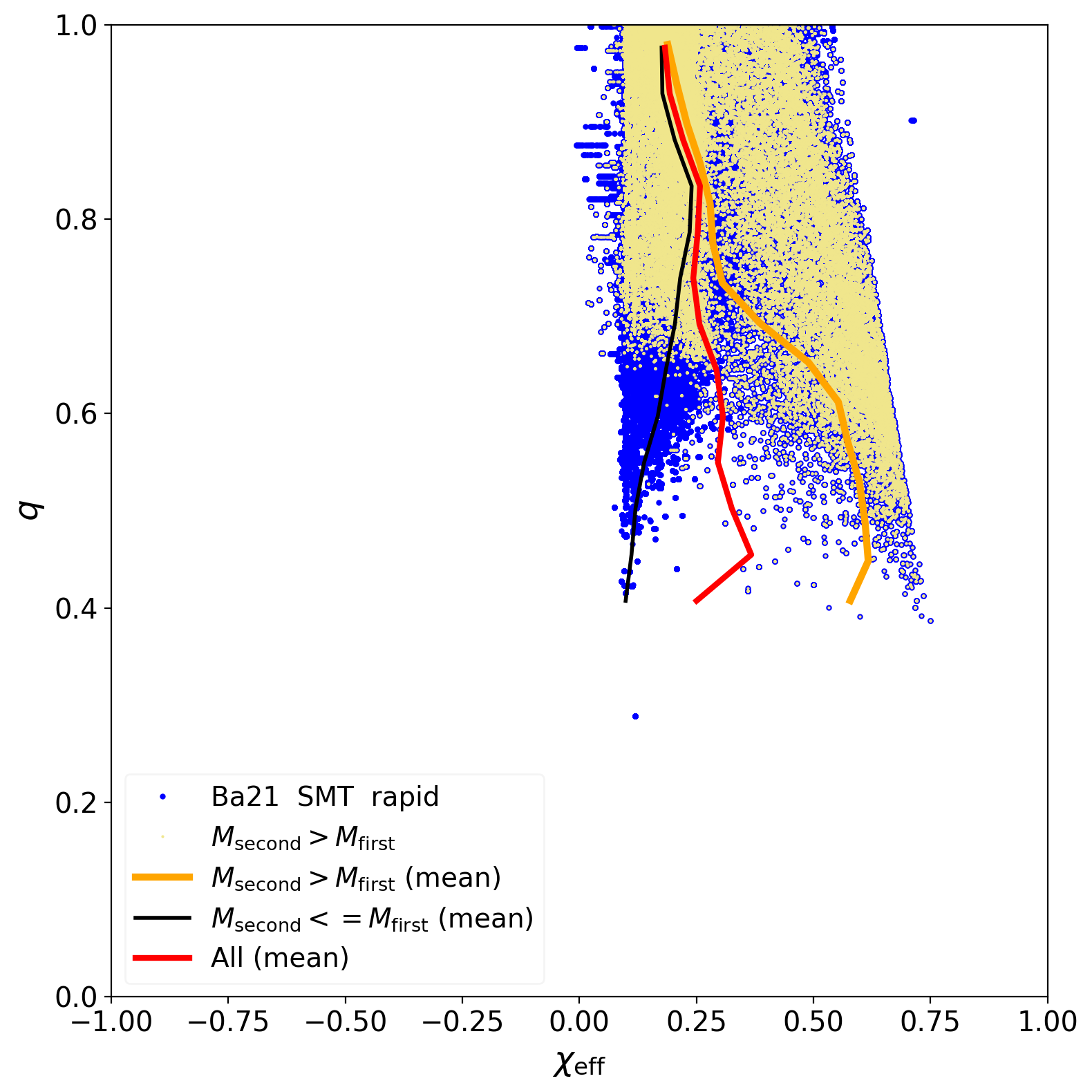}
\includegraphics[width = 8.5 cm, angle=0.0]{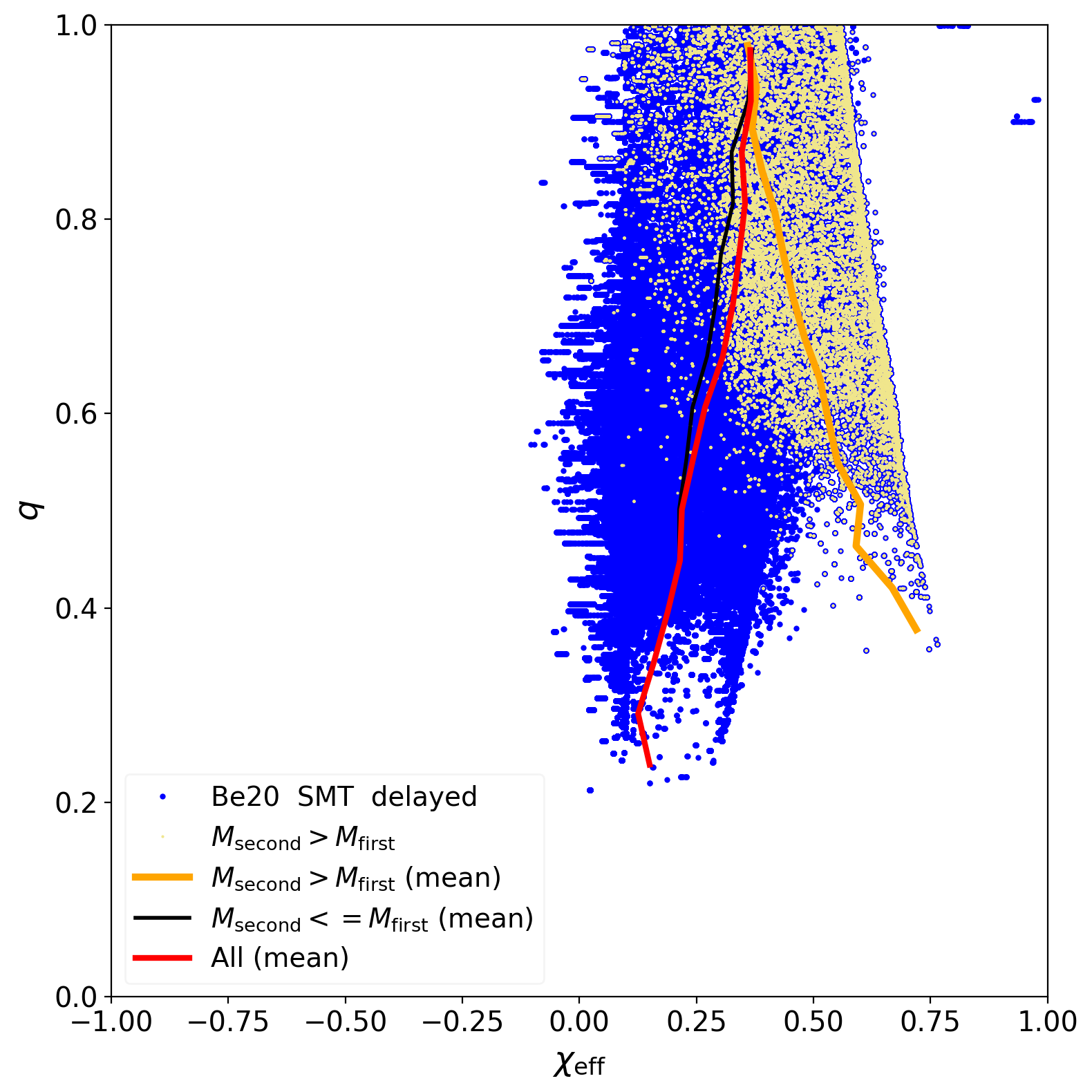}
\includegraphics[width = 8.5 cm, angle=0.0]{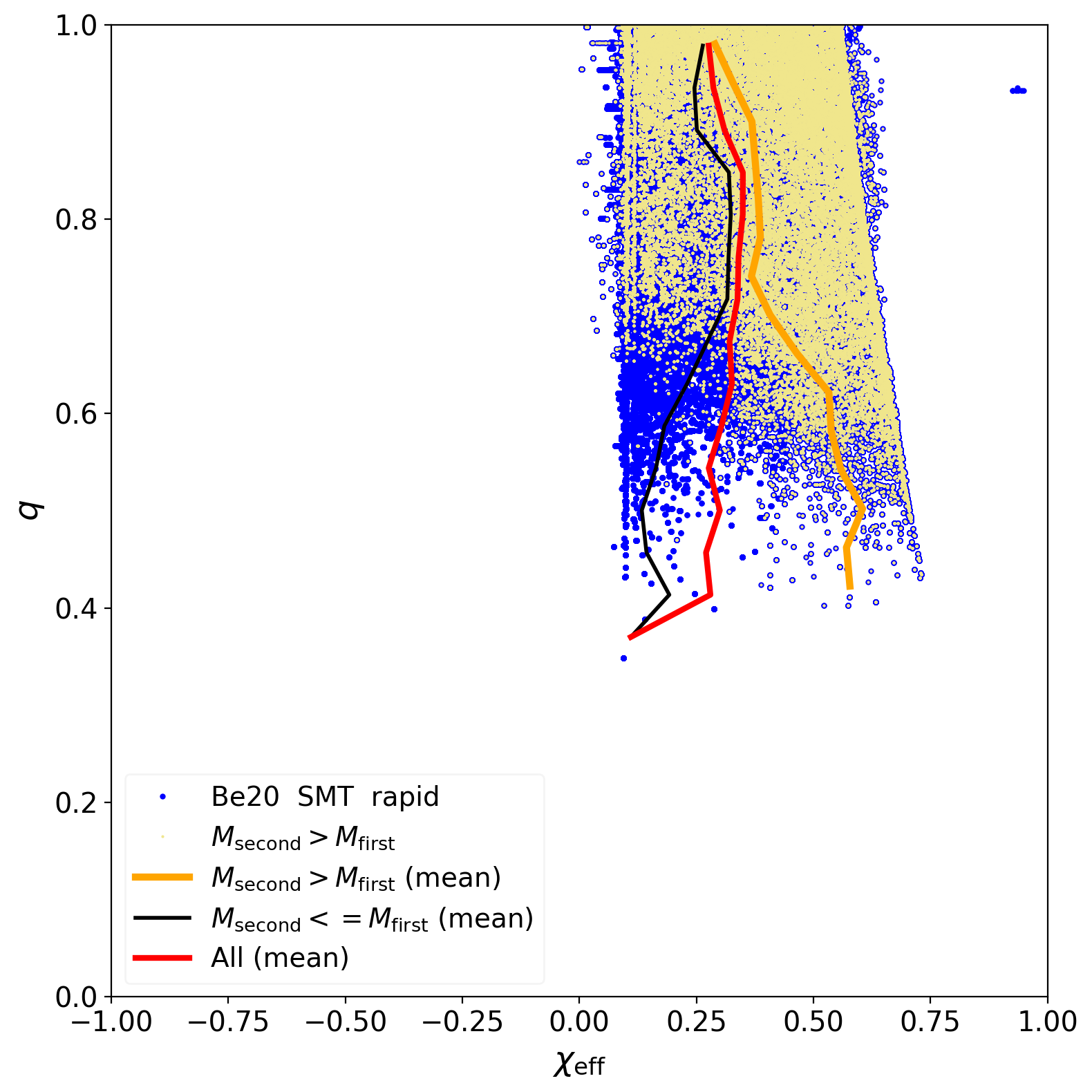}
	\caption{$\xeff-q$ scatter plots of present-day-observable BBH mergers occurring
	via the SMT channel, as obtained from
	isolated binary population syntheses in this study (see text). The choice of runtime BH-spin and remnant
	mass model combination is denoted in each panel's legend description.
	The content and the legend in each panel are the
	same as those in the lower four panels of Fig.~\ref{fig:dists_wrf2_qc-8_hgf-1} -- for detail, see
	the figure's caption.}
\label{fig:dists7_qc-8_hgf-1}
\end{figure*}

\subsection{Spin distribution of BBH mergers}\label{xfdist}

Fig.~\ref{fig:dists_wrf2_qc-8_hgf-1} (second row, solid lines) shows the $\xeff$ and $\abh$ distributions
for SMT-channel BBH mergers, with the Be20 $\abh-\porb$ relation and the WR progenitor's tidal spin-up check
treatment applied in the runtime (Sec.~\ref{bhspin})\footnote{In this study, the Be20 prescription, when applied in the runtime,
always incorporates the spin-up check. Therefore, for brevity, we shall hereafter refer to this treatment
in the text simply as Be20.}. Fig.~\ref{fig:dists_wrf2_qc-8_hgf-1} (second row, dashed lines) also
shows the post-processed $\xeff$ and $\abh$ distributions obtained from the mergers' WR-BH progenitors' initial configurations,
\ie, by applying the original approach in Be20 (Sec.~\ref{bhspin}).
It can be seen that both approaches yield nearly similar distributions at all metallicities,
except that for the runtime implementation, \ie, with tracking of the WR-BH evolution, the $\xeff$ distributions
exhibit a sharp peak between $-0.1\lesssim\xeff\lesssim0.1$. This is due to an excess population of non-spun-up
second-born BHs in the latter approach that are prevented from having a higher natal spin due to the spin-up check.

With the runtime $\abh-\porb$ relation of Ba21 (without spin-up check; see Sec.~\ref{bhspin}), the low-$\xeff$ peak
persists (Fig.~\ref{fig:dists_wrf4_qc-8_hgf-1}). For the Ba21 BH spin-up model, this is due to the
dependence of $\abh$ on $\mwr$ in addition to $\porb$ (Eqn.~\ref{eq:abh_ba21}). As the progenitor WR-BH binary evolves,
$\mwr$ declines and also the binary widens due to the strong wind of the WR star, both diminishing $\abh$.
Overall, with respect to the original Be20 treatment, the more updated (and general) Ba21 BH spin-up formulation produces
a $\xeff$ distribution that is peaked at a lower value ($\xeff\approx0.1$ for Ba21 versus $\approx0.3$ for Be20).
This characteristic is true irrespective of remnant scheme
(delayed or rapid; c.f., \eg, Fig.~\ref{fig:dists_wrf4_qc-8_hgf-1} \& Fig.~\ref{fig:dists_nsf3_wrf4_qc-8_hgf-1})
or formation sub-channel (SMT or CE; c.f., \eg, Fig.~\ref{fig:dists_wrf4_qc-8_hgf-1}
\& Fig.~\ref{fig:dists_wrf4_qcdef_hgfdef}).

\subsection{On the $\xeff-q$ anti-correlation}\label{xeffq}

The lower four panels of Fig.~\ref{fig:dists_wrf2_qc-8_hgf-1} and of its counterparts presented in this paper
show $\xeff$ (X-axis) vs $q$ (Y-axis) of the isolated-binary BBH mergers as scatter plots (blue-filled circles),
for the metallicities and cases as indicated in the respective legends. In each scatter plot, those mergers
are highlighted (with smaller, khaki-filled circles within blue-filled circles) where the second-born BH, with mass
$\msecond$, is more massive than the first-born BH of mass $\mfirst$. See the caption of Fig.~\ref{fig:dists_wrf2_qc-8_hgf-1}
for further details of the figures' description. In practically all cases explored in this paper,
the $\msecond>\mfirst$ subpopulation shows a characteristic $\xeff-q$ anti-correlation, especially
at low metallicities. This property generally applies to Be20 and Ba21 BH-spin models
(\eg, Fig.~\ref{fig:dists_wrf2_qc-8_hgf-1} \& Fig.~\ref{fig:dists_wrf4_qc-8_hgf-1}), delayed
and rapid remnant models (\eg, Fig.~\ref{fig:dists_wrf2_qc-8_hgf-1} \& \ref{fig:dists_nsf3_wrf2_qc-8_hgf-1}
or Fig.~\ref{fig:dists_wrf4_qc-8_hgf-1} \& \ref{fig:dists_nsf3_wrf4_qc-8_hgf-1}), and
SMT and CE channels (\eg, Fig.~\ref{fig:dists_wrf4_qc-8_hgf-1} \& Fig.~\ref{fig:dists_wrf4_qcdef_hgfdef}).

The overall notion of single star evolution suggests that a more massive stellar remnant would form
earlier since its more massive parent star would have a shorter lifetime \citep[\eg,][]{Kippenhahn_2012}. 
However, a $\msecond>\mfirst$ BBH system can occur in binary massive star evolution due to mass ratio reversal during
the first SMT phase in the binary, between the two BH-progenitor stars.
Generally speaking, a lower mass ratio (\ie, higher mass asymmetry) of the BBH implies a larger extent of
mass gain of the second-born BH's progenitor star, which happens to be a more massive donor during the
second SMT to the first-born BH (see illustrations in \citealt{Olejak_2021}). In turn, the greater is the mass
asymmetry during this second SMT, the greater is the orbit shrinkage until the WR-BH formation (via H-envelope stripping
of the donor). A shorter $\porb$
leads to a larger tidal spin-up of the WR star and hence a more spun-up (and more massive) second-born BH. Also, the
second-born BH receives a small spin-orbit tilt, due to its low natal kick owing to significant fallback in a
massive-progenitor SN
\citep[\eg][]{Banerjee_2020} and due to the tightness of the WR-BH binary ($\porb\lesssim1$ day)
\citep{Hurley_2002,Tauris_2017}. That way, a more asymmetric merger generally possesses a higher $\xeff$,
for the $\msecond>\mfirst$ subpopulation.

If the second RLO is unstable so that the binary enters
a CE phase instead (the CE channel), the anti-correlation would weaken,
as seen in Fig.~\ref{fig:dists_wrf4_qcdef_hgfdef}. With the Ba21 BH-spin model applied in this case,
the mass-ratio-reversed subpopulation still exhibits a mild anti-correlation, which is due to the mass dependence
of $\abh$ (Sec.~\ref{bhspin}; Eqn.~\ref{eq:abh_ba21}) --- a more massive second born BH from a WR star of
larger $\mwr$ generally receives both a lower natal kick/tilt and a higher $\abh$.
Notably, the morphology of the $\xeff-q$ scatter for the reversed population in Fig.~\ref{fig:dists_wrf4_qcdef_hgfdef}
agrees qualitatively with the counterpart in \citet[][their Fig.~9]{Broekgaarden2022}.
The latter BBH-merger population has been obtained from population syntheses using the
fast binary evolution code {\tt COMPAS} \citep{Riley_2022} and, as in Fig.~\ref{fig:dists_wrf4_qcdef_hgfdef},
corresponds to $\abh$ based on Ba21 and the CE formation channel that allows CE for HG donors (the ``optimistic CE'' channel).
Since, in the present work we by default adopt a small, non-zero $\abhzero$ (Sec.~\ref{bhspin}) instead of $\abhzero=0$ as
in \citet{Broekgaarden2022}, the vertical accumulation of non-spun-up mergers is
at a small, positive $\xeff$ value in our case instead of being exactly at $\xeff=0$.
Of course, if $\abhzero=0$ is chosen instead, this vertical accumulation shifts to $\xeff=0$
as well for the present binary-evolutionary model. This is demonstrated in Fig.~\ref{fig:dists_bhf4_wrf4_qcdef_hgfdef}.
In this context, see also \cite{Korb_2024}.

Notably, the overall BBH merger population often doesn't exhibit an anti-correlation
but a positive $\xeff-q$ correlation instead (the red line), unlike the $\msecond>\mfirst$ subpopulation. This is driven mainly
by the positive trend of the complementary $\msecond\leq\mfirst$ subpopulation (the black line). Among all
the cases presented in this paper, only the cases with rapid remnant model and low metallicity ($Z\leq0.001$)
exhibit the $\xeff-q$ anti-correlation also for the overall BBH-merger population
(see illustrations in Appendix~\ref{dists_more}). Only in such cases,
the mass-ratio-reversed subpopulation is adequately abundant to cause an overall $\xeff-q$ anti-correlation.
On the other hand, the delayed remnant model is
of high interest since it naturally produces low mass (few $\Ms$) BHs and remnants within the lower mass gap (Sec.~\ref{mqdist}).
In fact, several observed GW events \citep{Unequal_masss_2020,LowerMassGap2024,GWTC4a_cat} and the
microlensing event OGLE-2011-BLG-0462 \citep{Lam_2022} have been associated with low-mass BH members or mass-gap objects.
However, depending on the carbon-oxygen core mass, a substantial fraction of the second-born BHs is
of low mass and high natal kick for the delayed scheme,
relatively suppressing the production of $\msecond>\mfirst$, aligned-spin BBH systems for this scheme.
Until now, based on the updated observed GW-event population, it is unclear whether an overall $\xeff-q$ dependency actually exists \citep{GWTC4a_pop}.
Hence, the above observation regarding the role of remnant-mass model in producing a $\xeff-q$ anti-correlation, although interesting
by its own right, does not prefer the rapid over the delayed remnant model.

Fig.~\ref{fig:a2mean_qc-8_hgf-1} shows the metallicity dependence of the mean spin magnitude (Kerr parameter), $\abhmean$,
of the second-born BH (blue-filled circles) and the fraction, $\frev$, of mass-ratio-reversed BBHs (black-filled squares),
for isolated-binary BBH mergers. The case corresponding to each panel is indicated in the panel's title.
In all cases, $\abhmean$ declines with increasing $Z$: overall, a more metallic WR-BH binary,
owing to a larger wind mass loss, widens to a larger extent until the second BH formation,
leading to a weaker tidal spin up of the WR member and hence a smaller spin of the second-born BH.
Qualitatively, this is in accordance with the predictions by \citet{Detmers_2008}.
However, the present isolated-binary models do not incorporate any rotation driven wind enhancement of the WR
member and the consequent widening of the binary, as also suggested by these
authors. As seen in Fig.~\ref{fig:a2mean_qc-8_hgf-1}, with the Be20 BH spin scheme, $\abhmean$ declines with $Z$
until $Z=0.005$ and then nearly plateaus for higher $Z$. On the other hand, with the Ba21 scheme,
the decay of $\abhmean$ is stronger and continues until $Z=0.02$. The effect amplifies for the
Ba21 case due to the mass dependence of $\abh$ in this case (Eqn.~\ref{eq:def2}) unlike
Be20 (Eqn.~\ref{eq:abh_be20}): generally, a stronger
wind at a higher $Z$ results in a smaller $\mwr$ and hence a smaller $\abh$. In other words, based
on the underlying MESA models of Be20 and Ba21, the WR star's mass dependence has a stronger impact, on 
the merging BBH's $\abh$, than orbit widening of the progenitor WR-BH binary. As seen in Fig.~\ref{fig:a2mean_qc-8_hgf-1},
$\abhmean$ does not show a notable dependence on remnant mass model (delayed or rapid).

On the other hand, $\frev$ exhibits strong dependence on both metallicity and remnant model, as Fig.~\ref{fig:a2mean_qc-8_hgf-1}  
demonstrates. In particular, the rapid model yields a much higher $\frev$ among merging BBHs at high metallicities compared
to that from their delayed-model counterparts. This essentially translates into a higher extent of $\xeff-q$ anti-correlation
in the BBH merger population from the isolated binaries incorporating the rapid model, as discussed above.

It is important to note that the $\xeff-q$ trend for the mass ratio reversed merger population, as obtained here,
is a general trend and not a property of only the most massive and asymmetric mergers. This is demonstrated
in Fig.~\ref{fig:dists_nsf3_wrf4_qc-8_hgf-1_mu300} that shows a case where PPSN/PSN is disabled to allow
for BH formation within the `upper mass gap' of $\approx45\Ms-\approx120\Ms$, at low metallicities.
This case exhibits an anti-correlation to a similar extent as its counterpart with PPSN/PSN-enabled
(Fig.~\ref{fig:dists_nsf3_wrf4_qc-8_hgf-1}; as in all other runs in this paper). Fig.~\ref{fig:dists_nsf3_wrf4_qc-8_hgf-1_fa}
demonstrates that although a reduced accretion efficiency, $f_a=0.5$ (Sec.~\ref{qcr}), during the first (star-star) mass
transfer expectedly diminishes the relative population of $\msecond>\mfirst$ mergers, the
anti-correlation among such mergers still holds. All other runs in this paper use
$f_a=1$, as defaulted in $\bse$.

Fig.~\ref{fig:dists2_wrf4_qc-8_hgf-1} shows the BBH-merger events in Fig.~\ref{fig:dists_wrf4_qc-8_hgf-1}
in the $\mtot-q$ and $q-\tdelay$ planes, $\mtot$ and $\tdelay$ being the events' total mass and
delay time, respectively. As seen, overall, for the $\msecond>\mfirst$ subpopulation, $\mtot$ and $q$ are
anti-correlated, whereas $\tdelay$ and $q$ are positively correlated. These trends and the
$\xeff-q$ anti-correlation of the reversed population are mutually consistent. Generally, 
for the reversed population, higher $\mtot$ mergers are more asymmetric (\ie, of lower $q$),
which property is driven by $\msecond$, that also amplifies the spun-up $\abh$,
hence having higher $\xeff$ as well (Eq.~\ref{eq:xeffdef}).
More massive mergers generally possess shorter $\tdelay$s, leading to the $q-\tdelay$ positive correlation.
Such correlations can be looked for in the current and forthcoming GW event population \citep[\eg][]{Fishbach_2021}.

All the properties of mass, mass ratio, and spin distributions of BBH mergers, as discussed above, map into 
the corresponding distributions of the present-day-observable BBH merger populations (Sec.~\ref{popsynth}). 
The $\mone$, $q$, $\xeff$, and $\abh$ distributions of the intrinsic population of mergers observable at
$z\approx0$ are shown in Fig.~\ref{fig:dists1_wrf4_qc-8_hgf-1_pop} (blue-lined histograms) and its various counterparts
presented in Appendix~\ref{dists_more}. The $\xeff-q$ scatter plots of the present-day-observable
population in selected cases are presented in Figs.~\ref{fig:dists7_qc-8_hgf-1} and \ref{fig:dists7_qcdef_hgf-1}.
In each figure, the case considered is indicated in the legend and it corresponds to that for one of the metallicity-wise
illustrations presented in this paper. As seen, the qualitative properties of the individual-metallicity
populations apply as well to the corresponding observable population. In
particular, the $\xeff-q$ anti-correlation always applies for the $\msecond>\mfirst$ subpopulation;
only for the case with Ba21 BH spin up and rapid remnant models, the overall population
is also anti-correlated in the $\xeff-q$ plane (Fig.~\ref{fig:dists7_qc-8_hgf-1}, upper right panel).

\subsection{Mixing in dynamical, isotropically oriented BBH sources}\label{dynmix}

\begin{figure*}
\centering
\includegraphics[width = 8.75 cm, angle=0.0]{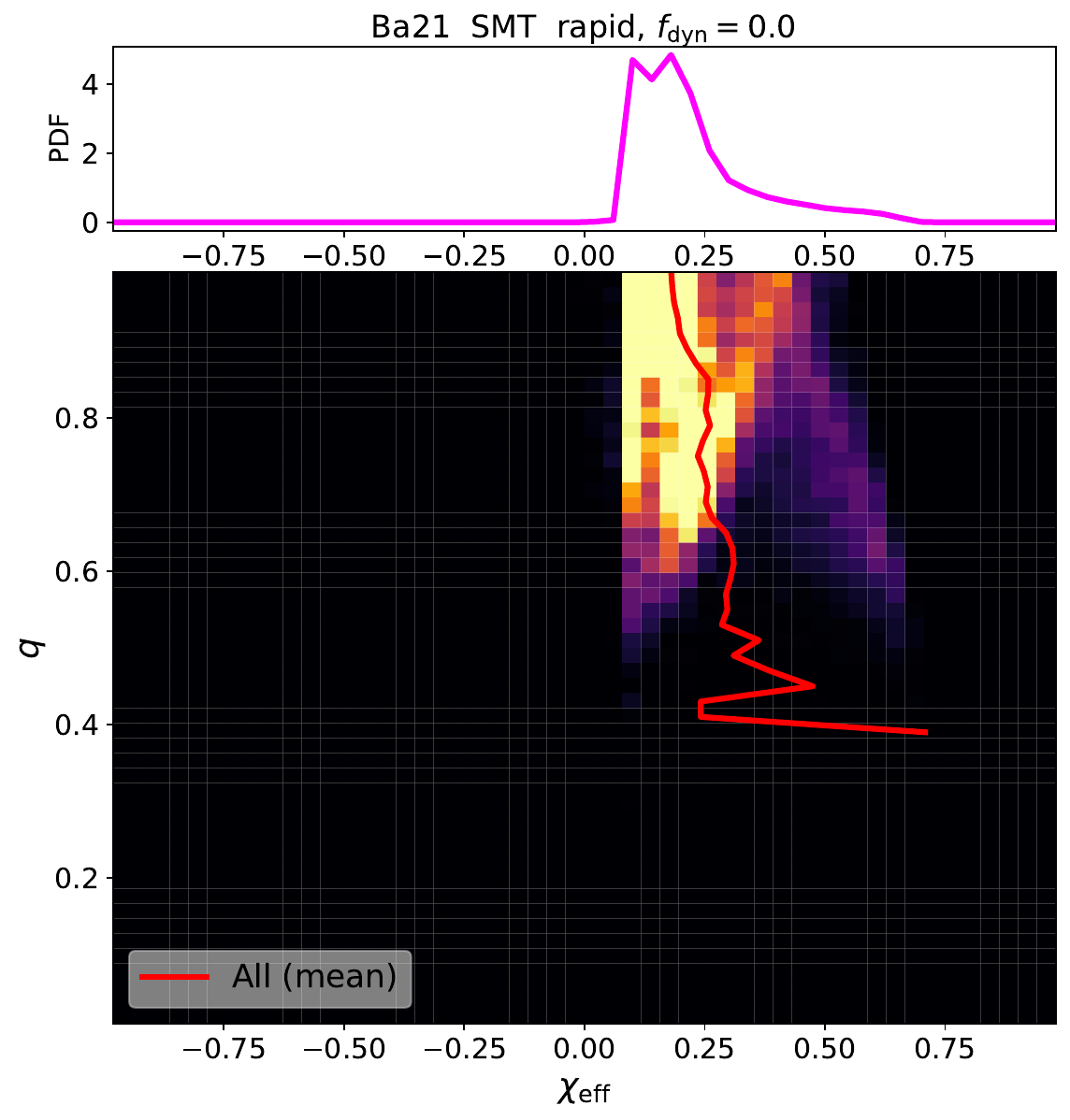}
\includegraphics[width = 8.75 cm, angle=0.0]{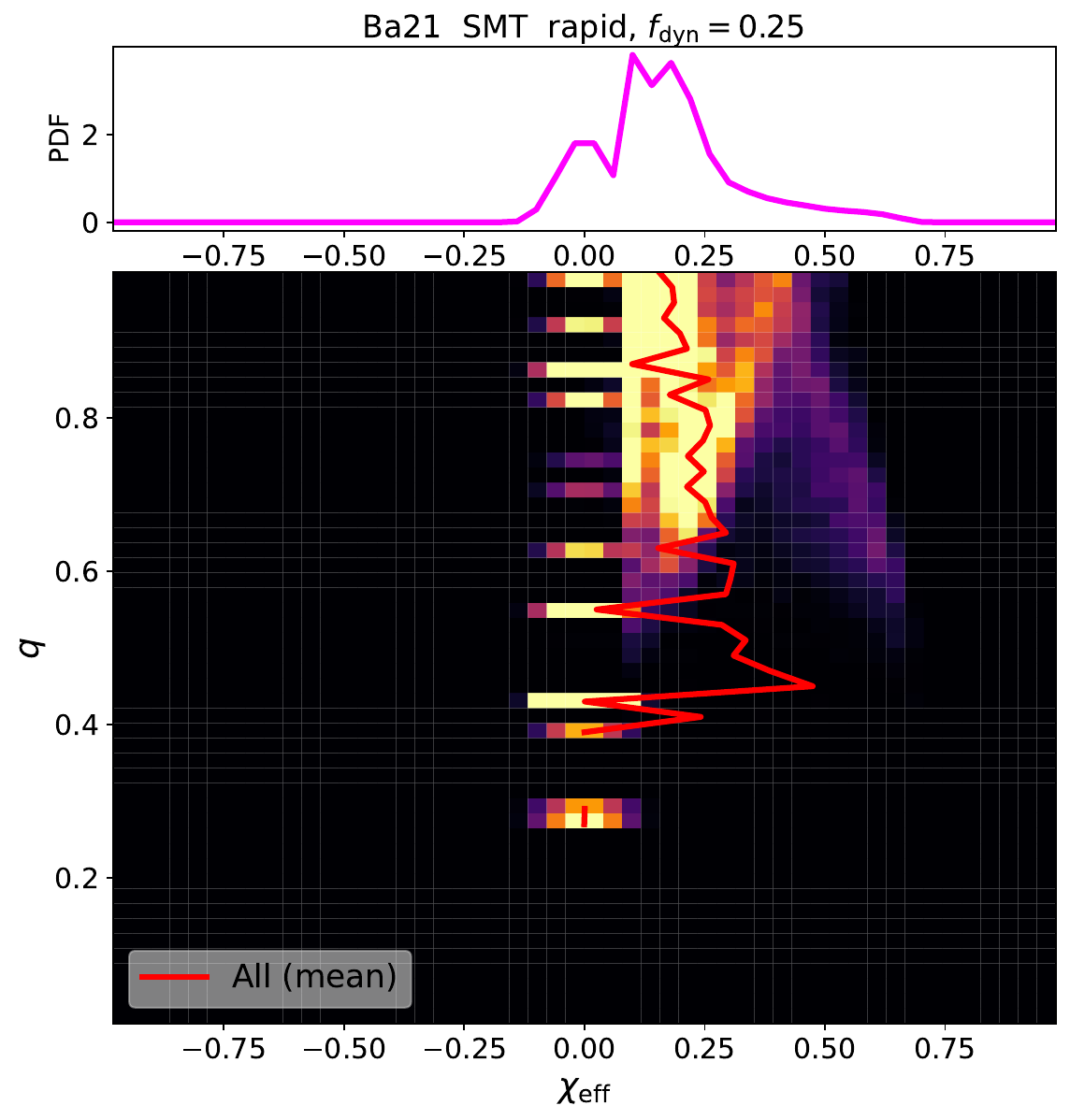}
\includegraphics[width = 8.75 cm, angle=0.0]{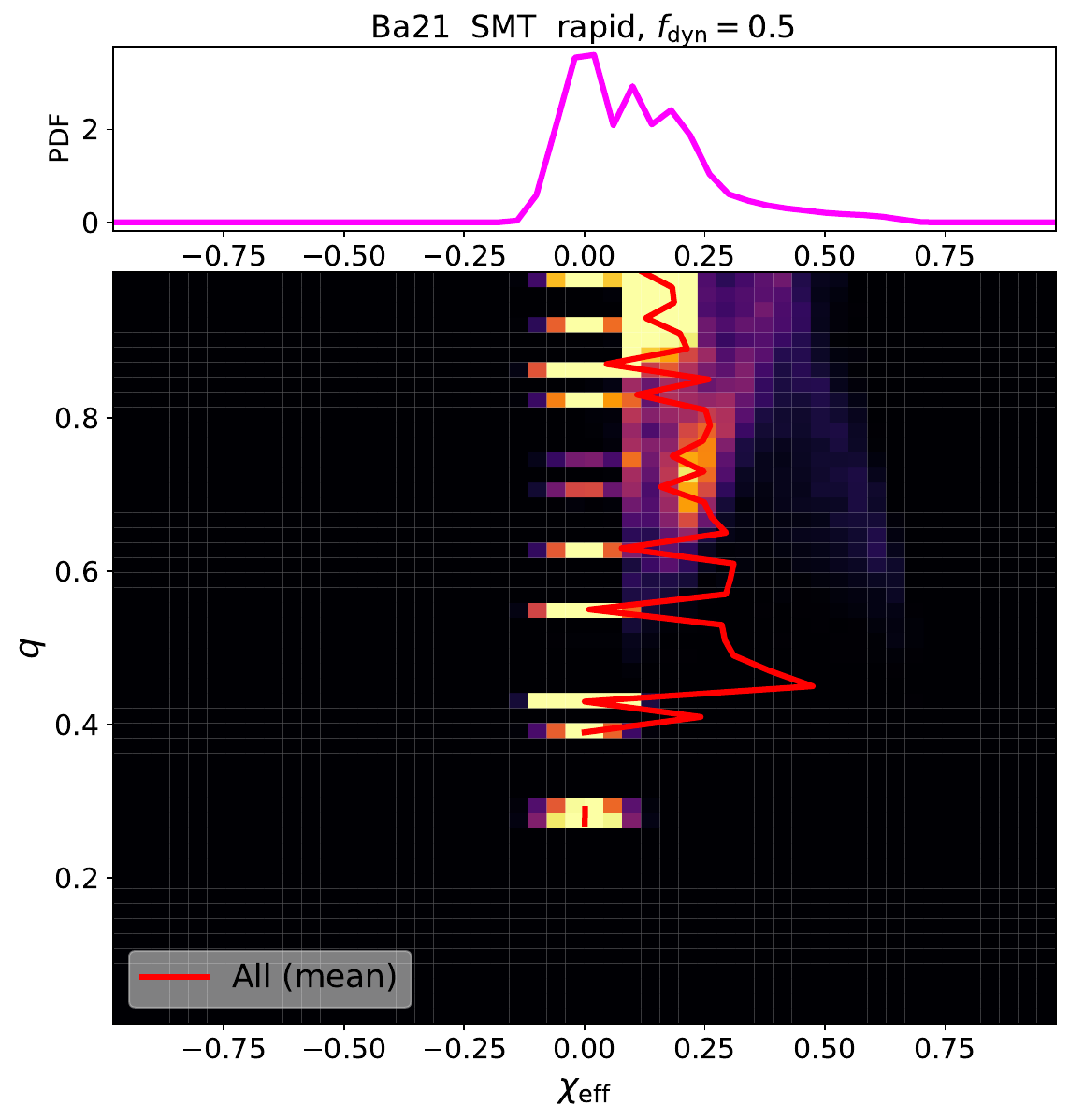}
\includegraphics[width = 8.75 cm, angle=0.0]{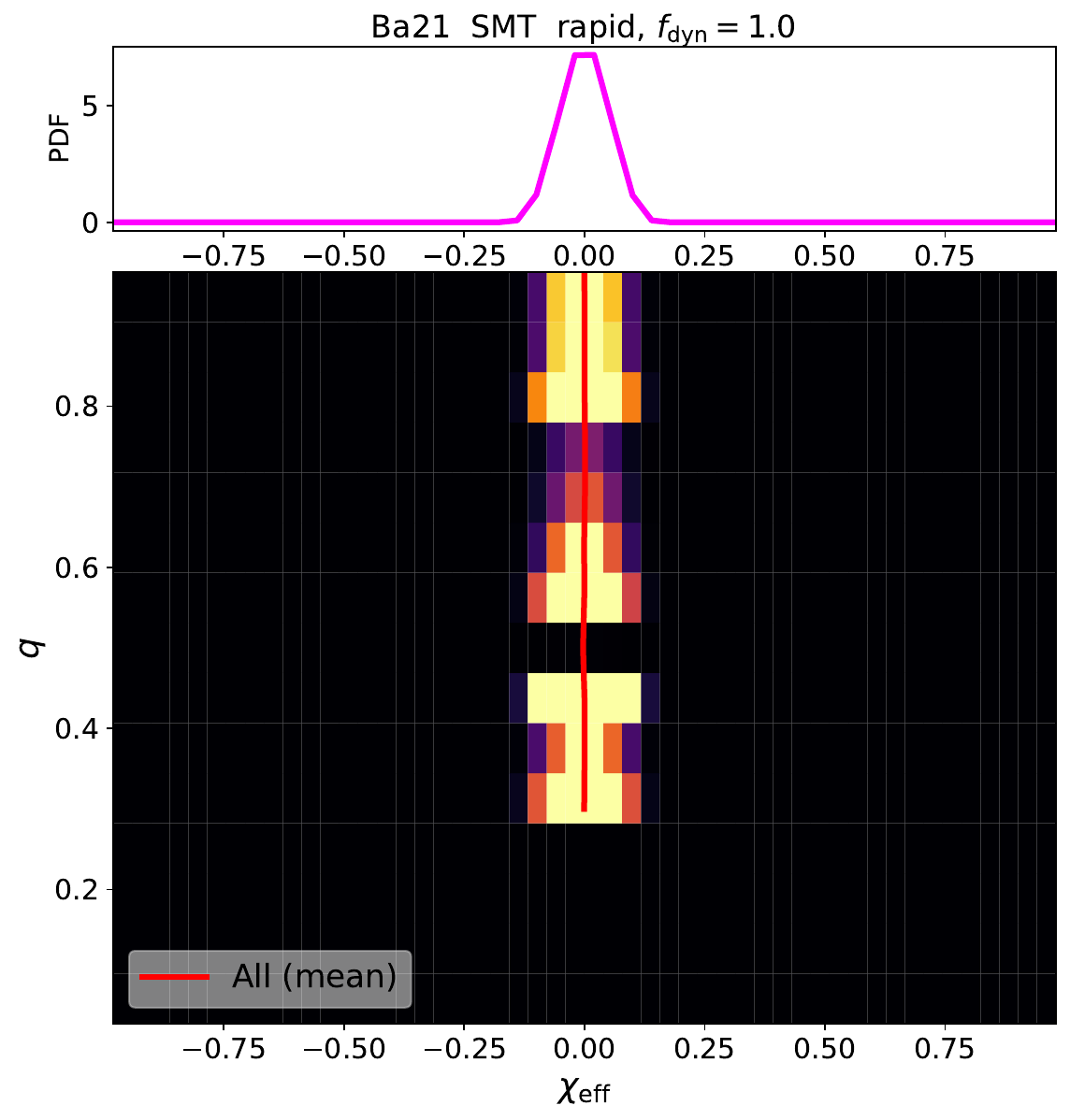}
	\caption{$\xeff-q$ density plots with varying fraction, $\fdyn$, of mixture with a
	the population of purely dynamically assembled BBH mergers (panel title).
	The subpopulation from the isolated binary channel
	corresponds to the BBH merger population shown in the upper right panel of
	Fig.~\ref{fig:dists7_qc-8_hgf-1}. The subpopulation of dynamically assembled BBH mergers
	is obtained from the N-body star cluster models of \citet{Banerjee_2022}.
	The marginal $\xeff$ distribution is shown for each
	panel (inset, magenta line). All the four density plots have the same relative
	colour saturation point based on the corresponding maximum density, such
	that the anti-correlated branch is clearly visible.}
\label{fig:dists8}
\end{figure*}

\begin{figure*}
\centering
\includegraphics[width = 8.5 cm, angle=0.0]{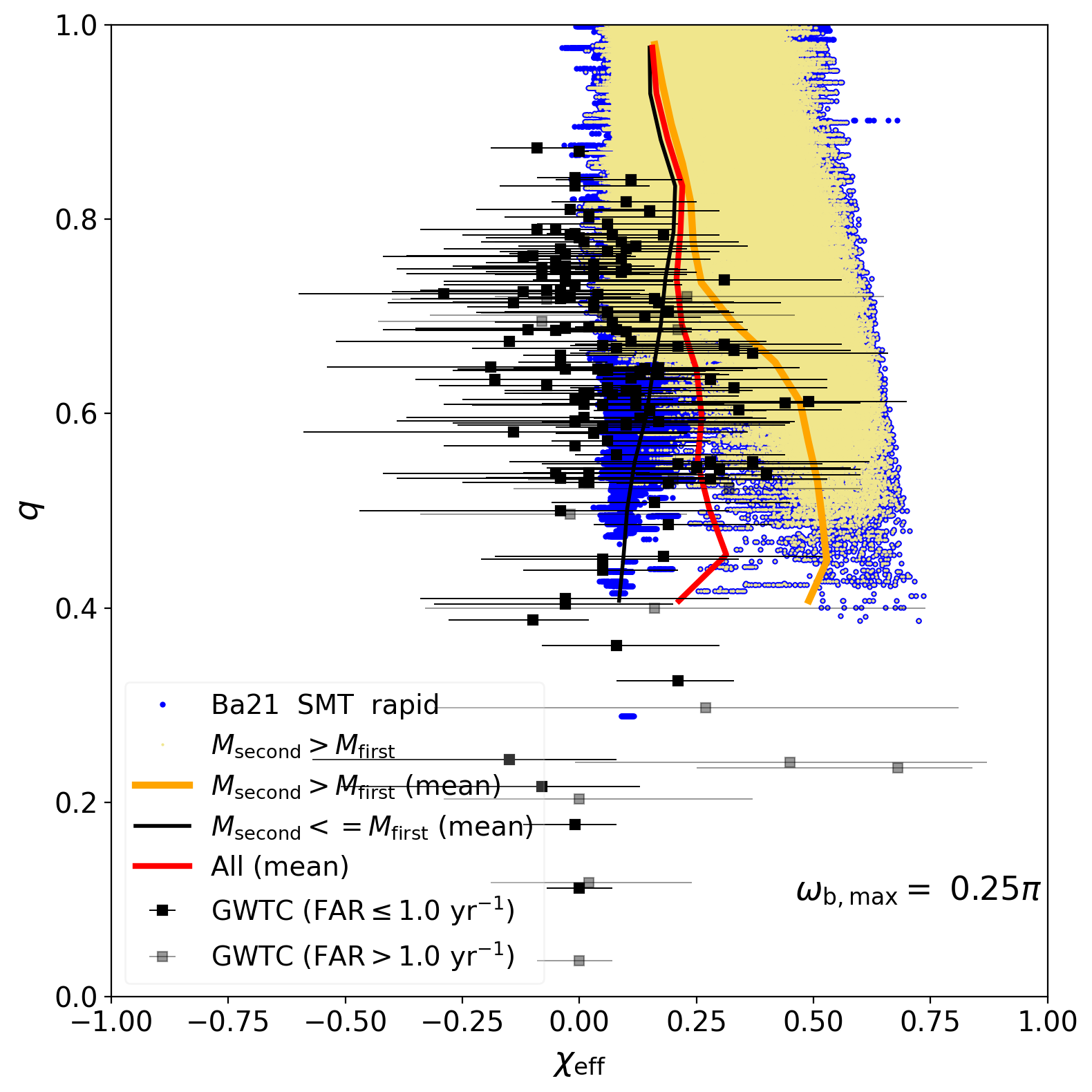}
\includegraphics[width = 8.5 cm, angle=0.0]{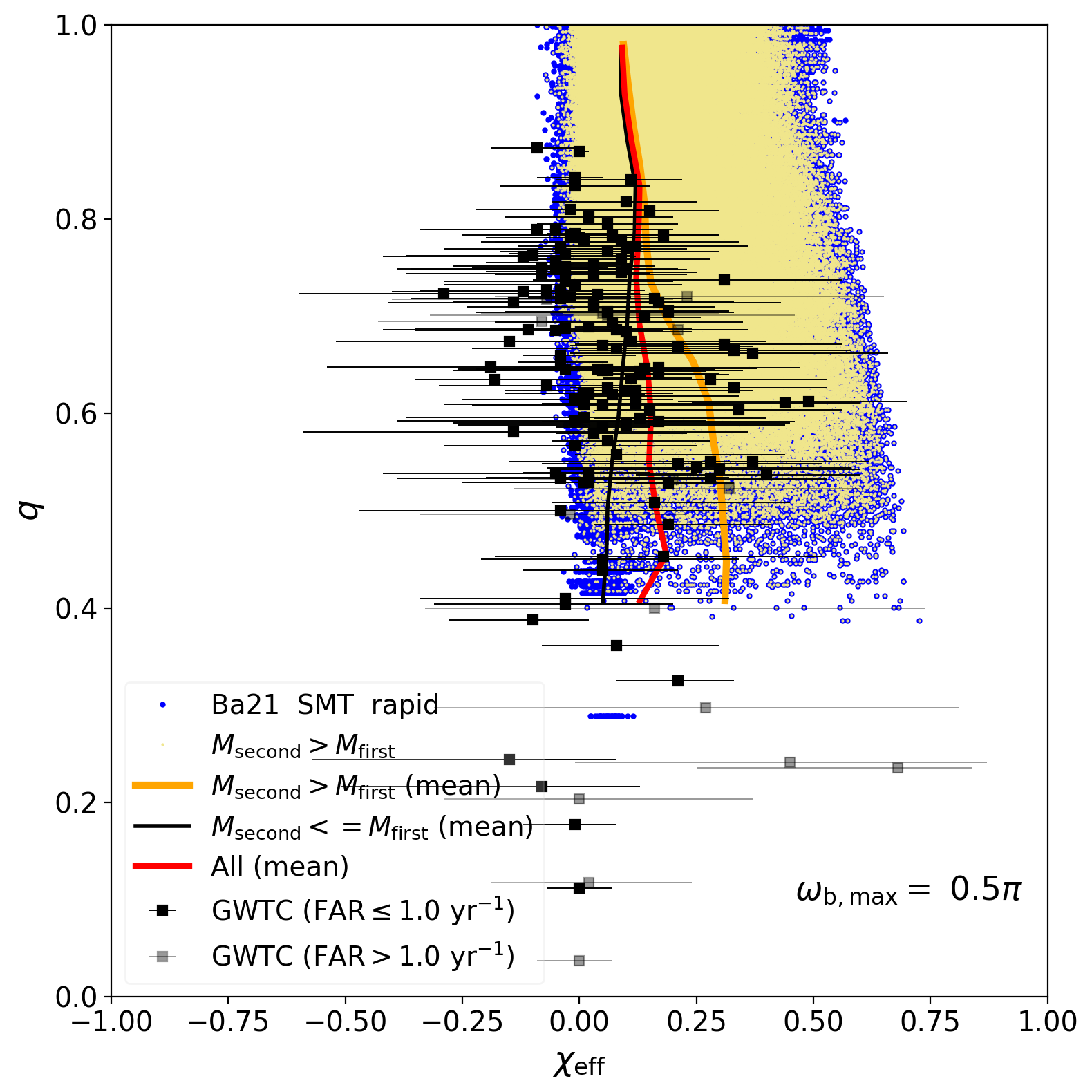}
\includegraphics[width = 8.5 cm, angle=0.0]{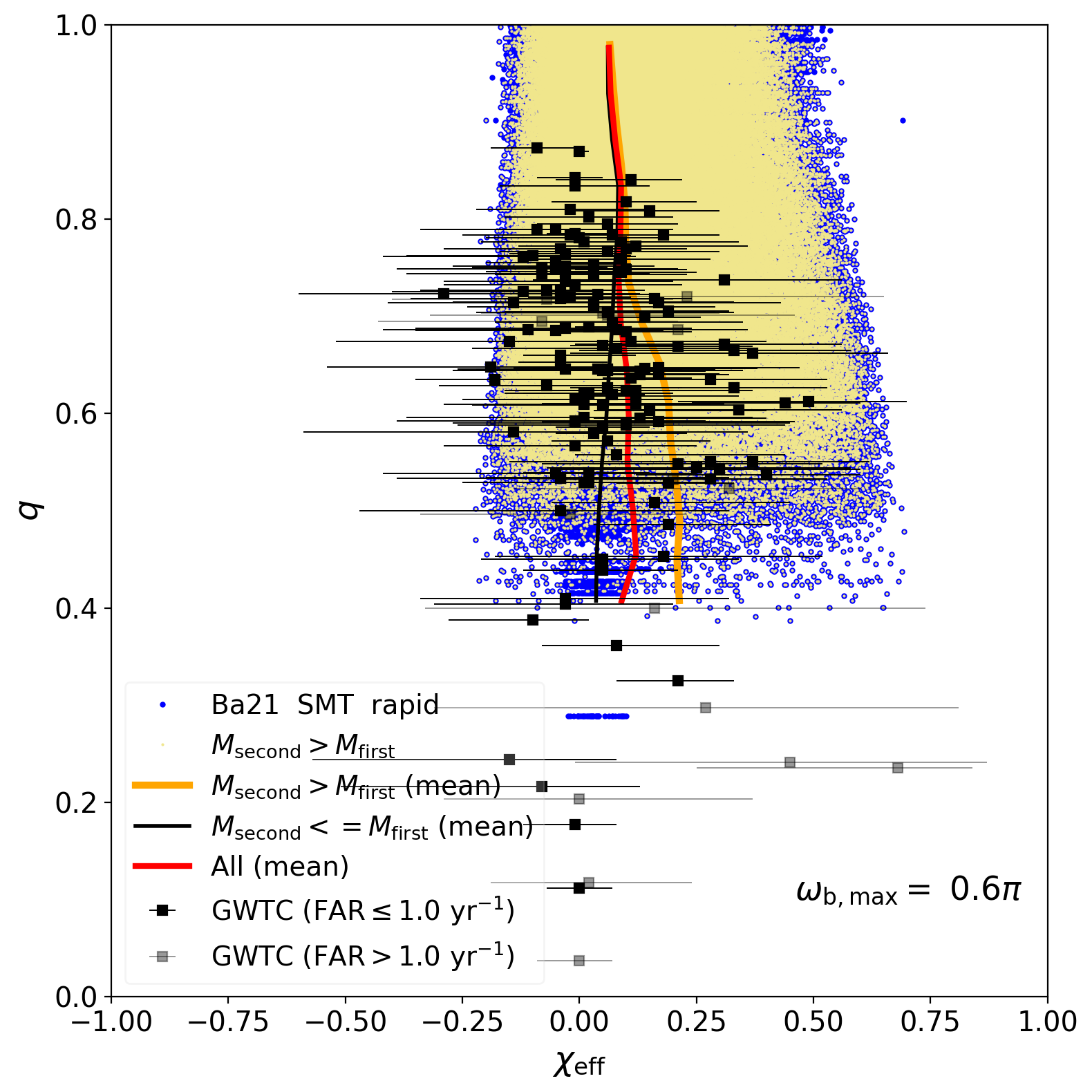}
\includegraphics[width = 8.5 cm, angle=0.0]{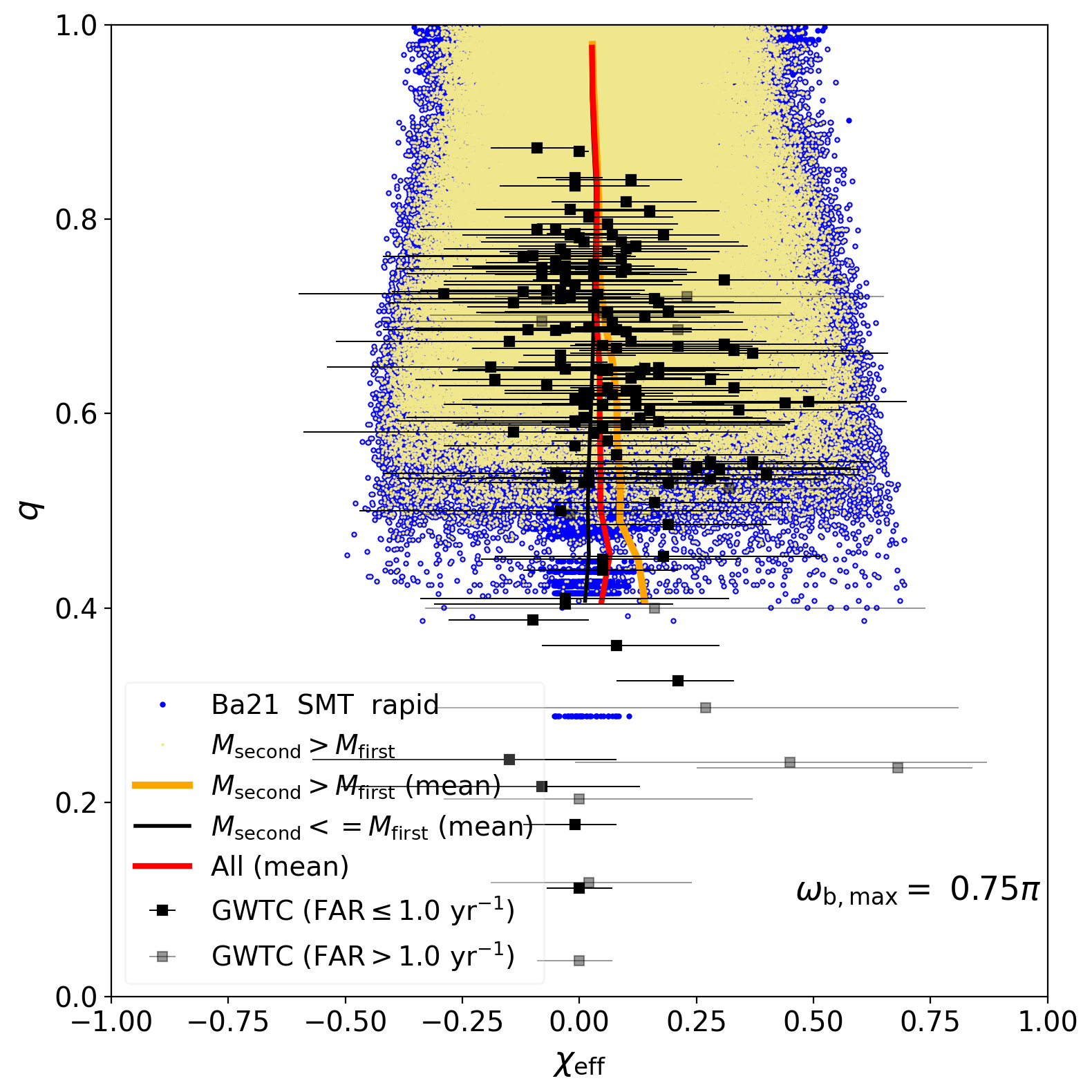}
	\caption{$\xeff-q$ scatter plots of present-day-observable BBH mergers occurring via the SMT channel,
	for the rapid remnant model and Ba21 BH-spin model. The legends are the same as in the
	previous counterpart figures. Each panel corresponds to a chosen maximum BH spin tossing polar angle,
	$\ombmax$, as indicated in the legend. The GWTC data points, \ie, the observed events are over-plotted
	(filled squares), which are distinguished based on their FAR. See text for the details.}
\label{fig:dists7_toss}
\end{figure*}

Notably, in all cases presented so far, the fraction of BBH mergers with $\xeff<0$ is vanishingly small.
This happens due to the fact that massive, spun-up second-born BHs often receive very small spin-orbit tilts 
(Eqn.~\ref{eq:tilts}), owing to their rather small natal kicks (Sec.~\ref{xeffq}).
In the present BH-spin scheme (Sec.~\ref{bhspin}), the massive, spun-up second-born BHs dominate
the $\xeff$ distribution's spread (c.f. Eqn.~\ref{eq:xeffdef}), since the non-spun-up
BHs possess very small spins. However, in \citet{Banerjee_2023},
it has been demonstrated that mixing with a population of dynamically assembled BBH mergers,
where the $\xeff$ distribution is isotropic and symmetric around zero
\citep[\eg,][]{Rodriguez_2018,Yu_2020},
can substantially enhance the negative $\xeff$ fraction, making it comparable to 
the LVK-determined value (of $\approx28$\%; \citealt{Abbott_GWTC3_prop}).
In this paper, we present a simple form of such an exercise, leaving a more elaborate analysis
for a future work.

The panels in Fig.~\ref{fig:dists8} show $\xeff-q$ density maps, where the $\xeff-q$ map
from present-day-observable, isolated-binary BBH mergers is combined with a varying fraction, $\fdyn$
(panel title), of that from dynamical BBH mergers. The isolated binary population corresponds
to the one presented in the upper right panel of Fig.~\ref{fig:dists7_qc-8_hgf-1}, that
is the most $\xeff-q$ anti-correlated among all the examples presented in this paper.
The dynamical population corresponds to that in \citet{Banerjee_2023}, which is
based on the N-body star cluster models of \citet{Banerjee_2022}. We choose only
the strictly dynamically paired BBH mergers and apply post-processing to these merging BHs' spins
in the same way as described in \citet{Banerjee_2023}. Both the isolated-binary
density function, $\phi_{\rm iso}(\xeff,q)$, and the dynamical density function,
$\phi_{\rm dyn}(\xeff,q)$, are individually normalized so that the plotted
combined density function,
\begin{equation}
\phi(\xeff,q) = (1-\fdyn)\phi_{\rm iso}(\xeff,q) + \fdyn\phi_{\rm dyn}(\xeff,q),
\label{eq:phidist}
\end{equation}
is also normalized. The marginalized, combined $\xeff$ distribution corresponding
to each panel in Fig.~\ref{fig:dists8} is plotted atop the panel (magenta line).

As can be seen in Fig.~\ref{fig:dists8}, the fraction of anti-aligned-spin ($\xeff<0$) BBH mergers
increases with increasing $\fdyn$. Generally, dynamical mixing tends to
erase the $\xeff-q$ anti-correlation. However, except for the pure dynamical channel
case ($\fdyn=1$; Fig.~\ref{fig:dists8}, lower-right panel), an anti-correlated branch is always
present and the anti-correlation continues to hold for the overall population
until at least $\fdyn=50$\%.
A detailed study of the extent of dynamical mixing, along the lines of \citet{Banerjee_2021b}, 
and its impact on merger spin distribution will be undertaken in the near future.

\subsection{Introducing BH spin-tossing}\label{toss}

The $\xeff$ distributions and correlations obtained so far rely on the assumption that the spin of a newly formed BH always aligns perfectly with that of its pre-SN progenitor star. However, it has been suggested that natal kicks can not only induce spin-orbit misalignment but also alter the direction of the BH’s spin relative to the pre-SN star’s spin through a process named “tossing” \citep{Tauris_2022}.

In this section, we investigate to what extent the $\xeff-q$ dependence is affected when the above perfect spin-alignment assumption is relaxed by allowing the BH natal spin's direction to toss \citep{Tauris_2022}. We assume that at the formation of a BH its spin makes a polar angle $\omega_b$ relative to the direction of the parent pre-SN-star's spin, $\hat{s}$, and an azimuthal angle $\zeta$ relative to the plane containing $\hat{s}$ and the post-SN orbital angular momentum. We continue to assume
that a stellar member is always spin-aligned with the orbital angular momentum so that
$\hat{s}$ is aligned with the pre-SN orbital angular momentum. In that case, the
polar angle, $\nu$, of the orbital angular momentum tilt due to the SN's natal kick (Eqn.~\ref{eq:nu})
and the BH's spin tossing, given by $\omega_b$ and $\zeta$, can be vector-combined in
a similar way as in Appendix.~\ref{angles}, to obtain the post-SN, resultant polar
angle between the BH's spin and the orbital angular momentum.

\begin{figure*}[!t]
\centering
\includegraphics[width = 8.75 cm, angle=0.0]{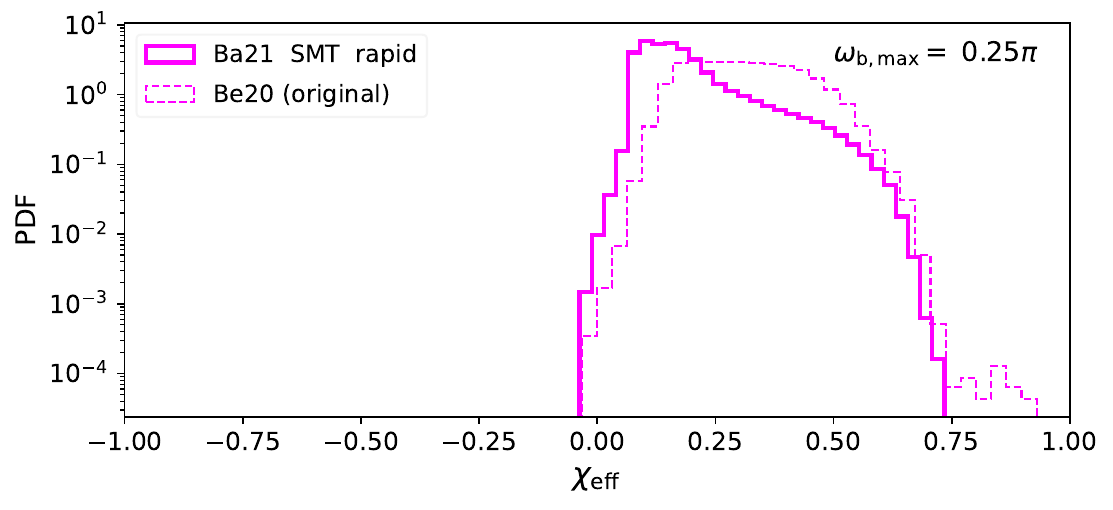}
\includegraphics[width = 8.75 cm, angle=0.0]{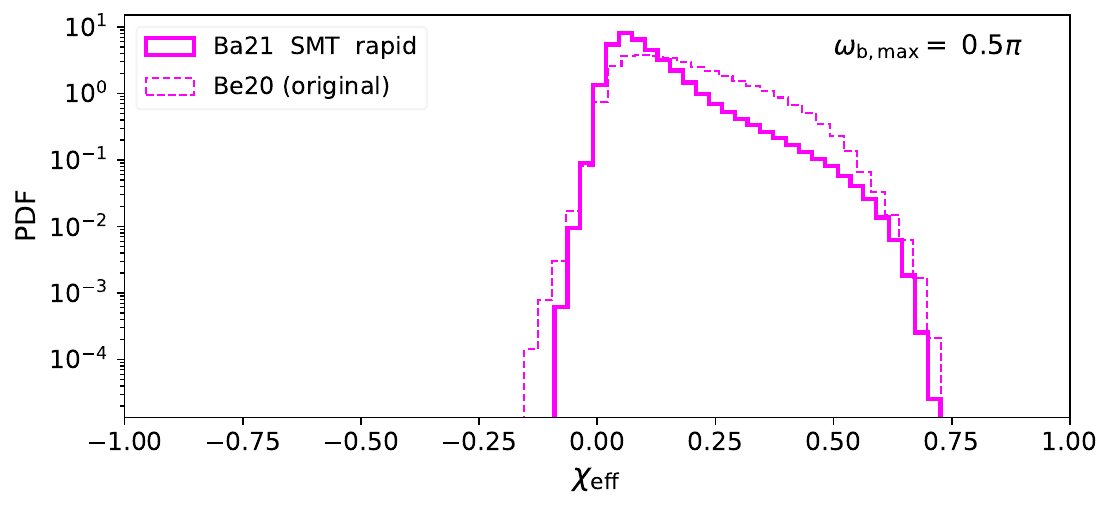}
\includegraphics[width = 8.75 cm, angle=0.0]{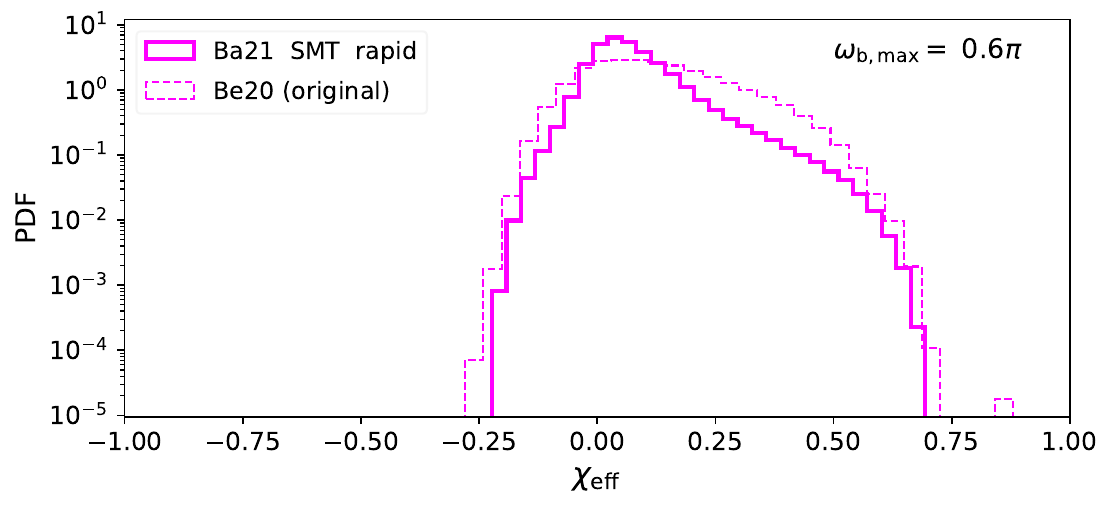}
\includegraphics[width = 8.75 cm, angle=0.0]{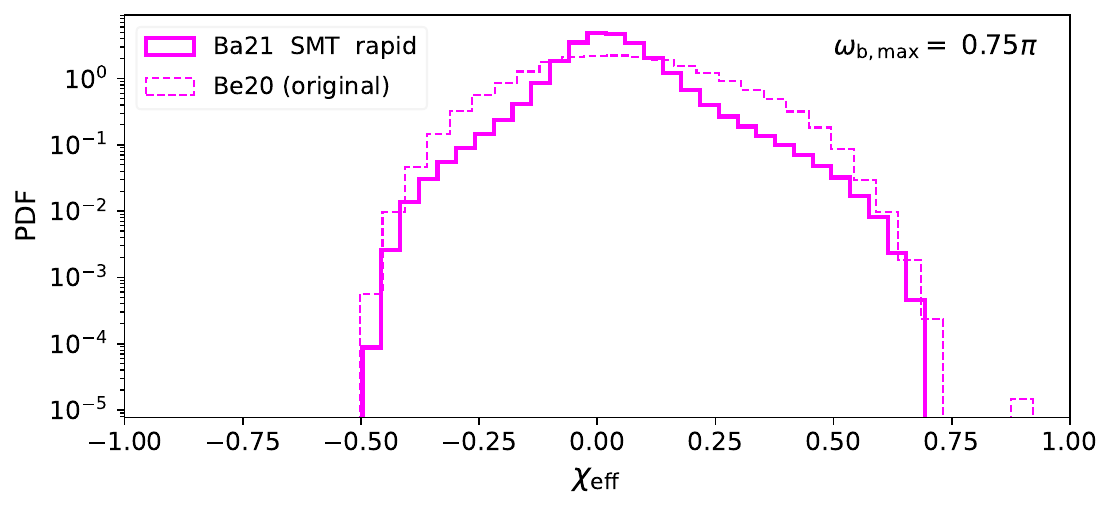}
	\caption{The marginal $\xeff$ distributions corresponding to the cases presented in
	Fig.~\ref{fig:dists7_toss}.}
\label{fig:dists111_toss}
\end{figure*}

Given that spin tossing is highly uncertain and unconstrained, we study its impact through a straightforward
parameterization approach. To that end, we assign $\omega_b$ randomly from an isotropic distribution
between zero and a maximum value, $\ombmax$, and $\zeta$ randomly from a uniform distribution
over $[0,2\pi]$. Fig.~\ref{fig:dists7_toss} shows the present-day-observable BBH mergers
for one of the computed binary population (Ba21, SMT, rapid; corresponding to the upper-right panel of
Fig.~\ref{fig:dists7_qc-8_hgf-1}) in the $\xeff-q$ plane, for $\ombmax=0.25\pi$, $0.5\pi$, $0.6\pi$, and $0.75\pi$.
The corresponding marginal $\xeff$ distributions are shown in Fig.~\ref{fig:dists111_toss}. 

As can be expected, the $\xeff-q$ anti-correlation of the mass ratio reversed population deteriorates with
increasing extent, $\ombmax$, of SN spin tossing. The $\xeff<0$ fraction increases correspondingly.
Nevertheless, as long as the extent of tossing remains partly aligned (\ie, $\ombmax\leq0.5\pi$)
the dependence sustains. On the other hand, with increasing anti-aligned tossing, the $\xeff-q$ anti-correlation
decays sharply, and the scatter becomes increasingly similar to that due to a population of dynamically
assembled, isotropically-oriented mergers \citep{Rodriguez_2018,Santini2023}.

In Fig.~\ref{fig:dists7_toss}, we also plot the latest GWTC event data points (mean values) and the corresponding
uncertainties (90\% credibility limits) in $\xeff$ (as obtained from the GWTC public catalogue; see Sec.~\ref{intro}),
highlighting the events with a false alarm rate (hereafter FAR)
of $\leq1\peryr$ (black-filled, full-shaded squares). Due to the large uncertainty in the $\xeff$ values in the currently
published GWTC, this comparison should be considered preliminary and qualitative. Also, due to the associated uncertainties
in the extent of the $\xeff-q$ anti-correlation (see, \eg, \citealt{Callister_2021}) or its absence,
we refrain from any quantitative comparison with the LVK data points.
As seen in Fig.~\ref{fig:dists7_toss},
a substantial BH spin-tossing ($\ombmax > 0.5\pi$) is required to encompass the observed data points,
for a purely isolated-binary origin of BBH mergers. This is due to the rather substantial (24\%-42\%)
$\xeff<0$ fraction in the population of the to-date-observed events \citep{GWTC4a_pop}.
Overall, this aligns with the recent findings of \citet{Tauris_2022}.

\section{Discussion of the adopted assumptions}\label{discuss}

\subsection{Comparison with other works}\label{compare}

All of the models tested in this work result in two characteristic BBH merger branches in the $\chi_{\rm eff}-q$ plane; see the figures in the main text and Appendix~\ref{dists_more}. In particular, one of the branches (khaki-filled circles) corresponds to a subpopulation of mass-ratio reversed binaries that follow the $\chi_{\rm eff}-q$ anti-correlation. The other, complementary branch (blue-filled circles) corresponds to BBH binaries for which the second-born BH is the less massive one. The latter branch consists of BBH mergers for which there is no clear dependence between $\chi_{\rm eff}$ and $q$ or there is a positive correlation between them, in contradiction to the detected population of BBH mergers. Although the general properties are consistent for the various tested models, the relative contribution of the two branches and the specific properties depend on metallicity and the physical assumptions. Our results are consistent with the results of \citet{Broekgaarden2022}, who also found similar subpopulations with their tested models. Their tested models differ by mass transfer stability criteria, so they consist of a significant fraction of post-CE systems in contrast to our models for only the SMT sub-channel. However, consistently with our work, their branch, which produce the $\chi_{\rm eff} -q$ anti-correlation, consists of reversed mass ratio binaries formed via SMT.

The models of \citet{Olejak_2024} also produce the branch of unequal mass, highly spinning BBH mergers via the mass ratio reversal SMT scenario. Their relation between time delays and mass ratio for SMT channels is consistent with our subpopulation of mass-ratio revered binaries. However, their models do not produce the second branch of BBH mergers (with less massive second-born BH) that exhibit a correlation between $\chi_{\rm eff} -q$. Moreover, their SMT channel results in a shortage of equal-mass binaries originating from Case A first mass transfer phase. Instead, equal mass and low-spinning binaries are outcomes of the CE sub-channel. These differences between our current models and the results of \citet{Olejak_2024} originate from two important differences in mass transfer physics models. First, the different adopted specific angular momentum loss during SMT with a non-degenerate companion in their models led to wider separations after the first system RLO. The second is the criteria for unstable mass transfer development in \citet{Olejak_2024} is more complex than this work's assumption on critical mass ratio limit. In particular, their criteria mimic the emergence of another possible type of instability in close binaries that often prevents binaries from surviving the case A or early B mass transfer phase and evolving into BBH mergers. However, the occurrence of this type of instability in the work of \citet{Olejak_2021a, Olejak_2024} is based on the small grid of simulated systems by \citet{Pavlovskii_2017} and, therefore, could overestimate the effect.

Some of the other recent studies, \eg, \citet{Zevin_2022}, due to less conservative criteria for CE development resulting in a relatively small contribution of SMT sub-channel (especially with unequal mass progenitors), did not find an equivalent correlation for $\chi_{\rm eff} -q$. In particular, in their simulations, BBH merger progenitors during the second post-MT phase tend to be too wide to tidally spin up the helium core that later evolves into the second-born BH.

In all tested models with BBH merger formation dominated by the SMT sub-channel, we find an overdensity of primary BHs with masses $\sim 10 M_{\odot}$, consistent with LVK detections. This feature has been already pointed out by \citet{vanSon_2022}, who explained how the critical mass ratio for SMT sets a lower limit on the primary BH mass. The peak in our synthetic BBH merger population is particularly pronounced once the rapid SN engine is adopted, which favors the formation of either low mass NSs or BHs of $M>5M_{\odot}$ via direct collapse.

Mass and mass ratio distributions of BBH mergers in our models are in good agreement with other similar works such as \citet{Broekgaarden2022,vanSon_2022,Dorozsmai_2024}.

\subsection{Caveats and future development}\label{caveat}

In this work, motivated by several recent studies, e.g. \citet{Pavlovskii_2017,Gallegos_2021,Shao_2022}, we explored an alternative stability criterion for mass transfer. In particular, in several tested models, we increased the critical mass ratio for CE development to a fixed value. This approach strongly reduces the relative contribution of the CE subchannel to the SMT channel. Increasing the critical mass ratio to such a high, fixed value is, however, an approximate and simplified approach to capture the main trend found in revised studies of the stability of mass transfer in massive binary systems \citep{Pavlovskii_2017,Gallegos_2021,Shao_2022}. In reality, the boundaries between stable and unstable mass transfer are much more complex and depend on the exact properties of the binary system at the onset of the second MT or other factors related to, e.g., orbital evolution of the system \citep{Picco_2024}. Models favoring the SMT subchannel have also been motivated by several other recent studies that indicate that the formation of BBH mergers via the CE subchannel is vastly overestimated for additional reasons. In particular, the distinction between giants with radiative and convective envelopes is not well defined by most of the rapid population synthesis codes \citep{Romagnolo2024}. Even if unstable mass transfer develops, the successful CE ejection is most likely possible under very strict conditions for the donor star \citep{Klencki_2021}.

In this work, we do not vary assumptions on specific angular momentum loss of the non-accreted mass. The angular momentum loss can be higher than the default values adopted in this work and can greatly affect the fate of BBH progenitors as well as the final properties of the synthetic BBH mergers population, see, e.g. \citet{Willcox2023,vanSon_2022,Dorozsmai_2024,Olejak_2024}.

The size of WR stars' radii (the post-mass transfer, stripped helium cores) in our simulations (and in other rapid codes based on \citealt{Hurley_2002} prescriptions) can be underestimated. This is a consequence of a few factors: a) the assumptions on the overshooting in the underlying stellar models b) tracking the evolution of helium core only till the end of core helium burning c) the helium cores are only partly stripped, so there is still a layer of hydrogen that remains after the mass transfer \citep{Klencki_2022,Ramachandran2023}. The compactness of WR companions affects the fraction of systems that later evolve into BBH mergers, as some WR stars may not fit their Roche Lobe size (e.g., Klencki et al. in prep.), leading to another mass transfer episode or a binary system merger. Systematically, larger WR-star radii would also decrease the number of BH-WR binaries that go through a tidal spin-up, especially those that reach the maximal spin value. That, on the other hand, could lead to a more observationally consistent spin distribution of BBH mergers, as our current isolated-binary models tend to produce a more positively biased distribution with higher $\chi_{\rm eff}$ values than the one inferred from GW detections.

Although we explore a number of cases, the present work should still be considered preliminary, carrying with it the oversimplifications of binary evolution modeling in a fast binary evolution code, particularly, those in $\bse$. However, the interesting results and consistencies with the properties of the observed GW-event population pave way for improvements in the near future. Such improvements can be in, \eg, the mass transfer stability criteria by considering more details on donor stellar structure and properties of the system (\eg, metallicity, mass ratio, orbital evolution; see, \eg, \citealt{Picco_2024}) and the model of accretion onto BH by incorporating GR effects.
At present, $\bse$ either lacks or contains former implementations of these aspects
compared to those in other population-synthesis codes such as $\startrack$ \citep{Olejak_2021a} and
{\tt COMPAS} \citep{Riley_2022}. It is essential to keep $\bse$ updated and matched with recent population synthesis ingredients as it serves
as the main stellar- and binary-evolution engine in several widely used binary-evolution wrappers such as {\tt COSMIC} \citep{Breivik_2020}
and star-by-star N-body codes such as $\nbpp$, $\nbseven$, {\tt PETAR}, $\mocca$, $\cmc$ \citep{Spurzem_2023}.

An even more consistent and widely applicable implementation of the second-born BH's spin would
be through relating $\abh$ to the total angular momentum of the progenitor WR star (instead of the WR-BH binary
orbital parameters). Such relations can be derived from detailed stellar evolution models of fast-rotating WR stars existing in the literature and/or are to be computed newly. Another important direction for development is to implement $\abh$ analogously in star-by-star N-body codes,
which also adopt $\bse$ as their stellar- and binary-evolution engine (see above). Implementation of $\abh$ in N-body codes will
enable more consistent, detailed, and generalized assessments of dynamical mixing (Sec.~\ref{dynmix})
in population-synthesis studies. Such lines of study will be undertaken in the near future.

\section{Conclusions}\label{summary}

In this study, we demonstrate that alternatively to scenarios such as AGN channel \citep{Santini2023, Delfavero2024}, isolated binary evolution through the SMT subchannel can produce anticorrelation between effective spin and mass ratio $\xeff$ (Sec.~\ref{intro}) along with other features of the detected BBH population. 

We examine the role of binary tidal interactions during the intermediate Wolf-Rayet (WR)–black hole (BH) phase in shaping the final BH spin distribution of BBH mergers formed through isolated binary evolution. To do so, we use an updated version of the fast binary evolution code $\bse$, incorporating prescriptions for elevated spins of the second-born BH, the former tidally spun-up WR companion (Sec.~\ref{bhspin}).
This implementation enables us to consistently account for two competing effects: (a) the spin-up of the WR star and the shrinking of the binary orbit due to tidal synchronization, and (b) the spin-down of the WR star and orbit widening caused by wind-driven mass and angular momentum loss.

We explore a range of uncertain physical assumptions, including variations in remnant mass prescriptions (so-called rapid vs. delayed engines by \citealt{Fryer_2012}), BH spin models, and the efficiency and stability of mass transfer (\ie, relative contribution of SMT and CE subchannels). We investigate the impact of metallicity by analyzing separate populations across different chemical compositions as well as properties of combined intrinsic BBH populations observable at redshift $\approx0$.

The main results of this study are as follows:
\begin{itemize}

\item All BBH merger populations formed through the SMT channel exhibit a prominent peak at $\mone\approx10\Ms$ for the primary mass of merging binaries (Sec.~\ref{mqdist}). This feature aligns well with the observed primary mass distribution reported by the LVK collaboration, suggesting a significant contribution of the SMT channel to the detected GW events. The peak is particularly pronounced in models employing the rapid supernova (SN) engine.

\item Among the cases examined in this study, the SMT channel, combined with the delayed remnant mass model, produces the most unequal-mass BBH mergers. In this scenario, mass ratios can reach as low as $q\approx0.2$  with the BBH mergers $q$-distribution displaying a broad peak spanning $0.2\lesssim q \lesssim0.8$ (Sec.~\ref{mqdist}). This extent of unequal-mass mergers is comparable to the outcomes predicted by models of dynamically assembled BBH mergers, thereby providing a potential alternative explanation for the origin of highly unequal-mass GW detections, such as those described in, e.g., \citet{Unequal_masss_2020}.

\item Isolated binary evolution produces an asymmetric $\xeff$ distribution shifted towards positive values. We find that $\xeff$ distribution obtained via the procedure described in this work (within $\bse$), and that derived from post-processing of WR-BH binary parameters at its formation produce similar $\xeff$ distributions (Sec.~\ref{xfdist}). This is because the overall orbital evolution of the WR-BH systems is relatively small, owing to the short lifetime of the WR phase. A WR-BH system, that undergoes significant orbital widening due to wind or coalesces due to Darwin instability (the process of runaway tidal spin-up of the WR star and orbital shrinkage),
does not produce a BBH that merges within a Hubble time.

\item In all studied cases, the subpopulation of mass-ratio-reversed BBH mergers, \ie, mergers with the second-born BH being more massive than the first-born BH ($\msecond>\mfirst$),
exhibits a distinct anti-correlation in the $\xeff-q$ plane (Sec.~\ref{xeffq}). The $\xeff-q$ anti-correlation is particularly prominent for SMT-channel mergers with the rapid remnant mass model, where a significant fraction ($\gtrsim40$\%; Fig.~\ref{fig:a2mean_qc-8_hgf-1}) of the BBH merger population is mass ratio reversed. This leads to an anti-correlation across the overall population. Conversely, BBH mergers with a less massive second-born BH exhibit the opposite trend. The characteristic $\xeff-q$ relation arises from the dependence of orbital shrinking during mass transfer and the Roche lobe size on the system’s mass ratio (Sec.~\ref{xeffq}). The to-date observed GW events do hint at an anti-correlation (Sec.~\ref{intro} and references therein), suggesting that a non-negligible fraction of BBH mergers may form through the mass-ratio-reversal scenario in isolated binary evolution.

\item The subpopulation of mass-ratio reversed BBH mergers disappears at high metallicities, causing the $\xeff-q$ anti-correlation to vanish as well. This trend supports the idea that the progenitors of observed BBH mergers predominantly originate from low-metallicity environments.

\item The $\xeff-q$ anti-correlation among $\msecond>\mfirst$ BBH mergers, as found in the isolated binary models presented here, is a robust feature of the mass-ratio-reversed subpopulation across a wide range of physical assumptions.

\item The fraction of anti-aligned ($\xeff<0$) systems increases with a growing contribution from dynamically assembled BBH mergers (Sec.~\ref{dynmix}). However, the dynamical mixing tends to erase the $\xeff-q$ anti-correlation in the population.
\end{itemize}

Our results demonstrate that the reported potential $\xeff-q$ anti-correlation in the to-date-observed GW events, along with the $10$ $M_{\odot}$ peak in the BH mass distribution, might be a signature of the isolated binary evolution channel's contribution with a significant fraction of mass-ratio-reversed systems.

Data availability: the isolated-binary GR merger data from the computations in this work are publicly available at this link: https://doi.org/10.5281/zenodo.20429516

\section*{Acknowledgements}
We thank the referee for their useful comments and criticisms that have helped to improve the manuscript significantly.
We would like to thank the late Krzysztof Belczynski for his invaluable motivation for this work, and Vishal Baibhav for his helpful discussions. 
SB acknowledges funding for this work by the Deutsche Forschungsgemeinschaft
(DFG, German Research Foundation) through the project ``The dynamics of stellar-mass black holes in
dense stellar systems and their role in gravitational wave generation''
(project number 405620641; PI: S. Banerjee).
All binary-evolution computations in this work were performed on the compute server Science,
located at the Argelander Institute f\"ur Astronomie (AIfA), University of Bonn and on the HPC facility
Marvin of the University of Bonn. SB acknowledges access to the Marvin facility.
This work has been benefited from extensive use of the {\tt python} packages
{\tt numpy}, {\tt scipy}, {\tt multiprocessing}, and {\tt matplotlib}.
We thank the developers and the python community for making these packages
freely available. Several coding and data management tasks related to this work
have been facilitated by the use of Claude Code (Opus 4.5/4.7 model).
SB acknowledges the generous support and efficient
system maintenance of the computing teams at the AIfA, HISKP, and University of Bonn. SB thanks
the AIfA administration for hospitality.
AO acknowledges funding from the Netherlands Organisation for Scientific Research (NWO), as part of the Vidi research program BinWaves (project number 639.042.728, PI: de Mink).\\


%
\bibliographystyle{yahapj}
\bibliography{bibliography/biblio.bib}

\onecolumngrid
\appendix
\newpage

\section{Spin evolution in isolated binaries}\label{spevol}

In this section, we present examples of the evolution of the spins of the binary members as modelled in $\bse$,
which evolution is the same as in the original version of the code \citep{Hurley_2002}.
See the figures' captions for the details. In our `spin-up-check' approach,
the BH spins are derived based on the spin, $\pspin$, of their
progenitor WR stars just before the BH formation, as discussed in Sec.~\ref{bhspin}.
We note, however, that in $\bse$, the WR star undergoes a boost in $\pspin$ beyond He-MS due to
core formation --- this is an artifact of the stellar-structural approximations in the code
\citep{Hurley_2000,Hurley_2002}.
Therefore, if the BH forms beyond He-MS, we use the value of $\pspin$ at the end of
He-MS (stellar type 7). That way, in applying the spin-up check , we omit tracking the WR-BH evolution
during the final phase with stellar type $>7$
that has a very short lifetime ($<<1$ Myr), in order to avoid spurious outcomes.

\begin{figure*}[!h]
\centering
\includegraphics[width = 8.9 cm, angle=0.0]{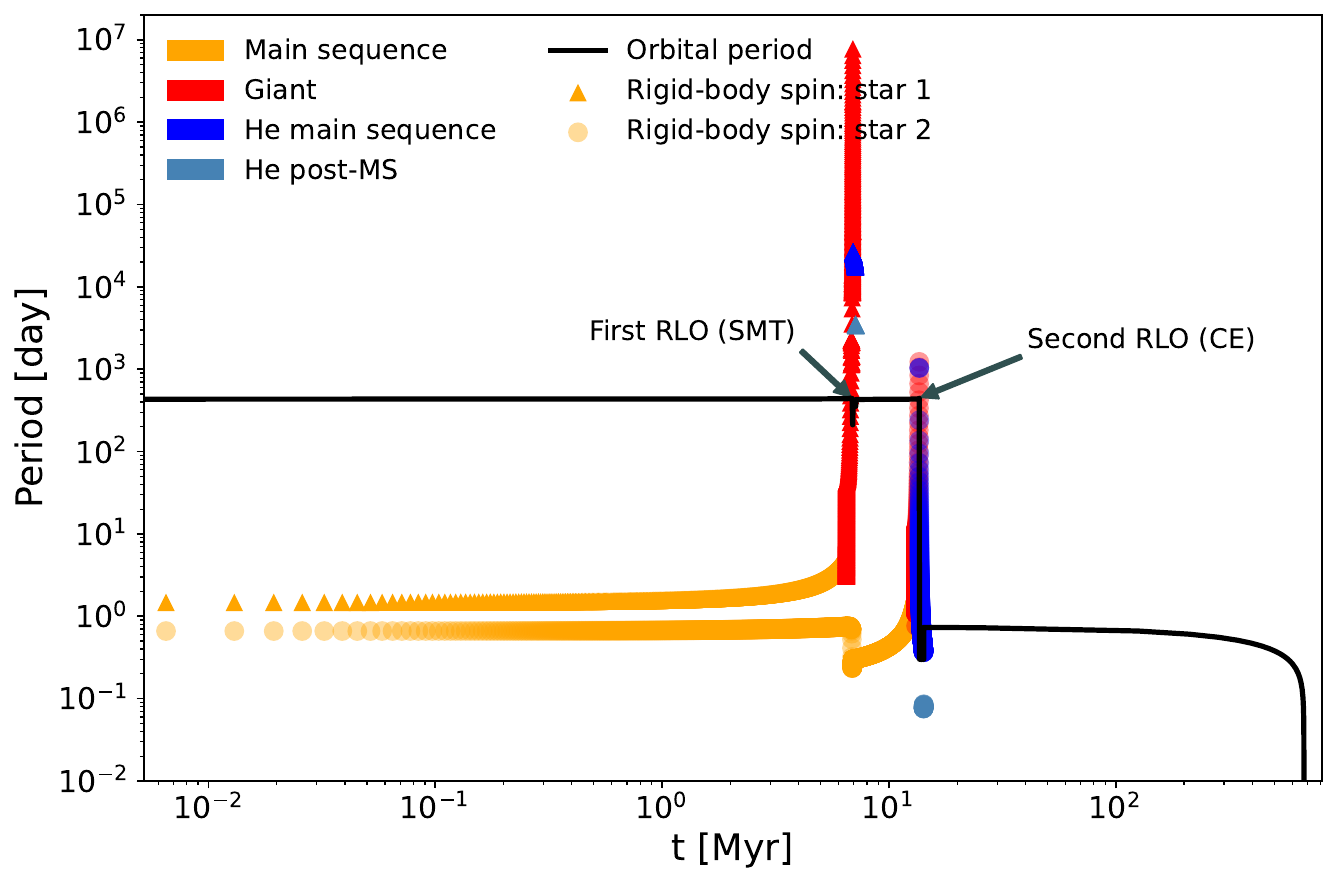}
\includegraphics[width = 8.9 cm, angle=0.0]{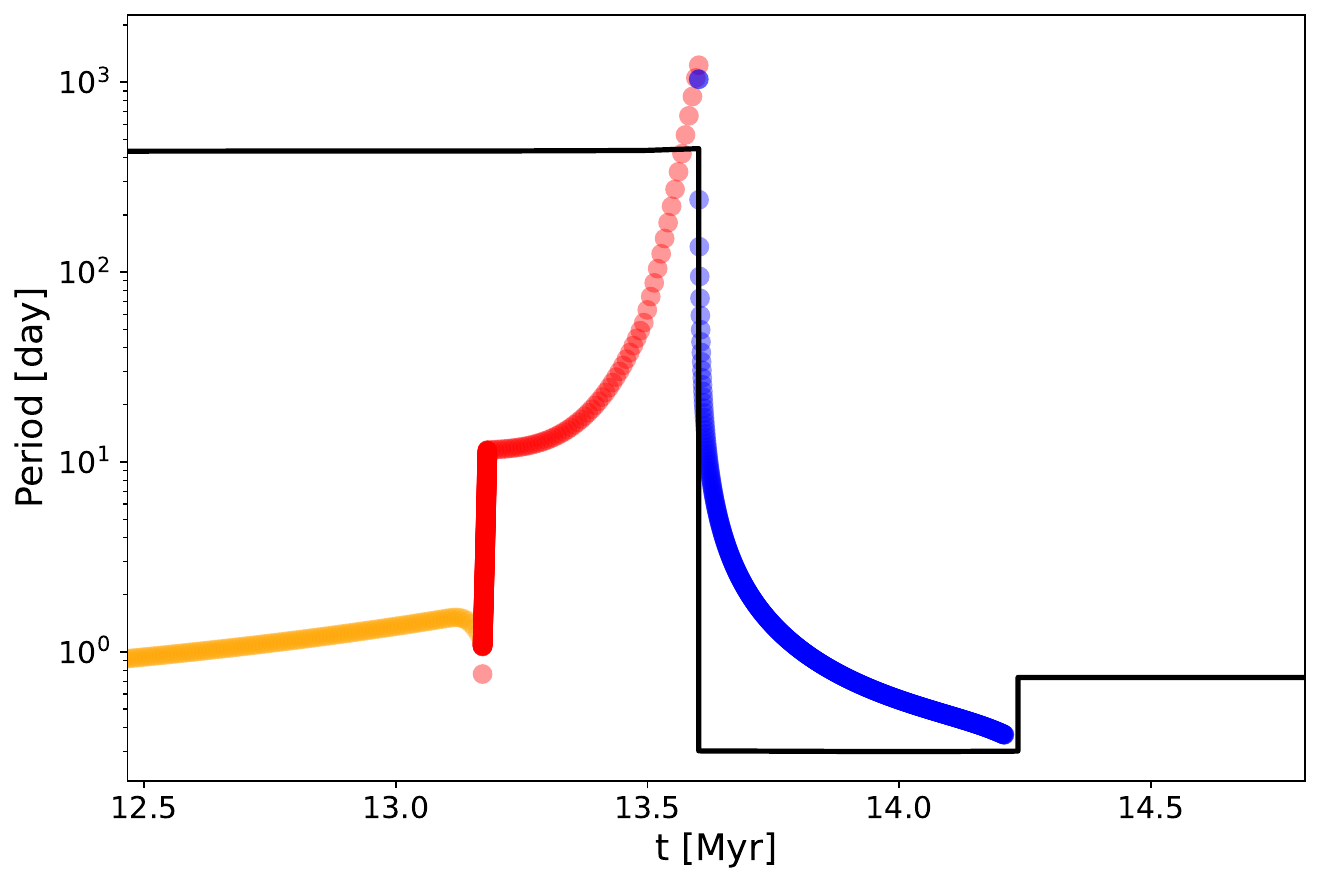}\\
\includegraphics[width = 8.9 cm, angle=0.0]{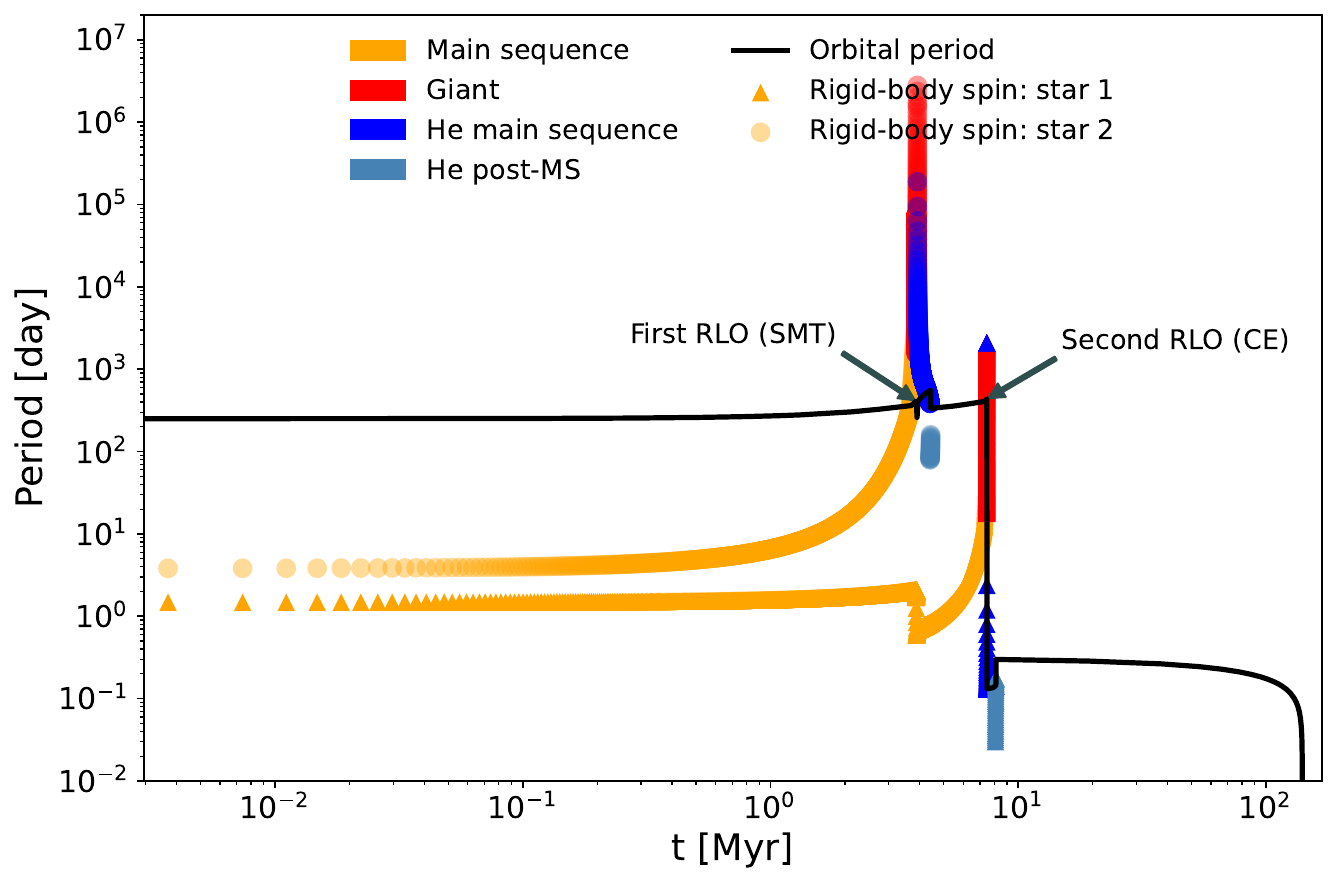}
\includegraphics[width = 8.9 cm, angle=0.0]{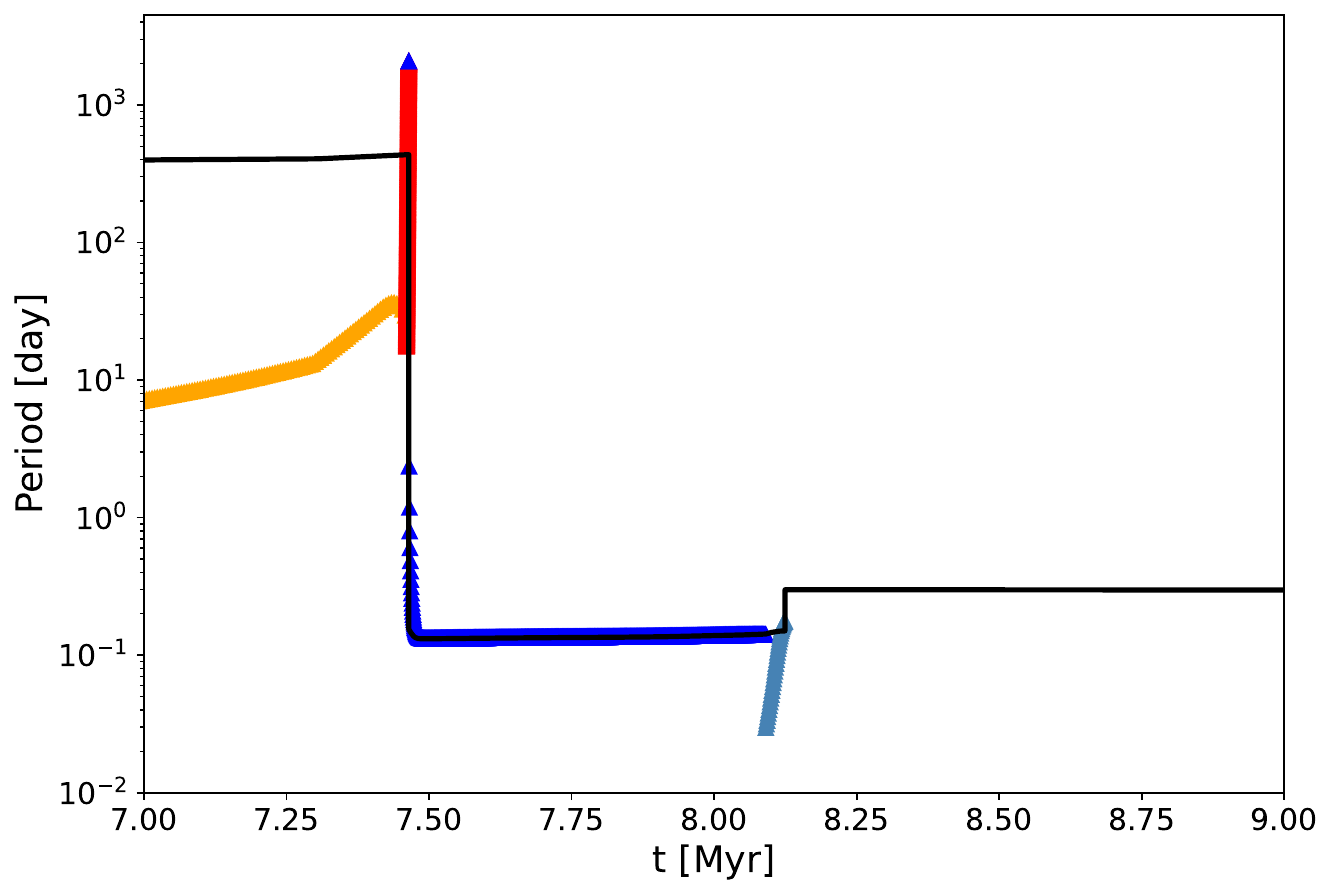}
\caption{{\bf Upper panels:} Time evolutions of the spin periods of the stellar members (filled circles and filled
	triangles, further distinguished by different color-alpha)
	and of the orbital period (black solid line), of a massive binary that produces a BBH merger via the CE sub-channel
	as modelled in {\tt BSE}. The binary-evolutionary age in Myr is plotted along the X-axis and the periods in days are
	plotted along the Y-axis. The members' spin periods are colour-coded according to the respective stellar-evolutionary
	stage (legend).
	Since a physical period cannot be associated with the spin angular momentum
	of a BH, the binary members' spins are plotted until just before the member becomes a BH. 
	The left panel shows the whole evolution of the binary starting from the MS-MS phase until the
	BBH inspiral and merger. The right panel zooms into the post-CE binary; the large decay
	in the binary's orbital period (and separation) at $\approx13.5$ Myr is due to the CE ejection resulting from the second
	RLO (between the first-born BH and its CHeB companion). It can be seen in the right panel that
	the WR member (a HeMS star) of the post-CE, tight WR-BH binary tidally spins-up (\ie, the spin period decreases)
	up to a $<1$ day spin period and close to the
	binary's orbital period, by the time it becomes a (second-born) BH (when the plotting of the filled circles terminates).
	Consequently, the second-born BH has a spun-up Kerr parameter, \ie, it has $\abh>\abhzero$ (Sec.~\ref{bhspin}). 
	The expansion of the binary orbit right before the beginning of the GR inspiral is due to the systemic
	mass loss and natal kick corresponding to the second BH formation. The panels also demonstrate the characteristic,
	artificial spin-up of the members, due to core formation and the associated reduction of the moment of inertia, at the
	beginning of the post-MS and post-HeMS phases. This example corresponds to a binary evolved at $Z=0.001$ with
	initial ZAMS member masses, orbital separation, and eccentricity being $30.7\Ms,10.5\Ms,831.6\Rs,0.0$, respectively.
	{\bf Lower panels:} The legends and the descriptions are the same as in the upper panels.
	The panels demonstrate a qualitatively
	similar example of BBH-merger-forming binary evolution as in the upper panels. Here, the WR star is
	fully spin-orbit synced during the WR-BH phase (right panel), forming a spun-up BH as it is spun
	up to $<1$ day period.
        This example corresponds to a binary evolved at $Z=0.02$ with
	initial ZAMS member masses, separation, and eccentricity being $20.7\Ms,67.8\Ms,744.3\Rs,0.0$, respectively.}
\label{fig:cecase}
\end{figure*}

\begin{figure*}[!h]
\centering
\includegraphics[width = 8.9 cm, angle=0.0]{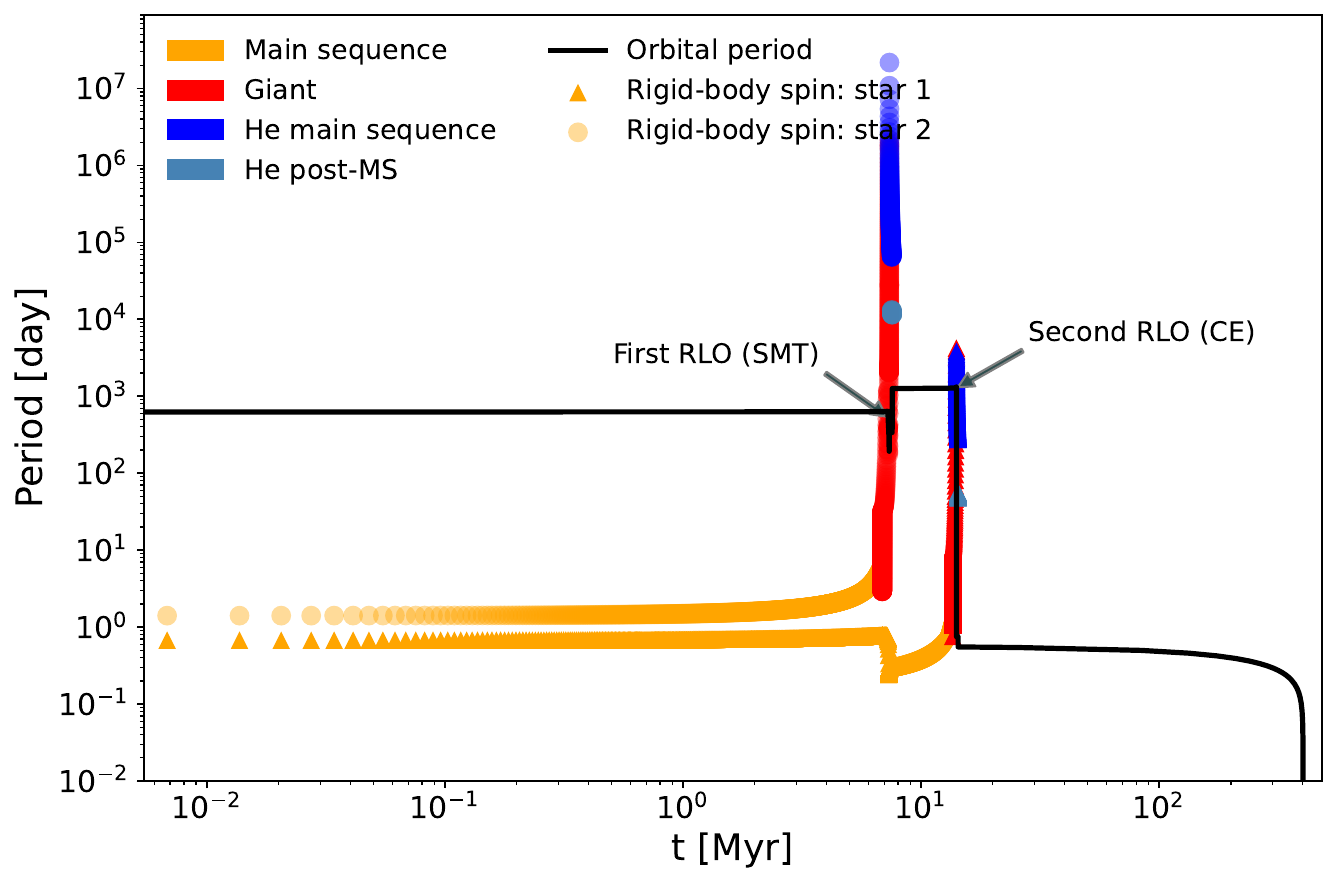}
\includegraphics[width = 8.9 cm, angle=0.0]{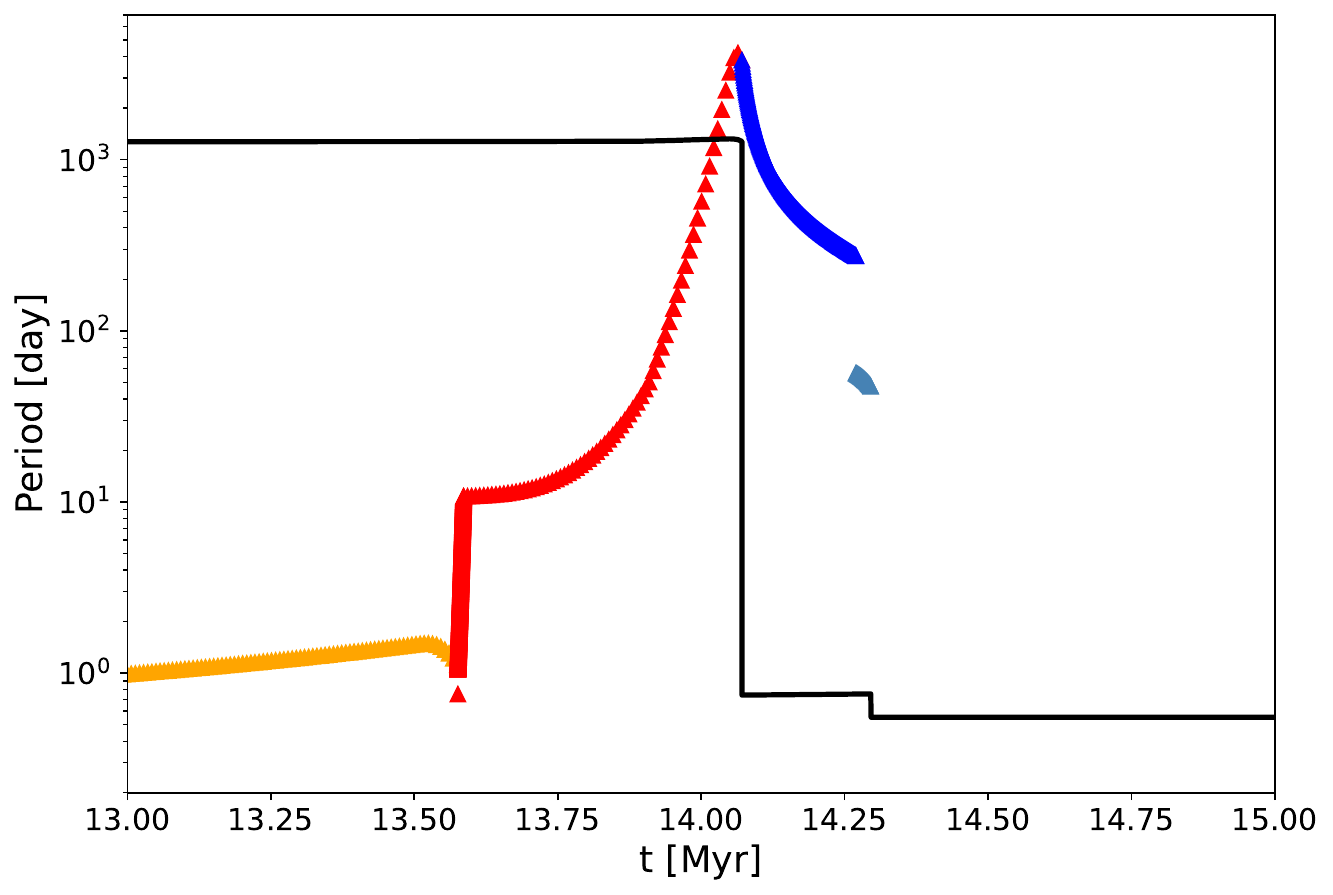}
	\caption{The legends and the descriptions are the same as in Fig.~\ref{fig:cecase}.
	This binary evolution example also forms a BBH merger.
	However, due to the relatively wide binary separation in the WR-BH phase and the correspondingly
	weak tidal interaction, the WR member fails to spin-up up to a $<1$ day spin period (right panel).
	Consequently, the merger would occur between two non-spun-up BHs if a spin-up check is applied (Sec.~\ref{bhspin}).
	On the other hand, since the orbital period in the WR-BH phase is still $<1$ day, the second-born
	BH would have been designated as spun-up had no spin-up check been applied. 
        This example corresponds to a binary evolved at $Z=0.001$ with
	initial ZAMS member masses, separation, and eccentricity being $10.9\Ms,28.8\Ms,1048.2\Rs,0.62$, respectively.
	}
\label{fig:cecase_nosync}
\end{figure*}

\begin{figure*}[!h]
\centering
\includegraphics[width = 8.9 cm, angle=0.0]{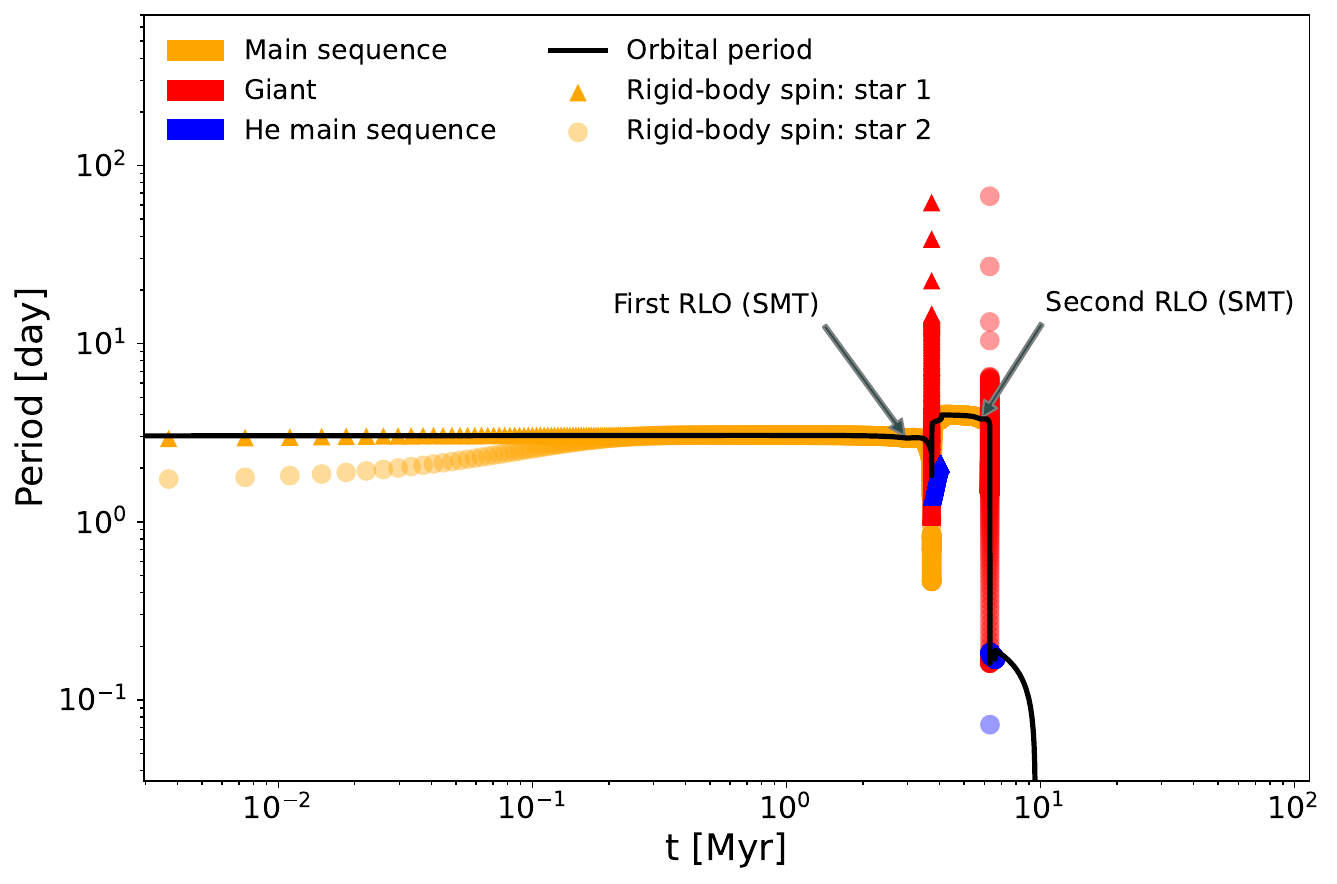}
\includegraphics[width = 8.9 cm, angle=0.0]{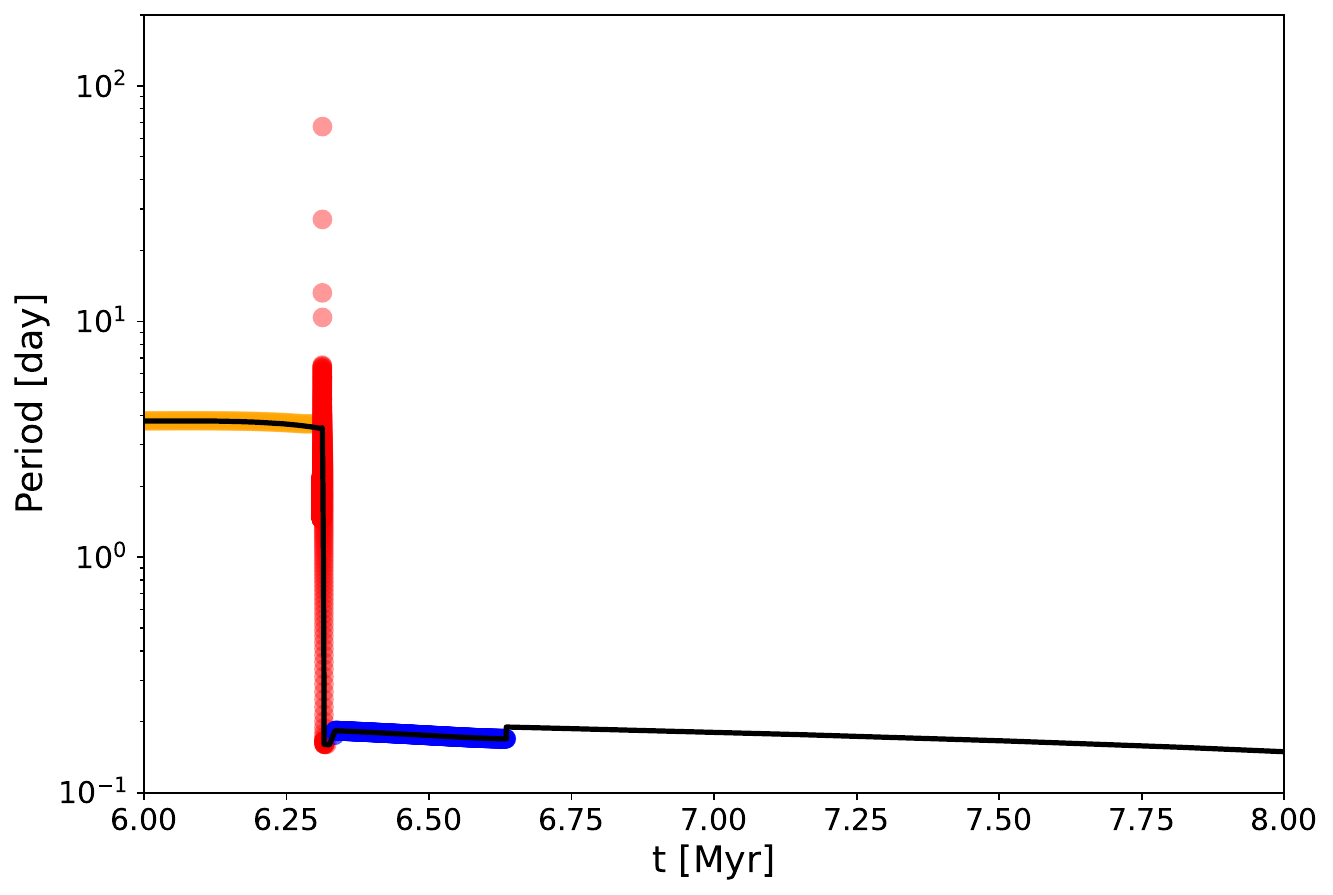}
	\caption{The legends and the descriptions are the same as in Fig.~\ref{fig:cecase}. In this example,
	the WR-BH binary formation and the BBH merger occur purely via SMT. Due to the 
	tight binary separation and the SMT episodes, the progenitor of the second-born BH
	remains spin-orbit synced for most of its life. Here, the orbit shrinkage before the
	WR-BH formation is due to dynamical-timescale mass transfer during the second
	RLO. This `second SMT' initiates between the first-born BH and its late-MS companion star,
	the companion being stripped of the H-envelope in its CHeB stage during the SMT to form the WR star.
	Due to the $<1$ day spin period ($=$ binary orbital period) of the WR star in the
	WR-BH phase, the second-born BH has a spun-up Kerr parameter (Sec.~\ref{bhspin}).
        This example corresponds to a binary evolved at $Z=0.001$ with
	initial ZAMS member masses, separation, and eccentricity being $82.0\Ms,37.2\Ms,43.3\Rs,0.0$, respectively.
	}
\label{fig:smtcase}
\end{figure*}

\begin{figure*}[!h]
\centering
\includegraphics[width = 8.9 cm, angle=0.0]{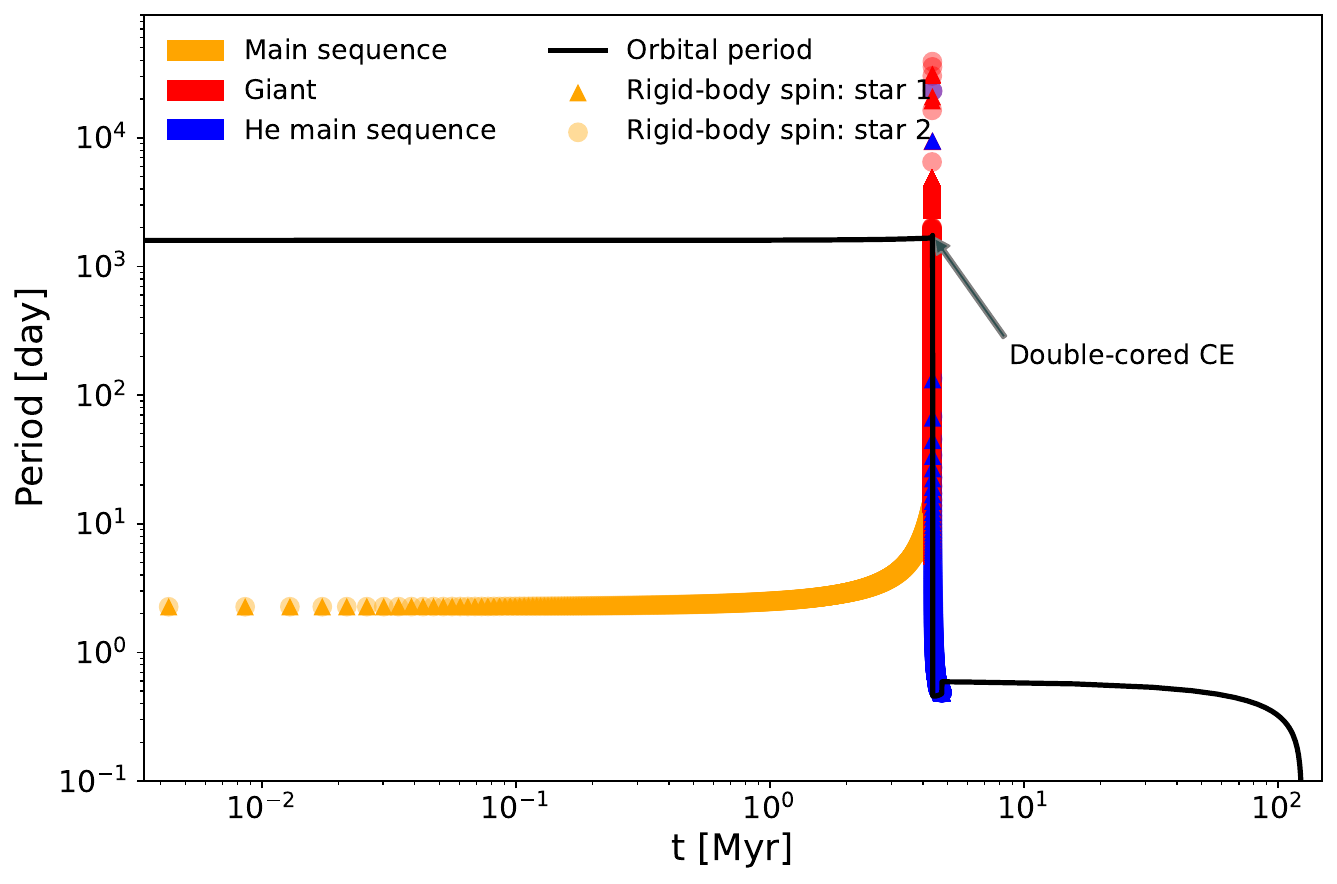}
\includegraphics[width = 8.9 cm, angle=0.0]{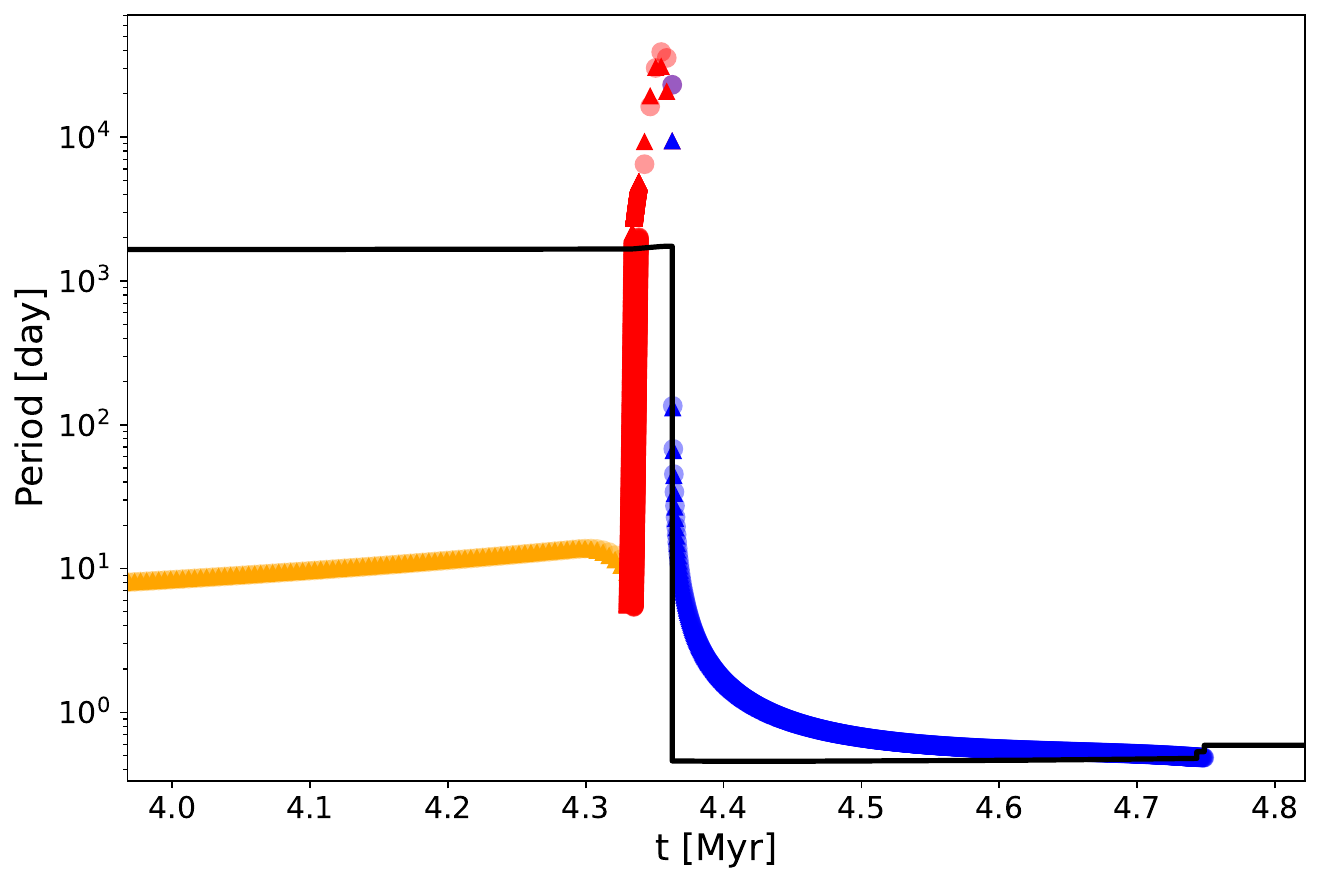}
	\caption{Formation of BBH merger via a double-cored CE evolution.
	The legends and the descriptions are the same as in Fig.~\ref{fig:cecase}.
	The nearly identically evolving binary members (mass ratio $\approx1$) comprise a contact binary
	when both of the members fill the Roche lobe during their CHeB
	phase. The resulting double-cored CE \citep{Hurley_2002} greatly shrinks the
	binary orbit, and the envelope ejection forms a tight WR-WR binary. 
	The WR members tidally spin up and catch up with the $<1$ day orbital period
	(right panel). In this case, both BHs are born spun-up (and nearly simultaneously)
	with $\abhfirst\approx\abh$. 
	The evolutionary tracks of the members being nearly identical,
	they lie onto each other so that only one track is visible.
        This example corresponds to a binary evolved at $Z=0.001$ with
	initial ZAMS member masses, separation, and eccentricity being $56.3\Ms,56.2\Ms,2768.8\Rs,0.5$, respectively.
	}
\label{fig:dcecase}
\end{figure*}

\clearpage




\section{Mass and spin distributions of isolated-binary BBH mergers for additional cases}\label{dists_more}

In this section, we present counterpart figures of several additional cases of isolated binary evolution,
as described in the figures' captions. 

\begin{figure*}[!h]
\centering
\includegraphics[width = 16.0 cm, angle=0.0]{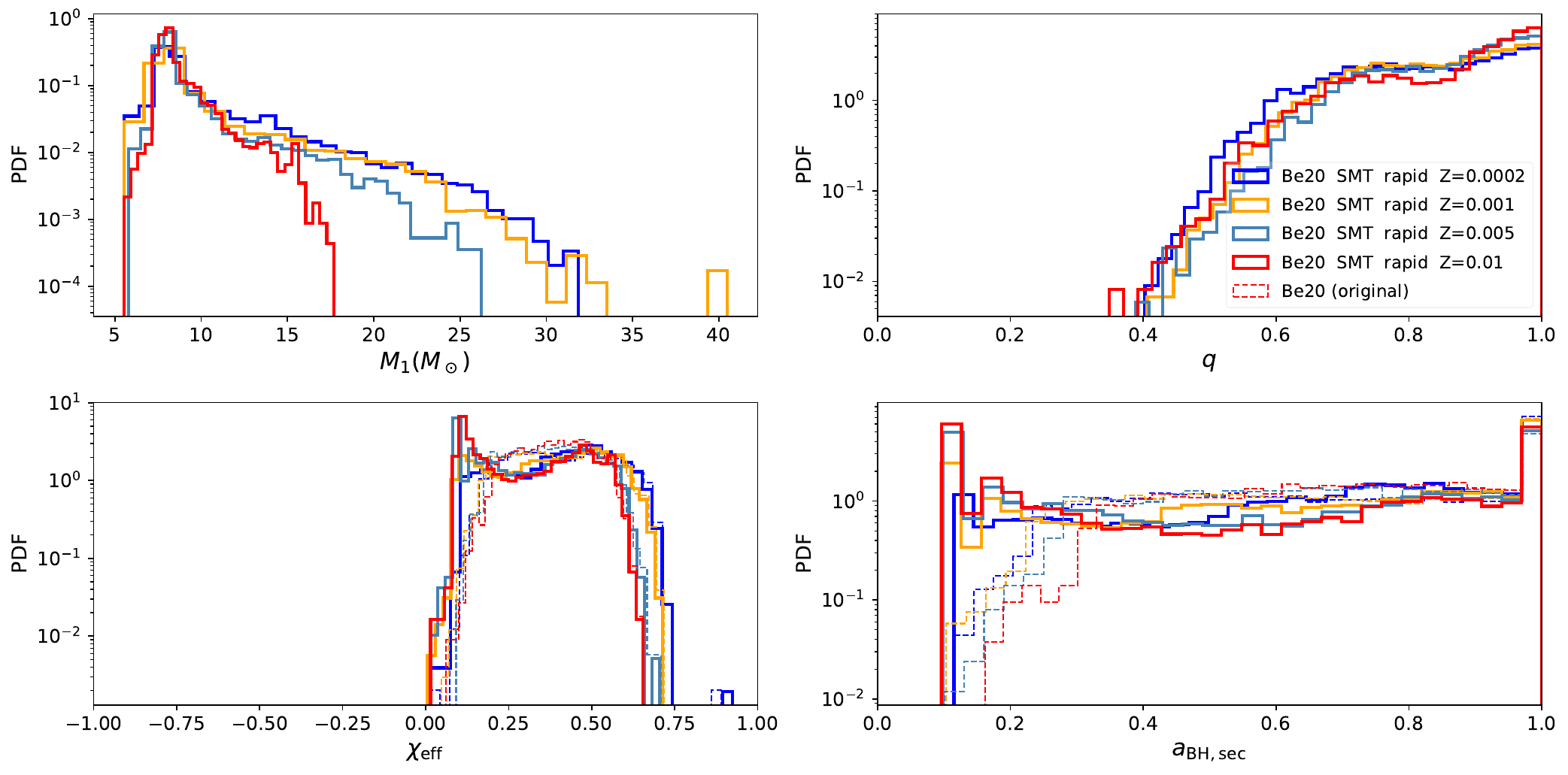}
\includegraphics[width = 16.0 cm, angle=0.0]{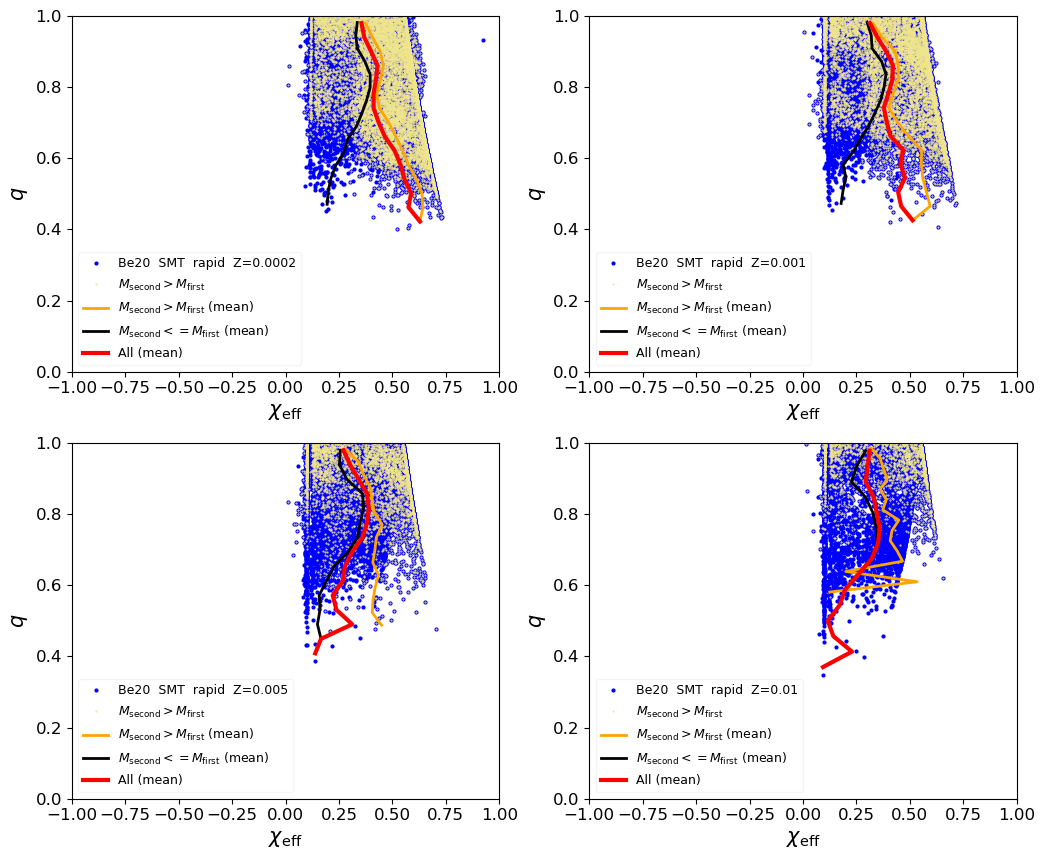}
	\caption{Same description as for Fig.~\ref{fig:dists_wrf2_qc-8_hgf-1} applies, 
	except that the `rapid' remnant mass model is applied.}
\label{fig:dists_nsf3_wrf2_qc-8_hgf-1}
\end{figure*}

\begin{figure*}
\centering
\includegraphics[width = 16.7 cm, angle=0.0]{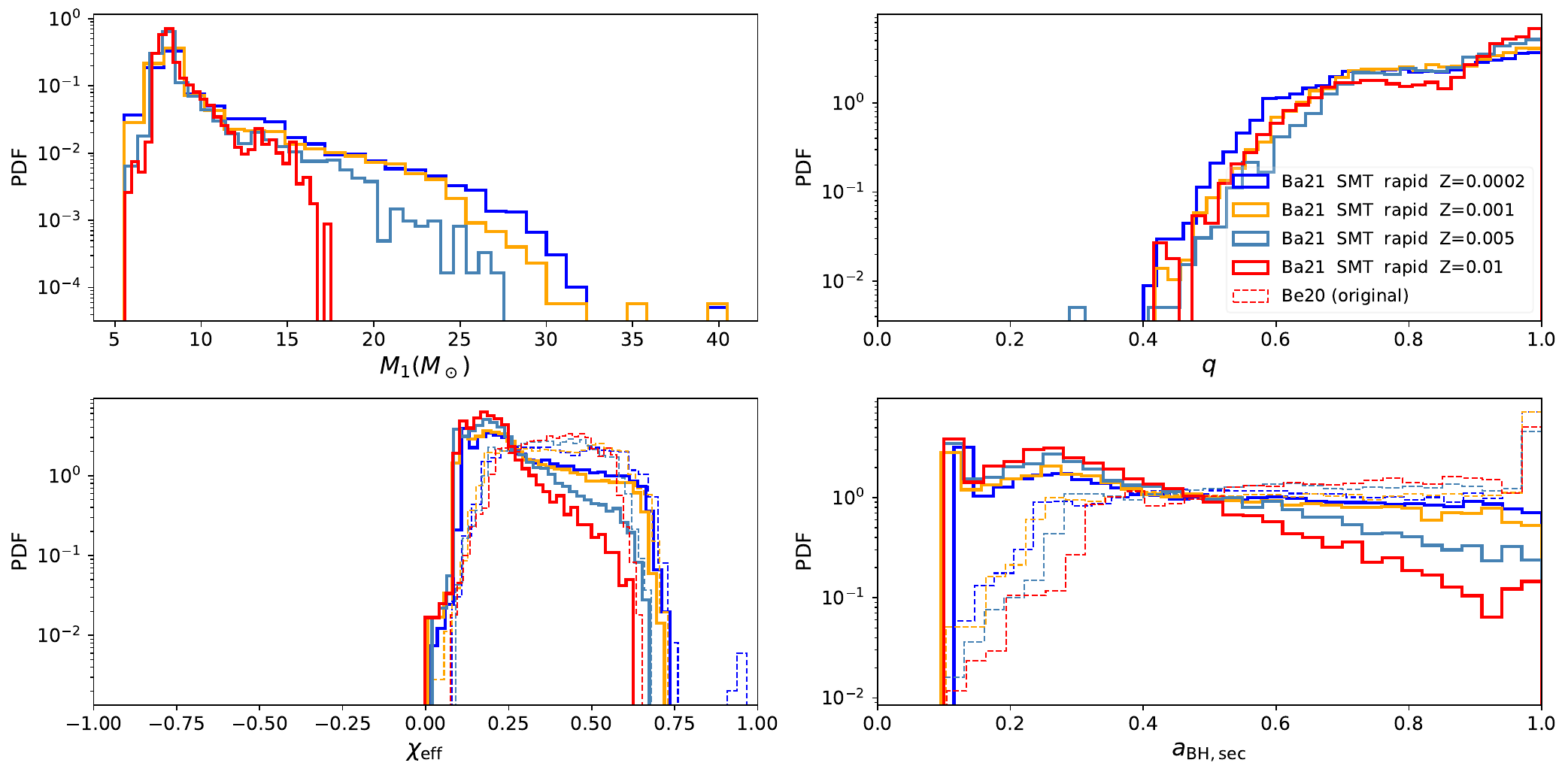}
\includegraphics[width = 16.7 cm, angle=0.0]{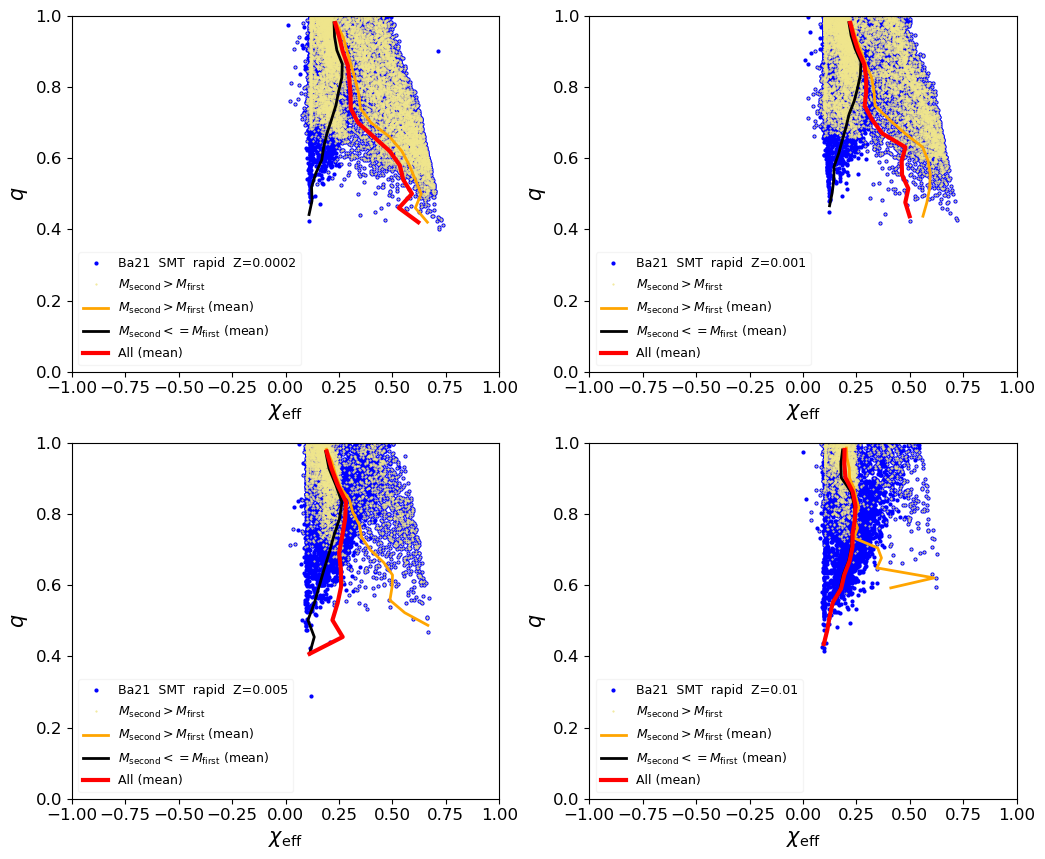}
	\caption{Same as in Fig.~\ref{fig:dists_wrf4_qc-8_hgf-1}, except that the `rapid' remnant mass
	model is applied. See the caption of Fig.~\ref{fig:dists_wrf2_qc-8_hgf-1} for a full
	description.}
\label{fig:dists_nsf3_wrf4_qc-8_hgf-1}
\end{figure*}

\begin{figure*}
\centering
\includegraphics[width = 16.7 cm, angle=0.0]{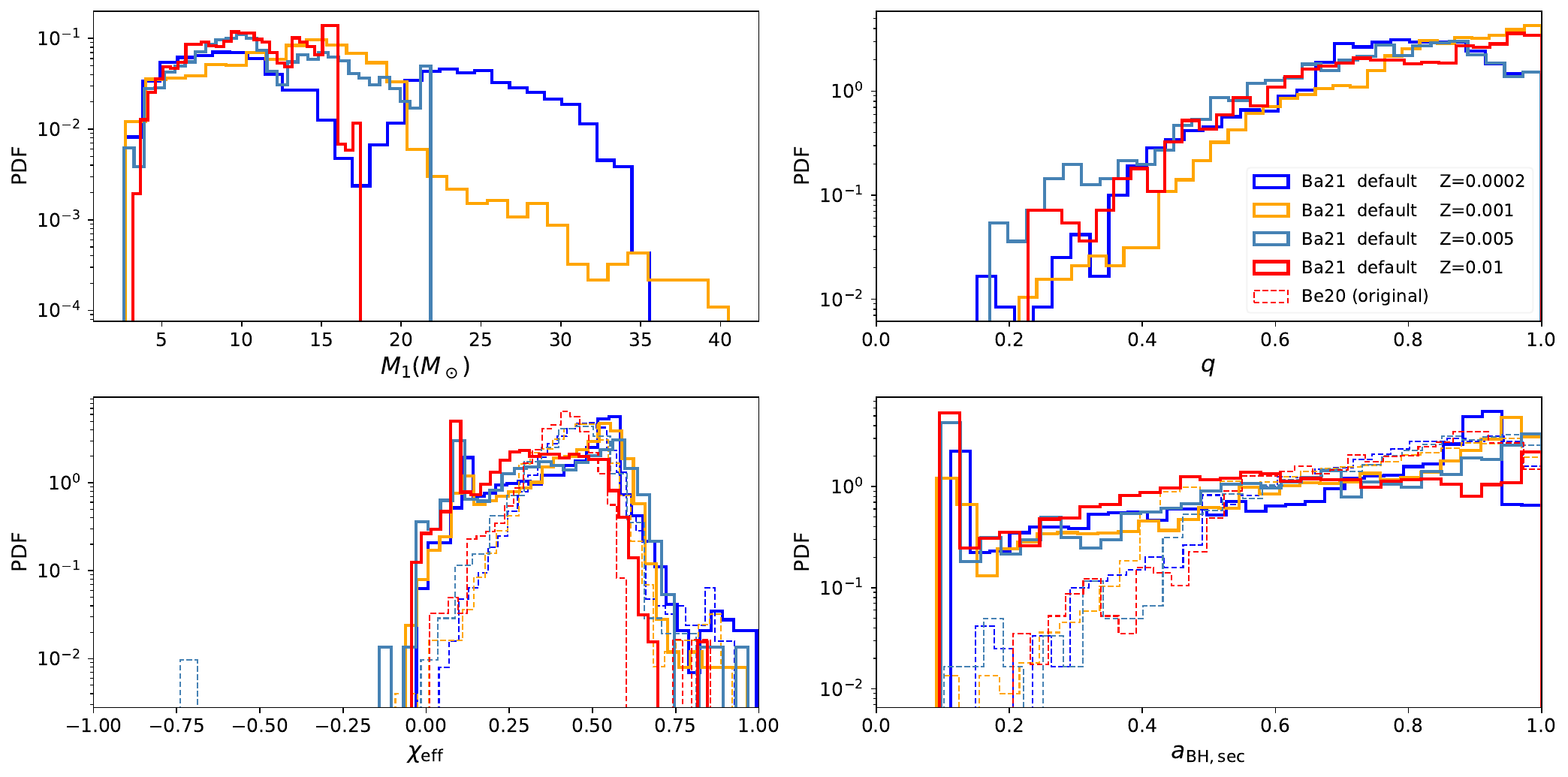}
\includegraphics[width = 16.7 cm, angle=0.0]{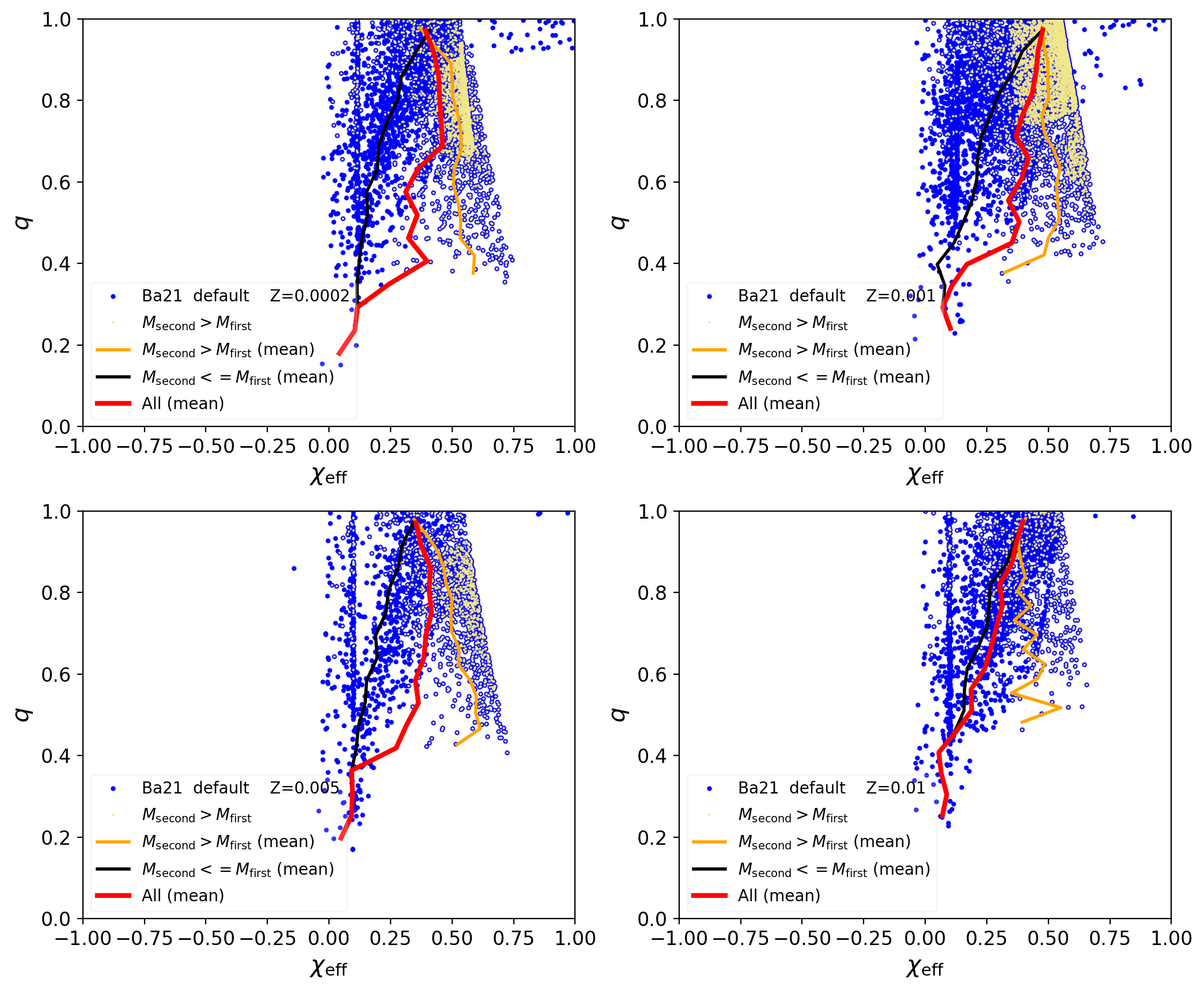}
	\caption{Same as in Fig.~\ref{fig:dists_wrf4_qc-8_hgf-1}, except that the default critical mass ratio
	criterion is applied and the HG donors are allowed to undergo CE (as defaulted in $\bse$).
	See the caption of Fig.~\ref{fig:dists_wrf2_qc-8_hgf-1} for a full
	description.}
\label{fig:dists_wrf4_qcdef_hgfdef}
\end{figure*}

\begin{figure*}
\centering
\includegraphics[width = 16.7 cm, angle=0.0]{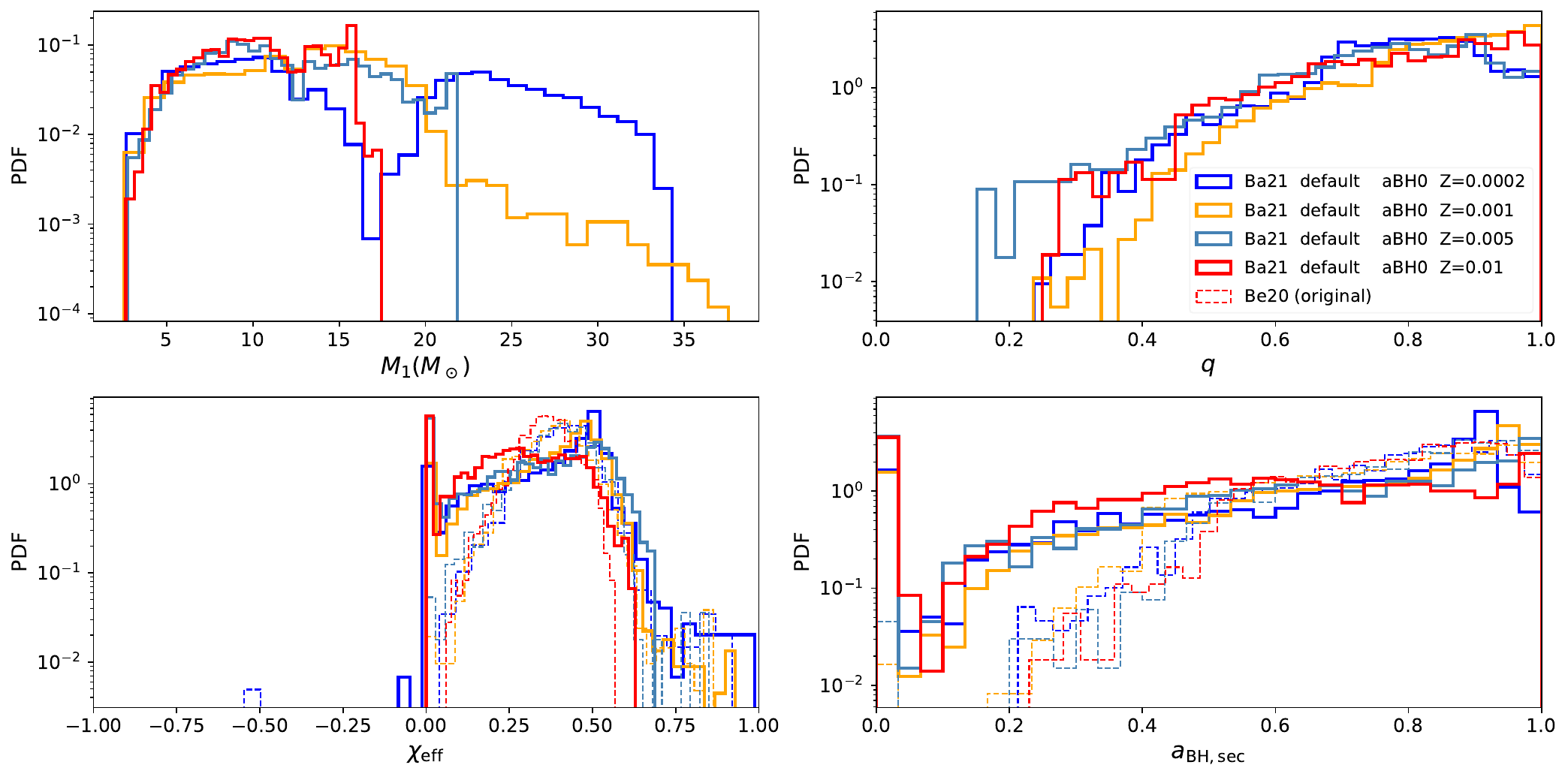}
\includegraphics[width = 16.7 cm, angle=0.0]{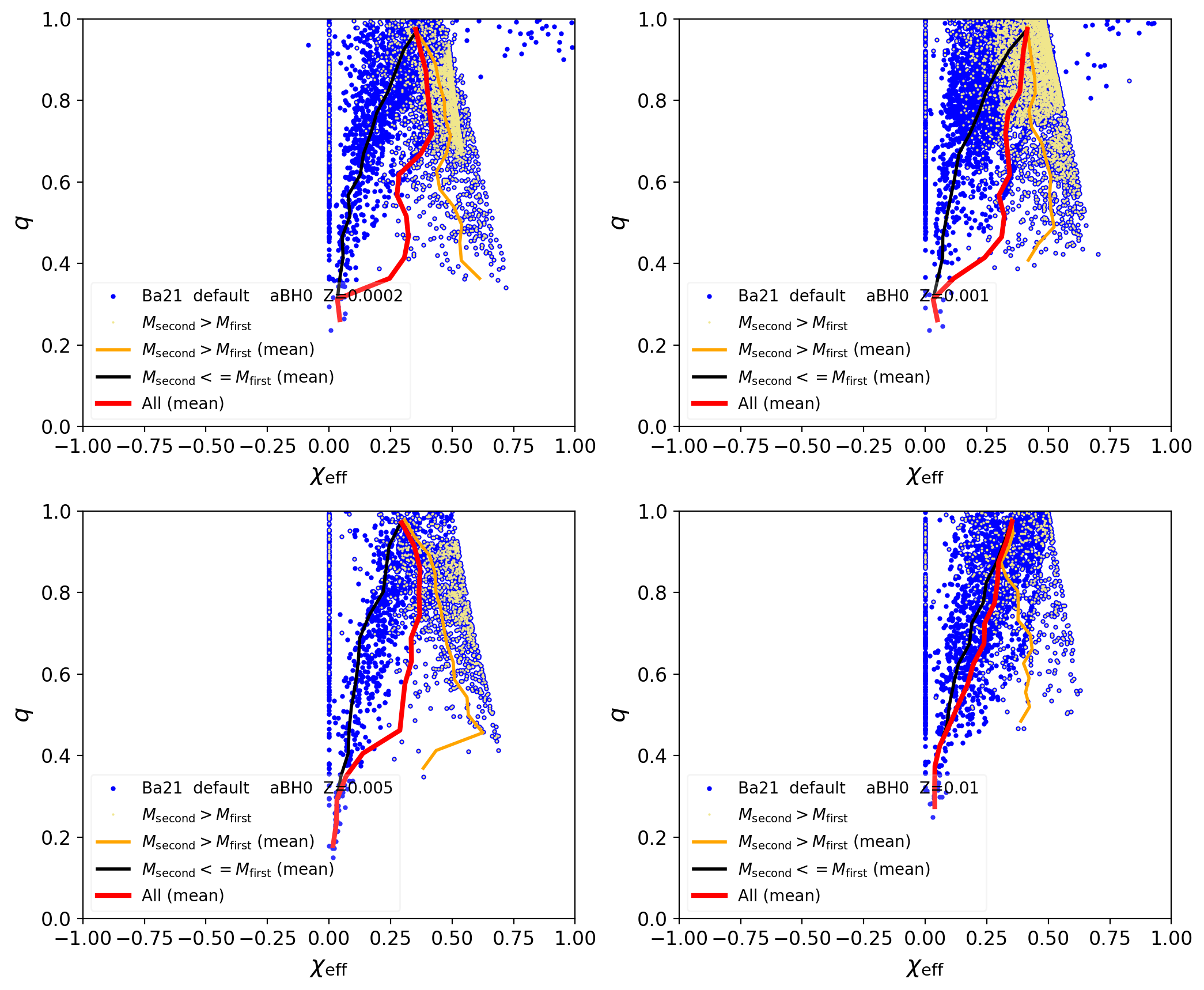}
\caption{Same as in Fig.~\ref{fig:dists_wrf4_qcdef_hgfdef}, but with the choice of zero natal spin
	for non-spun-up BHs, \ie, $\abhzero=0$. With this choice, the BH spin model matches with that in \citet{Broekgaarden2022}.}
\label{fig:dists_bhf4_wrf4_qcdef_hgfdef}
\end{figure*}

\begin{figure*}
\centering
\includegraphics[width = 16.7 cm, angle=0.0]{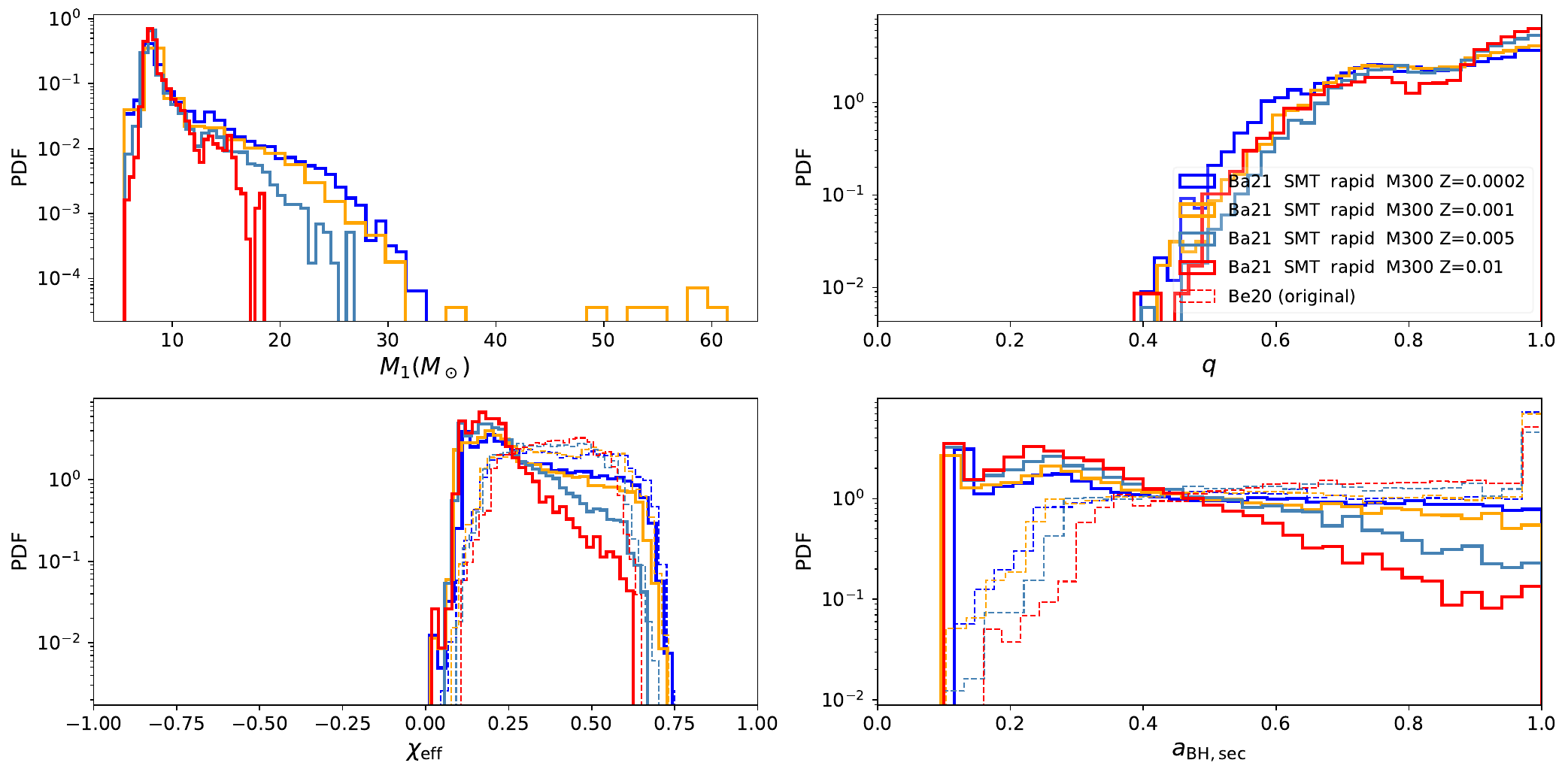}
\includegraphics[width = 16.7 cm, angle=0.0]{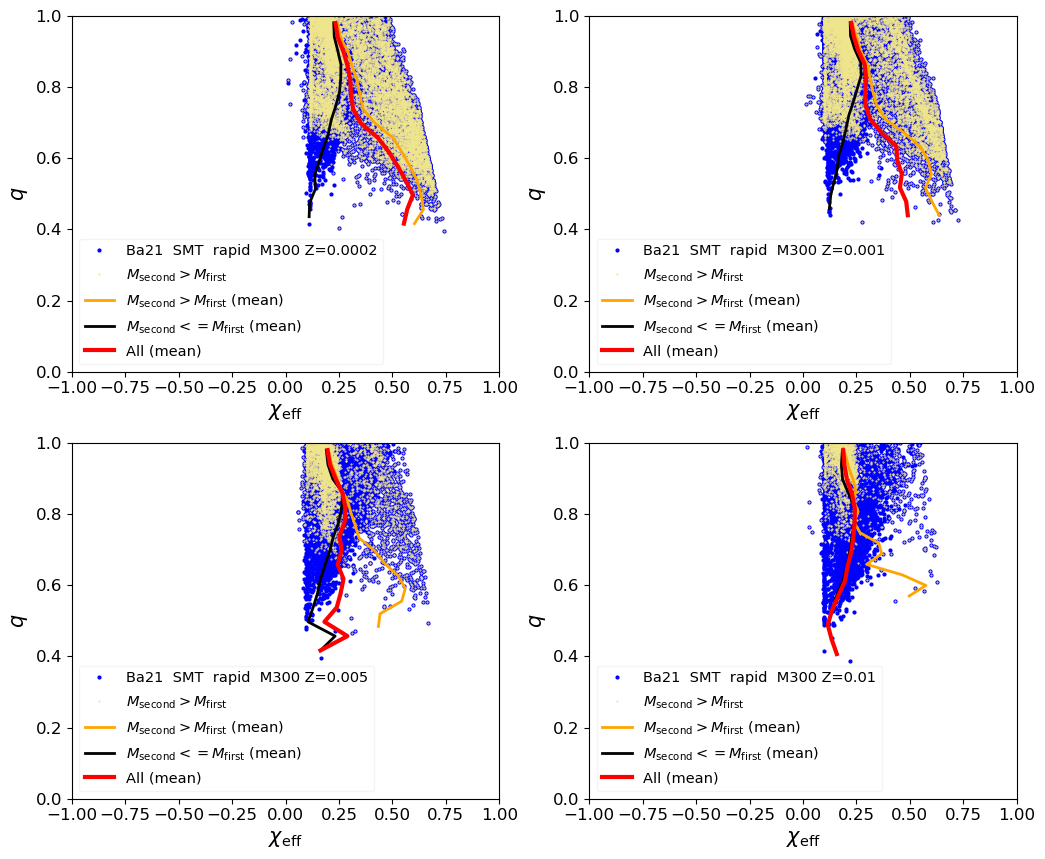}
	\caption{Same as in Fig.~\ref{fig:dists_nsf3_wrf4_qc-8_hgf-1}, except that the initial, ZAMS
	binary members have masses up to $300\Ms$ (instead of the default $150\Ms$ upper limit) and that
	PPSN/PSN is disabled. The latter allows BH formation within the canonical upper BH mass gap
	between $\approx 45\Ms - \approx 120\Ms$.}
\label{fig:dists_nsf3_wrf4_qc-8_hgf-1_mu300}
\end{figure*}

\begin{figure*}
\centering
\includegraphics[width = 16.7 cm, angle=0.0]{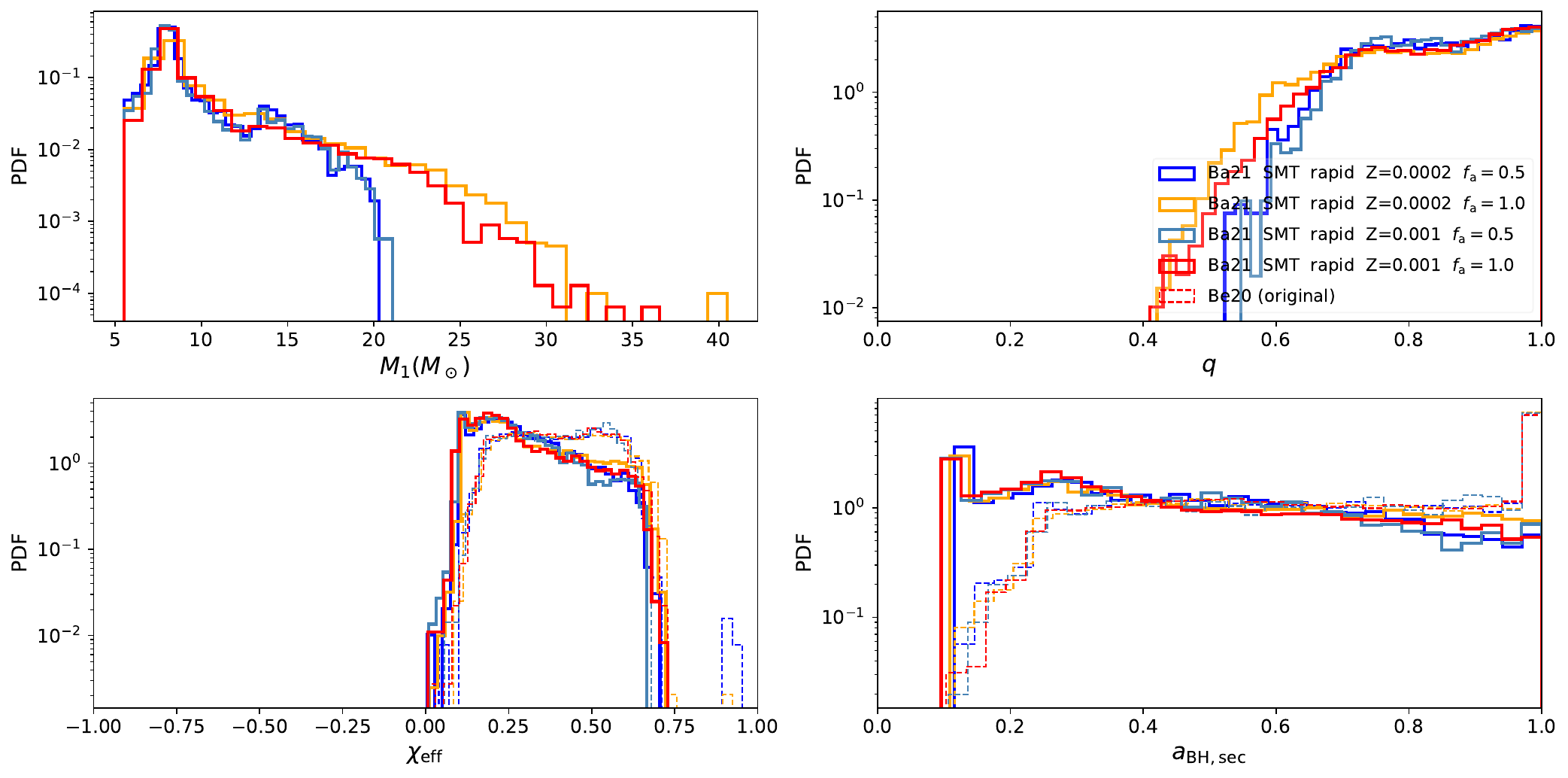}
\includegraphics[width = 16.7 cm, angle=0.0]{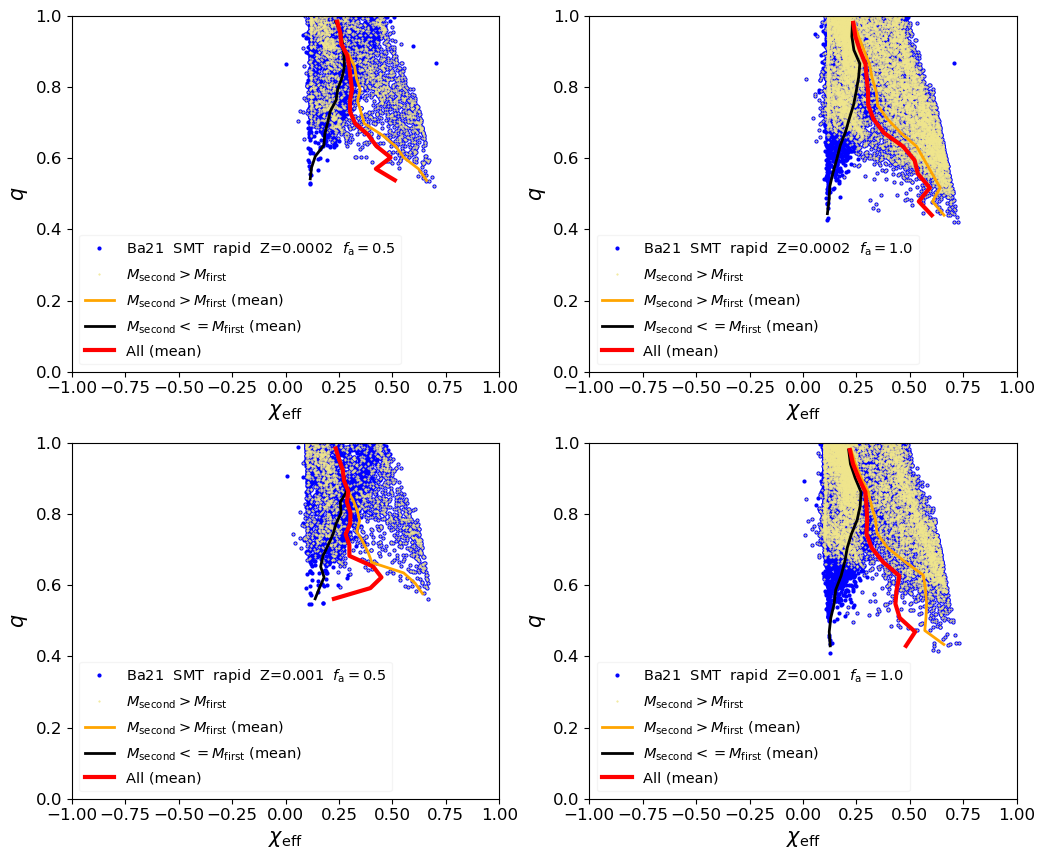}
	\caption{The format of the legend is the same as in
	Fig.~\ref{fig:dists_nsf3_wrf4_qc-8_hgf-1}. However, only two metallicities
	are considered here and the mass accretion efficiency, $f_a$, is varied as well (legend). For all other
        cases presented in this paper, the default value of $f_a=1$ is always applied.
	See text for further details.}
\label{fig:dists_nsf3_wrf4_qc-8_hgf-1_fa}
\end{figure*}

\begin{figure*}
\centering
\includegraphics[width = 17.5 cm, angle=0.0]{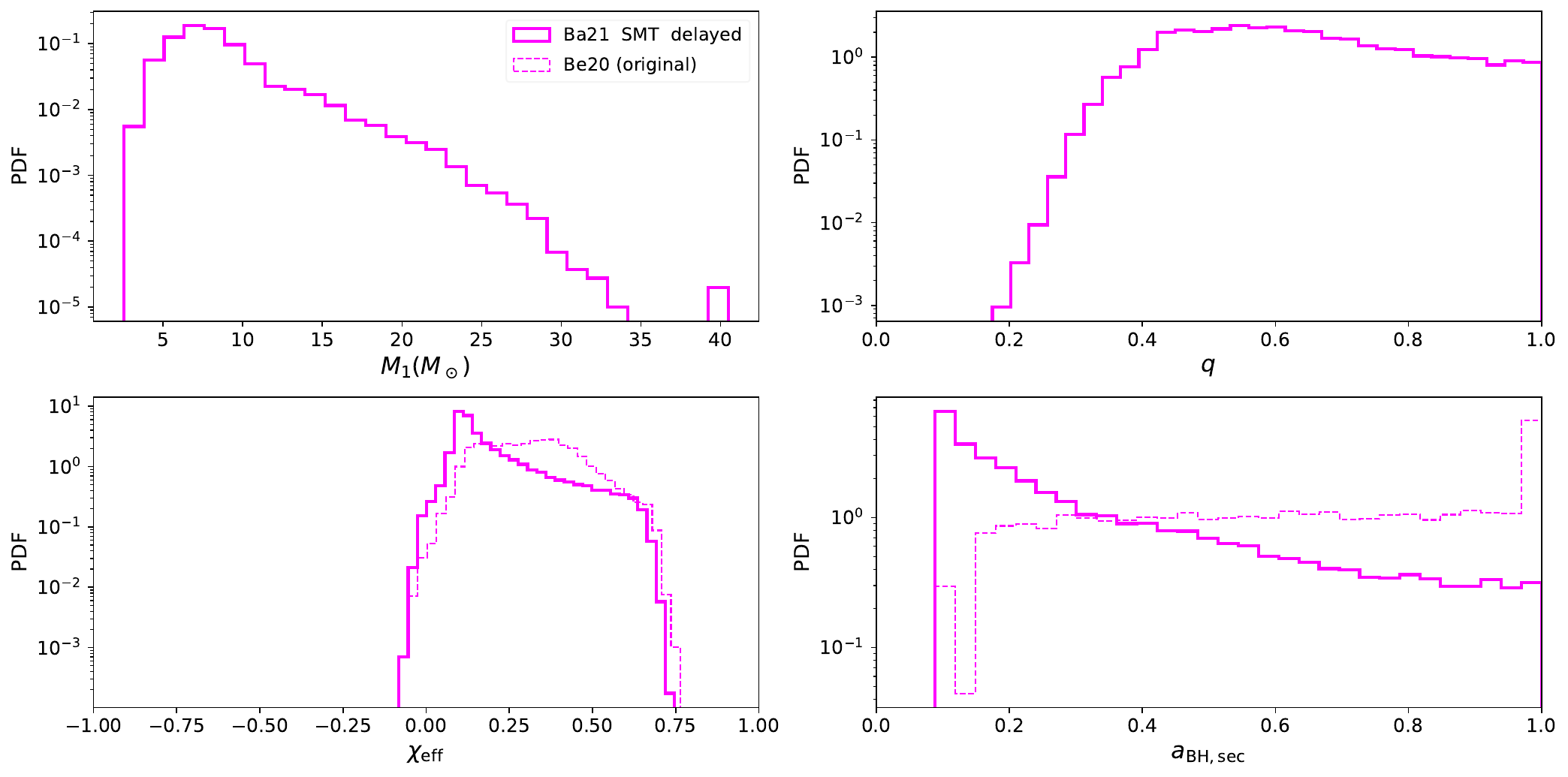}
	\caption{Intrinsic normalized distributions of $\mone$, $q$, $\xeff$, $\abh$ (X-axis label)
	of only the SMT-channel BBH mergers (see text) that are observable
	in the present epoch (redshift $\approx0$),
	as obtained from one of the isolated binary population syntheses in this study (see text) that assumes the
	delayed remnant-mass model and the Ba21 BH-spin prescription (without spin-up check).
	As in the previous figures, the dashed-lined $\xeff$ and $\abh$ distributions
	are obtained from BH spins determined by applying the original, post-processing approach of Be20.
	The solid-lined $\xeff$ and $\abh$ distributions are obtained from BH spins
	determined by $\bse$ at runtime (with the choice of the Ba21 prescription without spin-up check).}
\label{fig:dists1_wrf4_qc-8_hgf-1_pop}
\end{figure*}

\begin{figure*}
\centering
\includegraphics[width = 17.5 cm, angle=0.0]{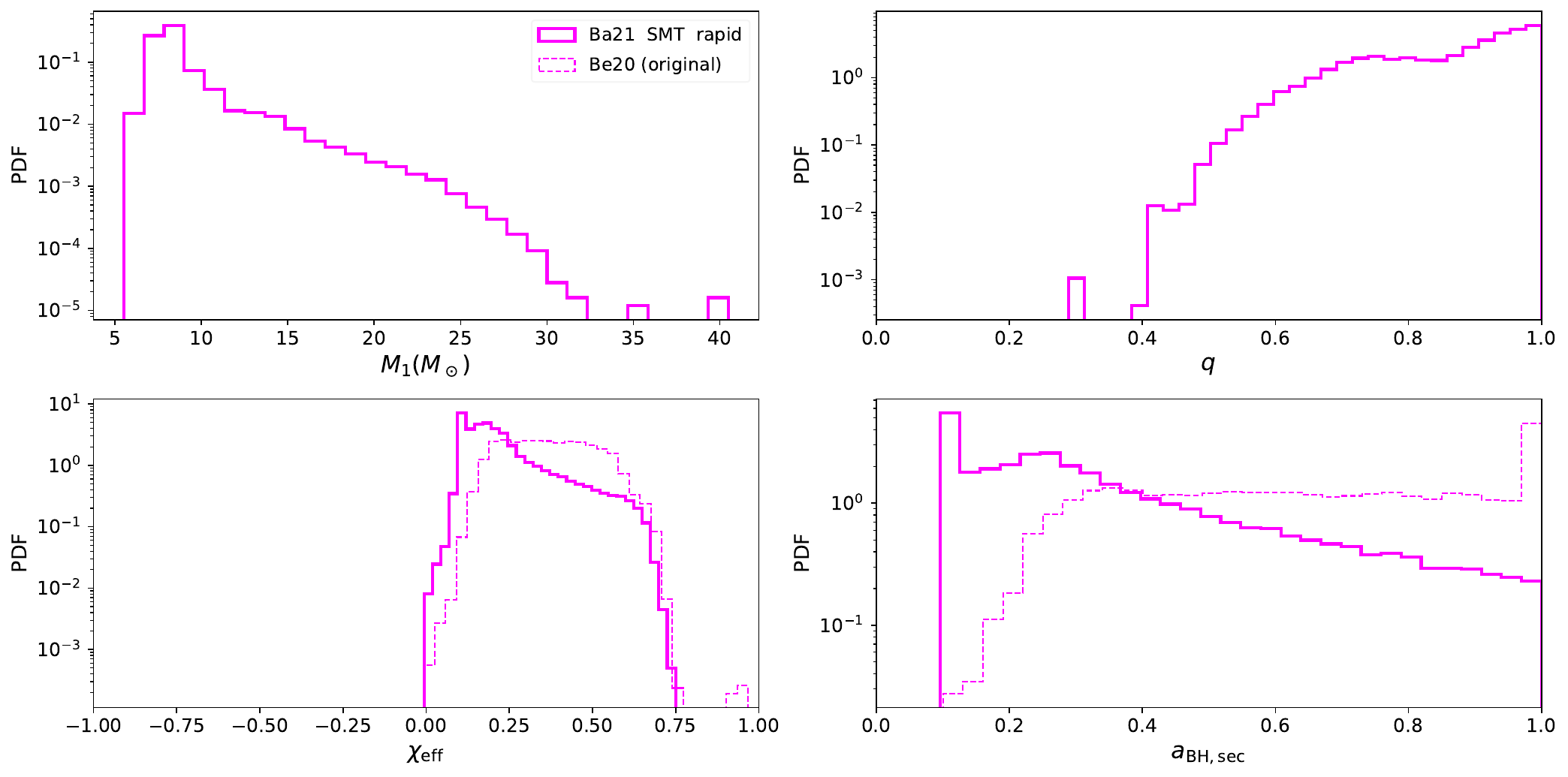}
	\caption{Same as in Fig.~\ref{fig:dists1_wrf4_qc-8_hgf-1_pop}, but applying the rapid remnant-mass
	model.}
\label{fig:dists1_nsf3_wrf4_qc-8_hgf-1_pop}
\end{figure*}

\begin{figure*}
\centering
\includegraphics[width = 17.5 cm, angle=0.0]{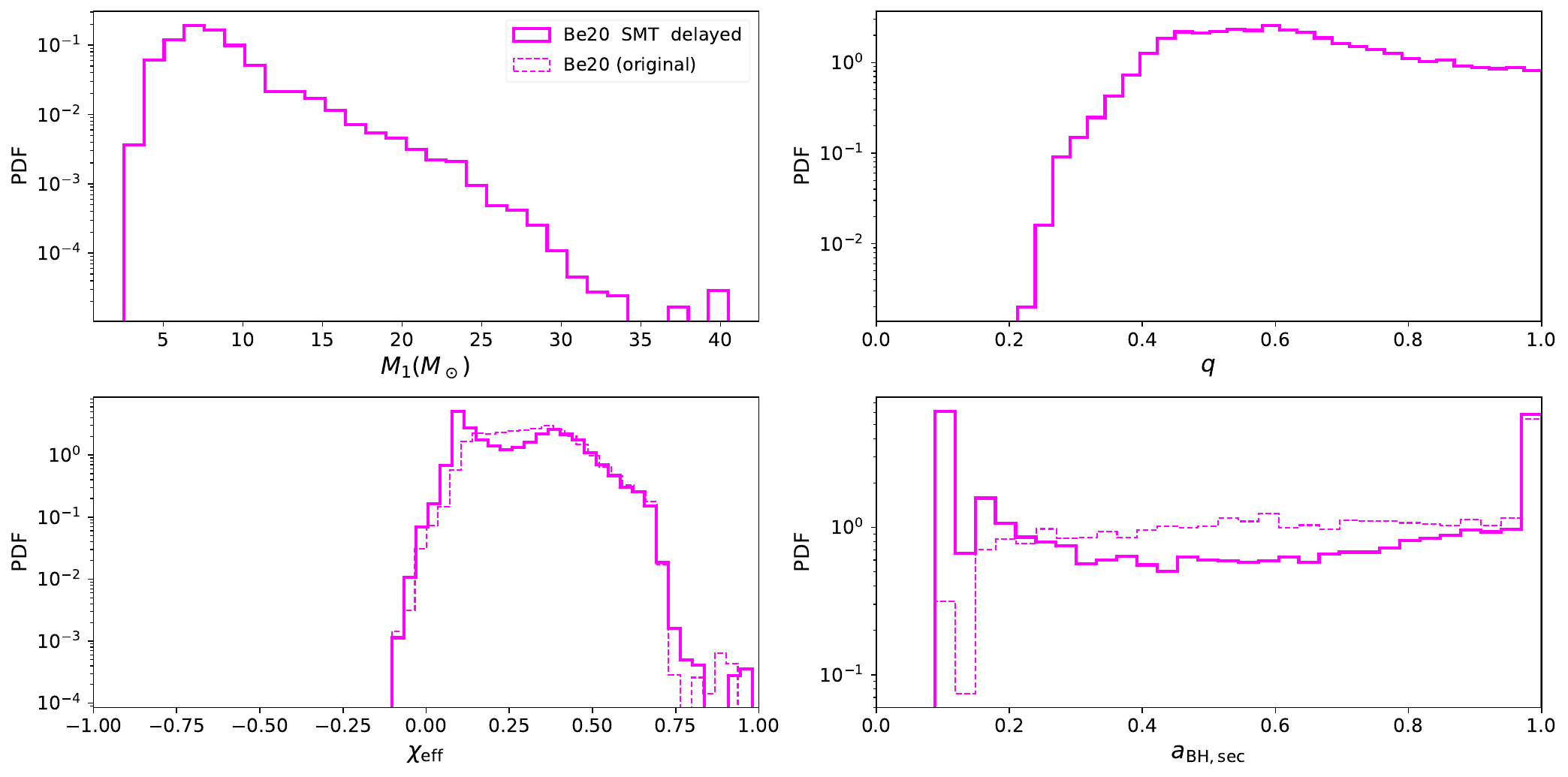}
	\caption{Same description as for Fig.~\ref{fig:dists1_wrf4_qc-8_hgf-1_pop} applies, except
	that the runtime BH-spin model is that of Be20 with spin-up check. See the caption of
	Fig.~\ref{fig:dists1_wrf4_qc-8_hgf-1_pop} for a full description.}
\label{fig:dists1_wrf2_qc-8_hgf-1_pop}
\end{figure*}

\begin{figure*}
\centering
\includegraphics[width = 17.5 cm, angle=0.0]{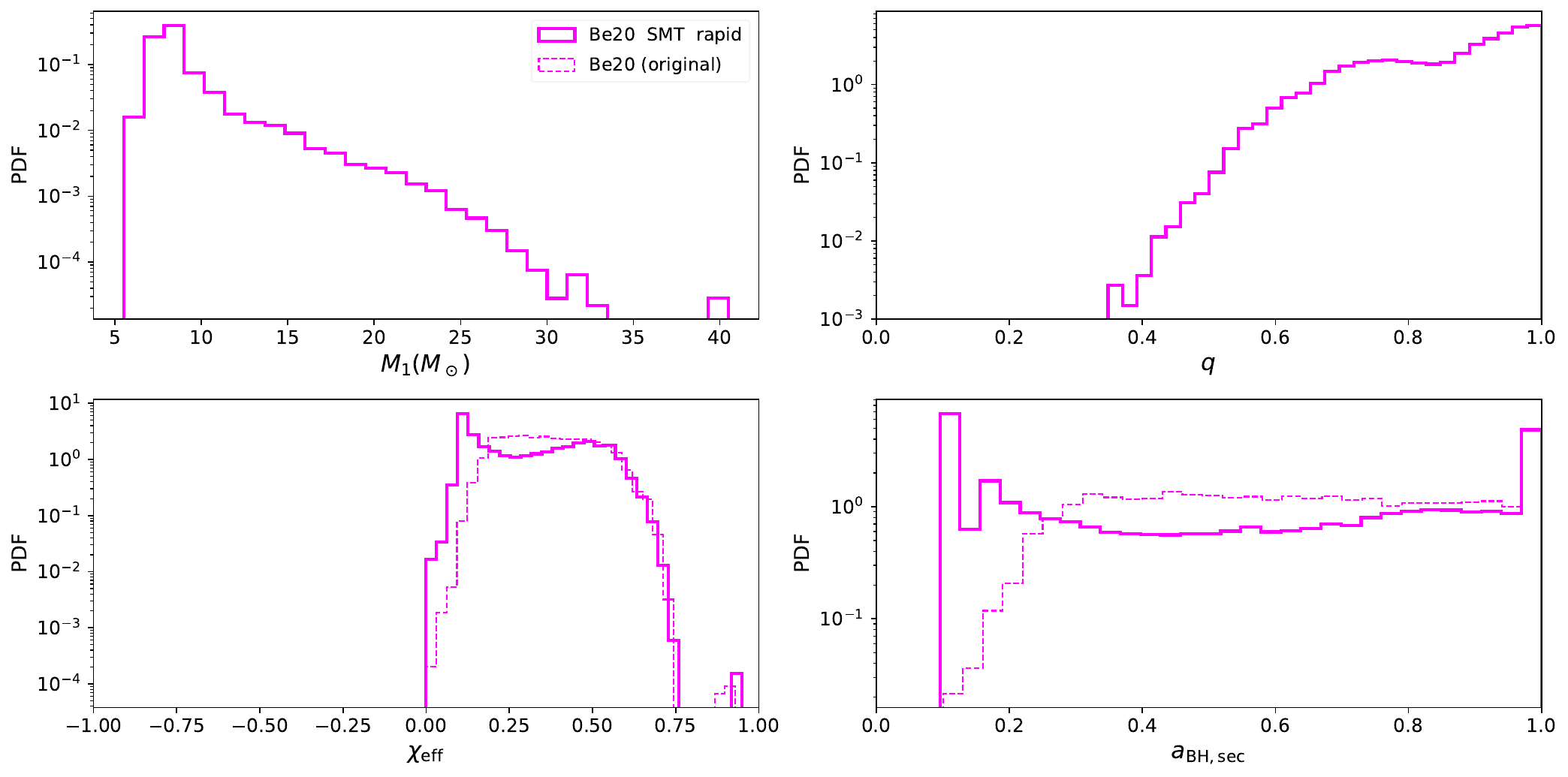}
\caption{Same as in Fig.~\ref{fig:dists1_wrf2_qc-8_hgf-1_pop}, but applying the rapid remnant-mass
	model.}
\label{fig:dists1_nsf3_wrf2_qc-8_hgf-1_pop}
\end{figure*}

\begin{figure*}
\centering
\includegraphics[width = 17.5 cm, angle=0.0]{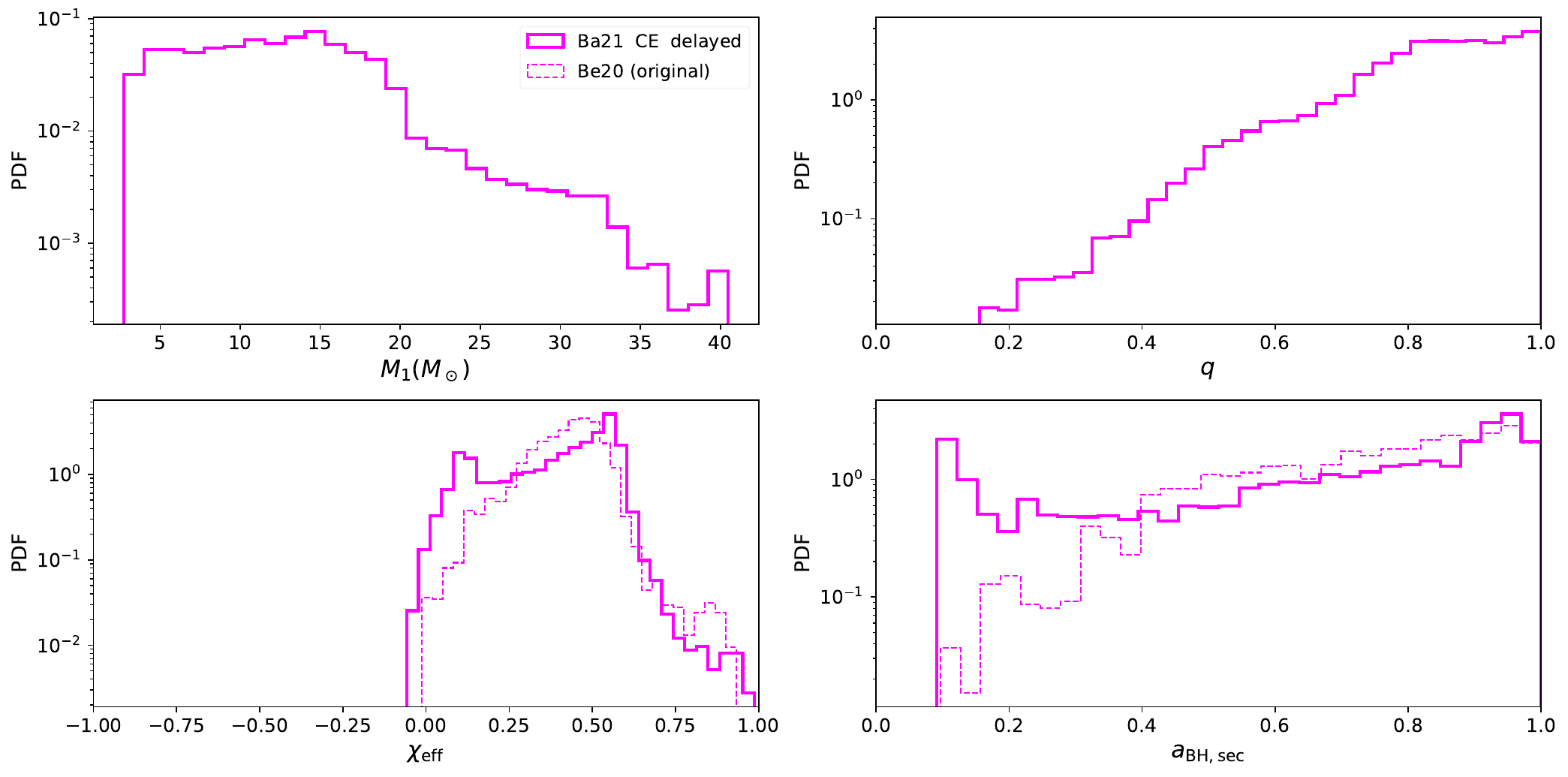}
\caption{Same description as for Fig.~\ref{fig:dists1_wrf4_qc-8_hgf-1_pop} applies,
	but for present-day-observable BBH mergers occurring via only the CE channel.}
\label{fig:dists1_wrf4_qcdef_hgf-1_pop}
\end{figure*}

\begin{figure*}
\centering
\includegraphics[width = 17.5 cm, angle=0.0]{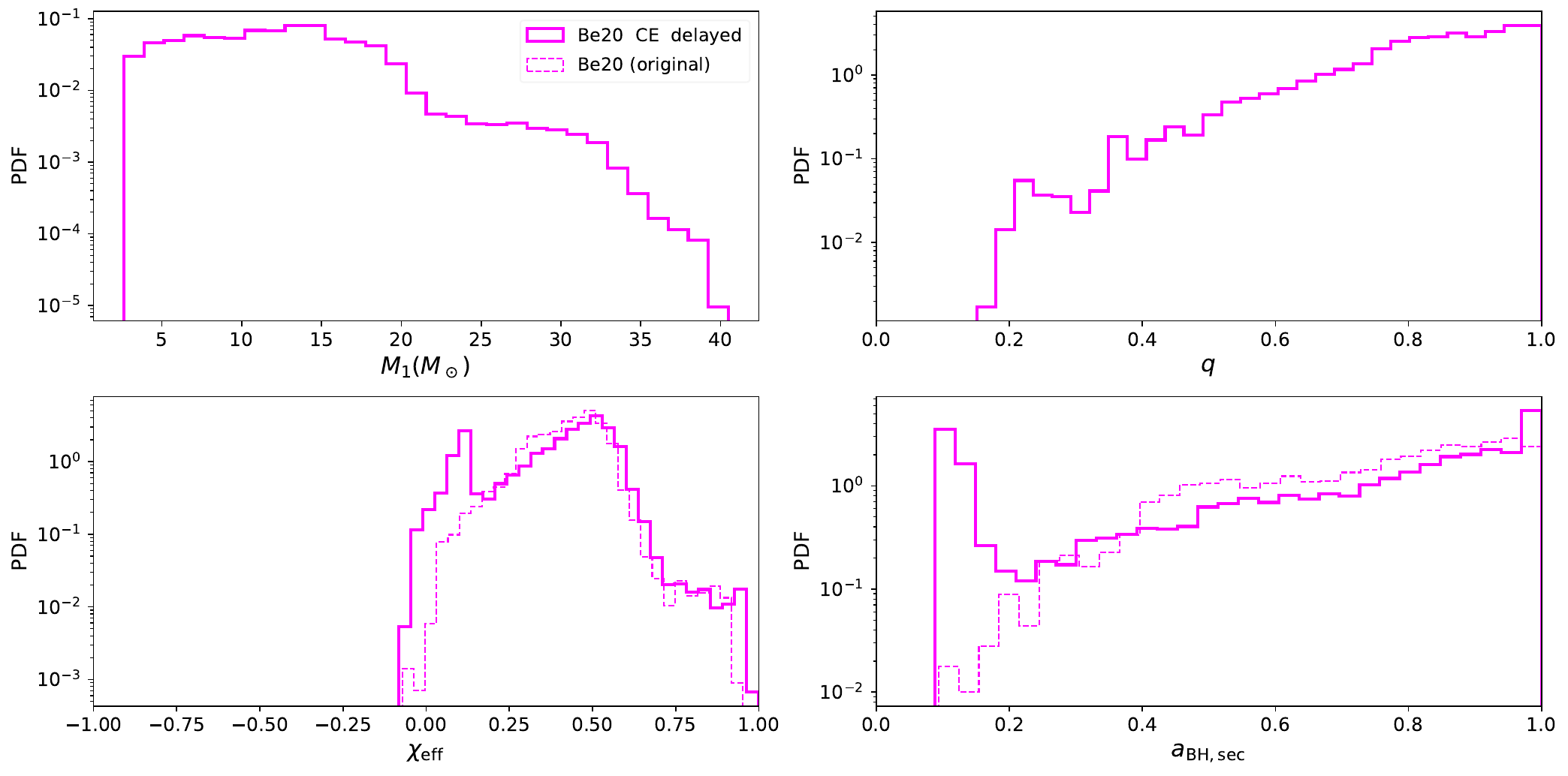}
	\caption{Same as in Fig.~\ref{fig:dists1_wrf4_qcdef_hgf-1_pop}, but for Be20 (spin-up check)
	runtime BH spin model.}
\label{fig:dists1_wrf2_qcdef_hgf-1_pop}
\end{figure*}

\begin{figure*}
\centering
\includegraphics[width = 17.5 cm, angle=0.0]{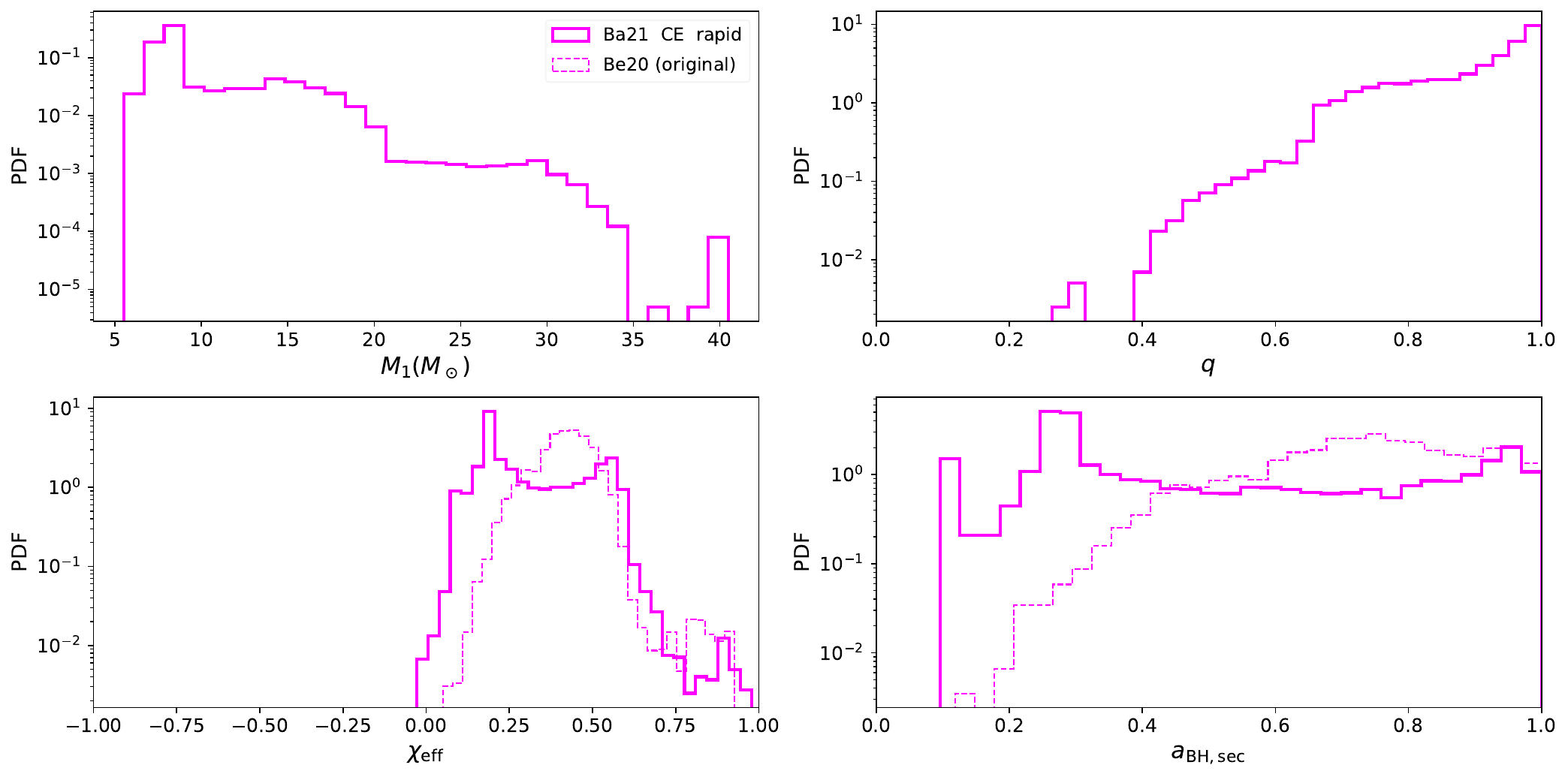}
\caption{Same as Fig.~\ref{fig:dists1_wrf4_qcdef_hgf-1_pop}, but for the rapid remnant mass model.}
\label{fig:dists1_nsf3_wrf4_qcdef_hgf-1_pop}
\end{figure*}

\begin{figure*}
\centering
\includegraphics[width = 17.5 cm, angle=0.0]{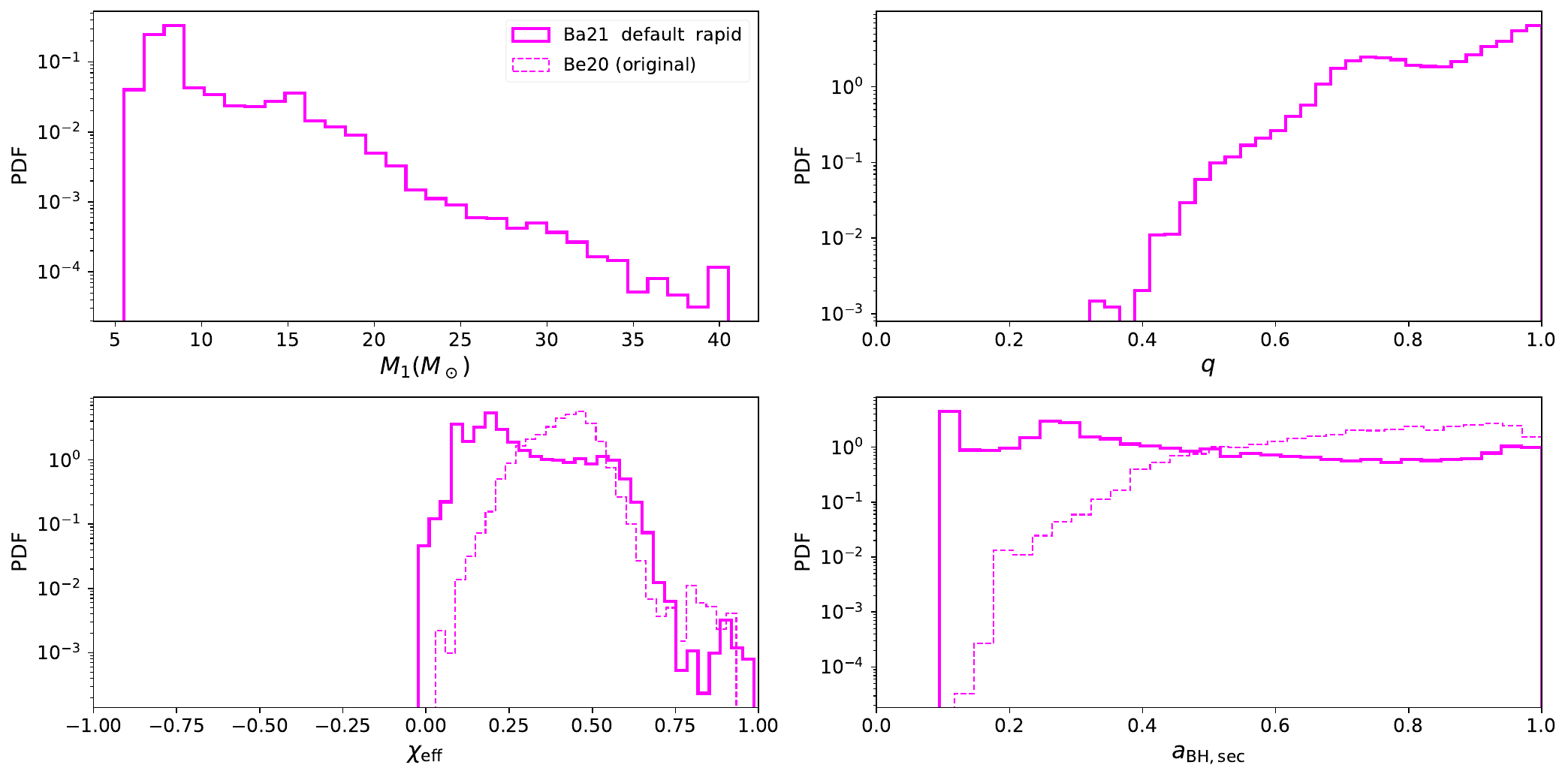}
\caption{Distributions for the Ba21 default rapid model population.}
\label{fig:dists1_nsf3_wrf4_qcdef_hgfdef_pop}
\end{figure*}

\begin{figure*}
\centering
\includegraphics[width = 8.5 cm, angle=0.0]{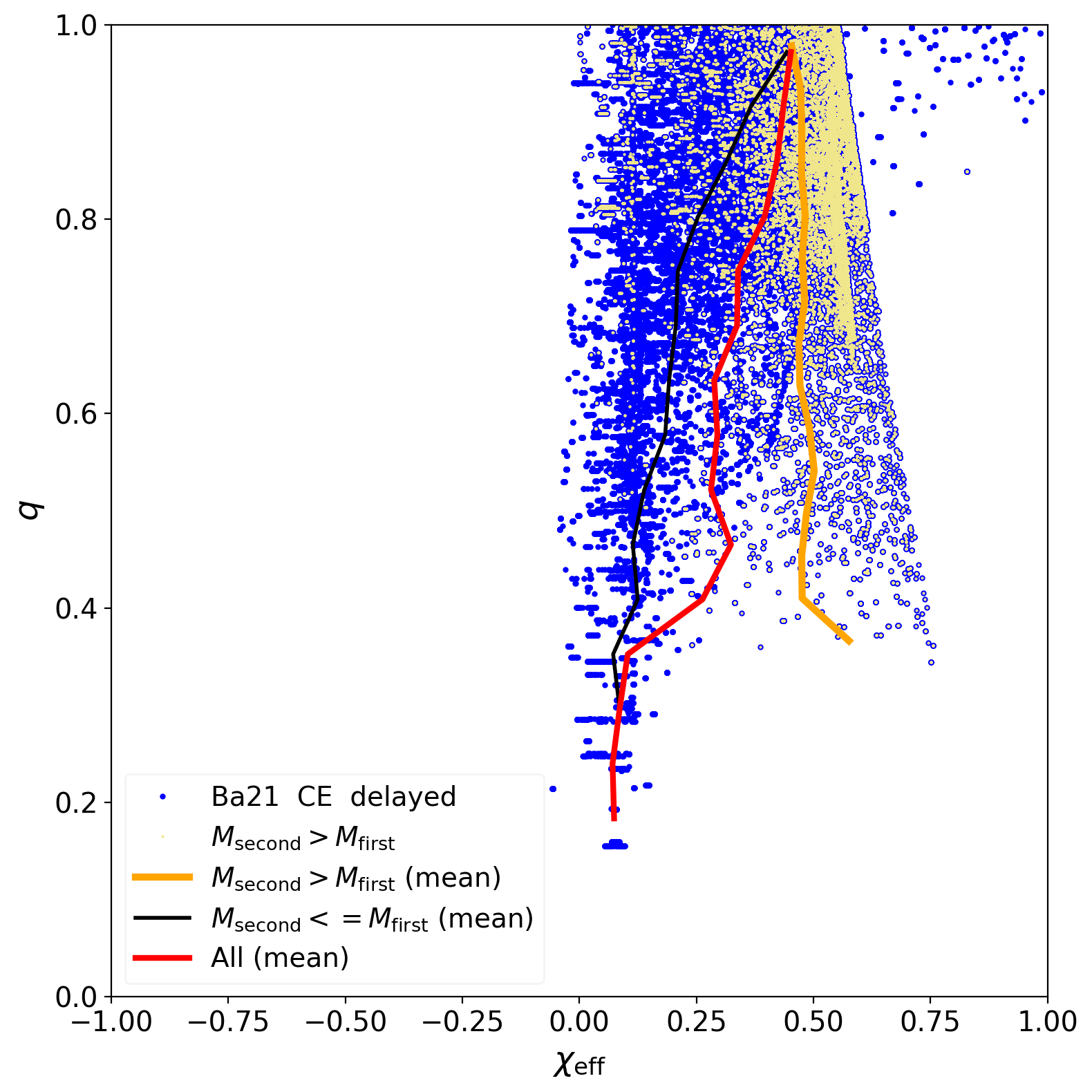}
\includegraphics[width = 8.5 cm, angle=0.0]{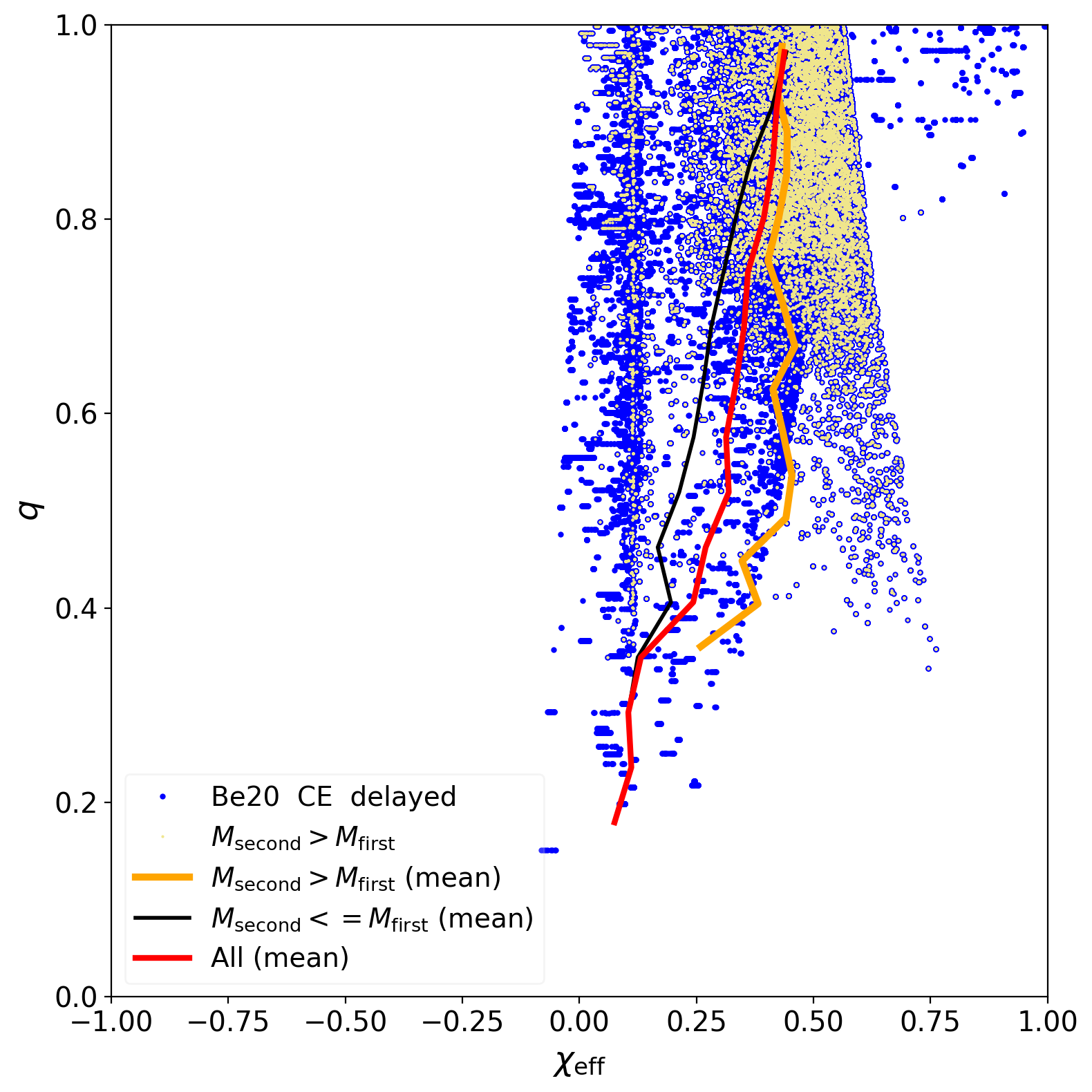}
\includegraphics[width = 8.5 cm, angle=0.0]{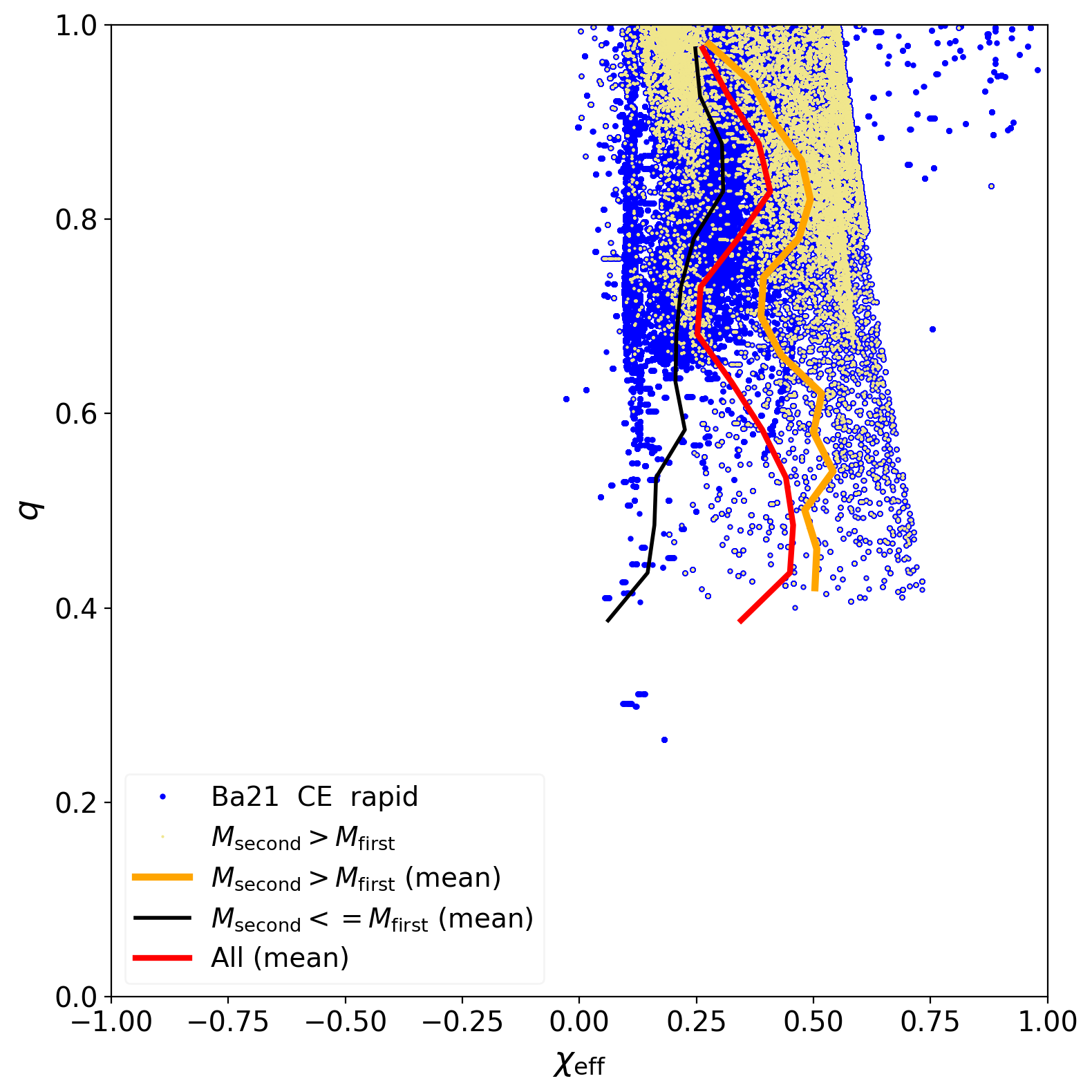}
\includegraphics[width = 8.5 cm, angle=0.0]{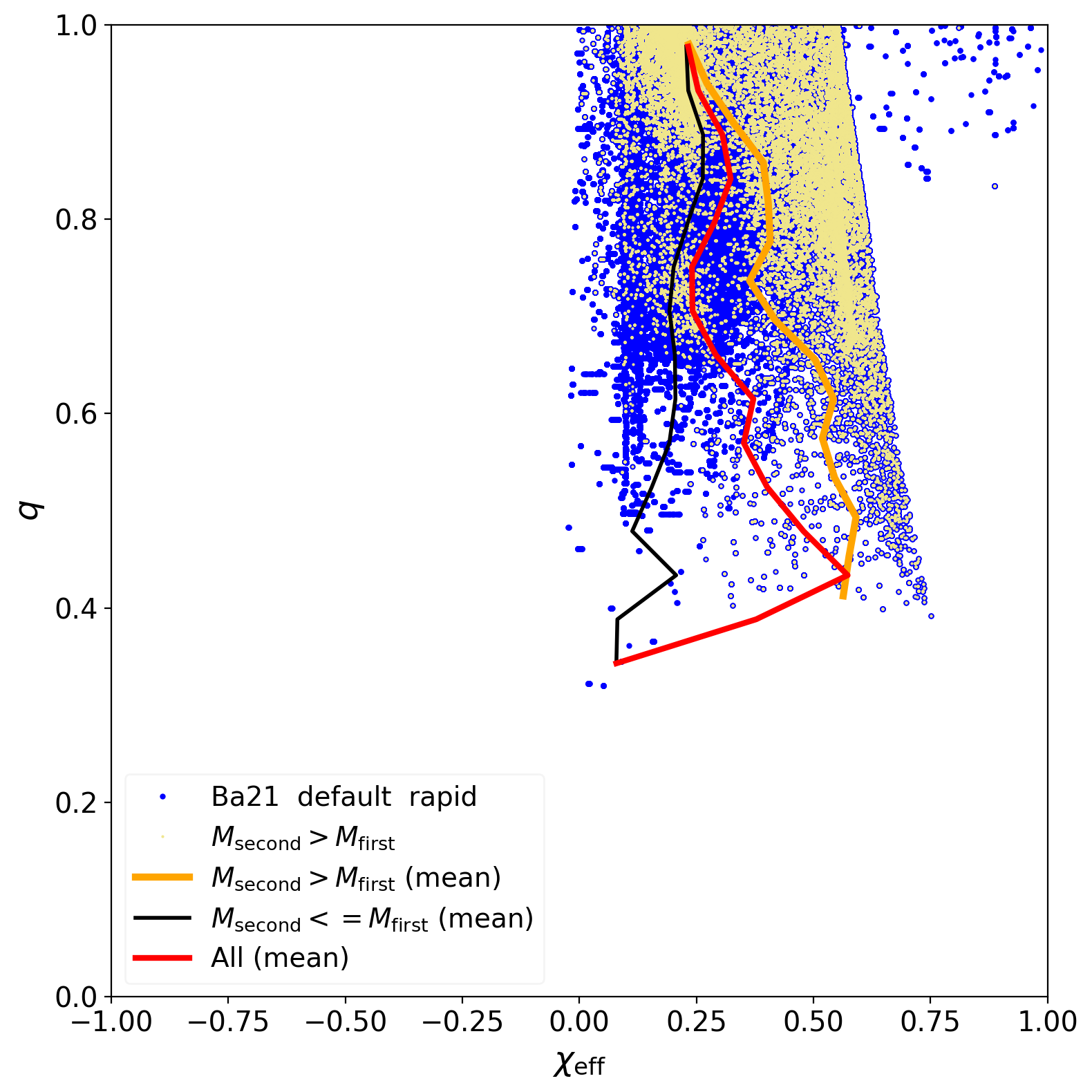}
	\caption{Scatter plots as in Fig.~\ref{fig:dists7_qc-8_hgf-1},
	but for present-day-observable BBH mergers occurring via only the CE channel (see text).}
\label{fig:dists7_qcdef_hgf-1}
\end{figure*}



\clearpage

\section{Implementation of BH birth spin and other updates in $\bse$}\label{bhspin_app}

In this section, we provide further details of how the natal spin of BHs is implemented in $\bse$,
which will help the reader follow the updated code linked below. After formation,
BH spin is assigned in the subroutine {\tt kick.f}, following the strategy for assigning
fallback-mediated natal kicks to stellar remnants \citep{Banerjee_2020}. Since the {\tt KICK} subroutine
is necessarily called right after a stellar remnant formation, it is generally a convenient place for assigning
natal quantities to newborn remnants.

As for non-spun-up BH natal spins (see Sec.~\ref{bhspin}), $\abhzero$, choice can be made between the prescriptions
`Geneva', `MESA' (B20, \citealt{Morawski_2018}), or FM19 (negligible spin), as described in \citep{Banerjee_2020c}.
These prescriptions are readily implemented, since the required carbon-oxygen core mass, $\mco$,
is already imported in {\tt KICK} from {\tt hrdiag.f} for implementing the natal kick prescriptions.

To obtain the spun-up spin, $\abh$, of a BH, the orbital period of the binary, $\porb$, and the spin, $\pspin$, and
other stellar properties (including mass and evolutionary stage) of the pre-core-collapse
stellar member are captured from {\tt evolv2.f}, just before the BH's formation. This is done by storing
these quantities in a dedicated {\tt COMMON BLOCK}, at the beginnings of the looping over the binary
components; specifically, before the calls of {\tt star} and {\tt hrdiag}.
This ensures that the recording happens before any change
of stellar type and/or parameters of the members occur within the loop. The value of $\abh$ is then obtained inside {\tt KICK}
(which is called always after {\tt star} and {\tt hrdiag}) by accessing this {\tt COMMON BLOCK}
and applying the analytical expressions of Be20 or Ba21 (Eqns.~\ref{eq:abh_be20}-\ref{eq:def1}),
as chosen in the input. Since {\tt KICK} is called right after the transformation of the progenitor
star into a BH (generally, into any type of stellar remnant) and before
the loop progresses, the state of the binary and of the BH-progenitor member (see above) right before
the BH formation is correctly captured inside {\tt KICK}.
For numerical continuity, the maximum between $\abh$ and $\abhzero$
is chosen (Eqn.~\ref{eq:lims}) as the final natal spin of a BH; $\abh > \abhzero$
implies the formation of a spun-up BH. For convenient code implementations, two new routines
{\tt wrbh} (file {\tt wrbh3.f}) and {\tt abh} (file {\tt abh.f}) are introduced, where the
analytical expressions and the conditions are implemented.

User-defined $\qcrit$ is introduced by allowing the user to override the default
$\qcrit$ in {\tt evolv2.f}, based on input values. This is done similarly as in
{\tt COSMIC} \citep{Breivik_2020} (and as hinted in the original version of {\tt evolv2.f}).

In the original {\tt evolv2.f}, the mass transfer rate during a star-star mass transfer
is capped based on the thermal or dynamical timescale of the donor and the recipient;
when the recipient a white dwarf, an NS or a BH,
the accretion rate is capped based on the Eddington limit \citep{Hurley_2002}.
For accretion onto a white dwarf, mass loss from the system due to nova eruption is also considered.
In the version of $\bse$ used in this study, the original mass transfer scheme has not been altered
and is always applied. To implement an adjustable accretion efficiency, $f_a<=1$, for
star-star mass transfer (Sec.~\ref{qcr}), $f_a$ is simply multiplied to the final total
mass that is to be accreted onto the recipient (the variables {\tt dm2} and {\tt dm22}).
The appropriate component-mass and angular-momentum updates to the binary
then follow automatically by the procedures for the same in {\tt evolv2.f} \citep{Hurley_2002}. 
That way, $f_a=1$ restores the default mass accretion rate; with $f_a < 1$, the mass
transfer is less conservative than the default mass transfer.
Note that the above arrangement with the $f_a$ parameter affects only the
accretion onto a normal star; the original $\bse$-treatments for accretion onto a compact
remnant or nova eruption remain unaffected by the choice of $f_a$. Note also that even
with the choice of $f_a=1$, \ie, the original $\bse$ mass transfer, the process is still non-conservative
\citep{Hurley_2002}. At present, there is no option in this version of $\bse$ to enforce
a strictly conservative mass transfer.

The user's option of allowing or disallowing CE for HG donors is implemented in {\tt evolv2.f}
by introducing an input-specified flag, {\tt hgflag}, in the conditional statement that is relevant
for a stellar type 2 (an HG star) donor, for deciding whether to be treated by the common envelope
routine, {\tt COMENV}. If {\tt COMENV} is not to be invoked for an HG donor,
then the case is arranged to be handled by the routine {\tt MIX} to produce a merger
product. Both single-RLO and contact-binary cases are dealt with in this way.
Depending on the choice of {\tt hgflag}, the collision matrix (routine {\tt INSTAR})
is adjusted, so that mergers involving HG stars are correctly treated in {\tt MIX}.
Notably, \citet{Giacobbo_2018a} have also disallowed HG CE in their version of $\bse$, namely, {\tt MOBSE}.

For further details of these implementations and testings,
the reader is encouraged to refer to the source code at the following URL:\\
{\tt https://github.com/sambaranb/updated-BSEv2}

\section{Evolving isolated binary populations with $\bse$}\label{popsynth_app}

In this study, several sets of $1.2\times10^7$ isolated binaries are evolved (Sec.~\ref{popsynth}).
To facilitate evolving a large number of isolated binaries with $\bse$, 
a self-developed wrapper script, written in the python programming language, is utilised.
The script invokes $\bse$ to evolve a number of independent individual binaries in parallel, on a specified number of CPU threads, and does the relevant post-processing of the outcomes (\eg, processing of output
strings, extraction of stellar remnants and their properties, GW merger search, preparing convenient output data),
also in parallel. The parallelization is achieved by utilizing the built-in python package
{\tt multiprocessing}.
To what extent acceleration can be achieved compared to a single-core run depends on the number
of parallel threads utilised and as well on the write speed of the storage drive where the output data
is written. With $6\times48$ parallel threads and SSD-type drives, a set of $6\times 2\times 10^6$
massive binaries could be evolved in approx. 10 hours.
With $6\times64$ threads and ultra-high-speed, HPC-quality (or `machine learning') drives,
such a set could be evolved within approx. 5 hours. The wrapper is currently under further development
but is available upon reasonable request to the corresponding author.

\section{Spin-orbit misalignment angles}\label{angles}

In this section, a derivation of Eqn.~\ref{eq:tilts} is presented. For convenience in notations,
in this particular section, the subscripts `fir'/`first' and `sec'/`second' are replaced by
`1' and `2', respectively.

Let $\lzerocap$, $\lonecap$, and $\ltwocap$ are the directions of the orbital angular momenta
of the original star-star binary, the star-compact binary after the first SN, and
the compact-compact binary after the second SN, respectively. Since the spin of a stellar member
in the close, interacting binary is assumed to be aligned with the orbital angular momentum
and also a newly formed BH's spin is assumed to be aligned with its progenitor star's spin (see Sec.~\ref{popsynth}),
we have
\begin{equation}
\aonecap = \lzerocap; {\rm ~~}\atwocap = \lonecap. 
\label{eq:id1}
\end{equation}

Also, the SN tilt angles, $\nu_1$ and $\nu_2$, are, by definition, 
\begin{equation}
\lzerocap\cdot\lonecap = \cos\nu_1; {\rm ~~} \lonecap\cdot\ltwocap = \cos\nu_2.
\label{eq:id2}
\end{equation}

From Eqn.~\ref{eq:tilts}, \ref{eq:id1}, and \ref{eq:id2} it then follows that
\begin{equation}
\theta_2 = \cos^{-1}(\atwocap\cdot\ltwocap) = \cos^{-1}(\lonecap\cdot\ltwocap),
= \cos^{-1}(\cos\nu_2) = \nu_2
\label{eq:th2}
\end{equation}
and that
\begin{equation}
\lonecap\cdot\aonecap =  \cos\nu_1.
\label{eq:id3}
\end{equation}

For convenience, let us take the Z-axis aligned with $\lonecap$ and the XZ-plane as the plane containing    
$\aonecap$ and $\lonecap$, so that
\begin{equation}
\lonecap = \kvec; {\rm ~~}\aonecap = \cos\nu_1\kvec + \sin\nu_1\ivec.
\label{eq:id4}
\end{equation}
Let $\ltwocap$ make an azimuthal angle $\Omega$ with this plane, \ie (c.f. Eqn.~\ref{eq:id2}),
\begin{equation}
\ltwocap = \cos\nu_2\kvec + \sin\nu_2\cos\Omega\ivec + \sin\nu_2\sin\Omega\jvec. 
\label{eq:id5}
\end{equation}
Hence, from Eqns.~\ref{eq:tilts}, \ref{eq:id4}, and \ref{eq:id5},
\begin{equation}
\theta_1 = \cos^{-1}(\aonecap\cdot\ltwocap)
	= \cos^{-1}(\cos\nu_1\cos\nu_2 + \sin\nu_1\sin\nu_2\cos\Omega).
\label{eq:th1}
\end{equation}

Since the tilt angles, $\nu_1$ and $\nu_2$, are orbit averaged (\ie, the exact phase of the
exploding star in the orbit at the time of the SN is not explicitly considered) and both the parent star-star
and star-BH binaries can generally be expected to be circular for tight, interacting binaries, we
can take $\Omega$ to be uniformly distributed over 0 to $2\pi$ \citep{Hurley_2002}. Hence, $\theta_1$ ranges over
$\theta_1\in[ \nu_1-\nu_2, \nu_1+\nu_2]$.

If we adopt the convention that an angle is always positive for rotation from the initial direction to the final
direction then, after replacing $\nu_1$ by $-\nu_1$,
\begin{equation}
\theta_1 = \cos^{-1}(\cos\nu_1\cos\nu_2 - \sin\nu_1\sin\nu_2\cos\Omega).
\label{eq:th1a}
\end{equation}

Notably, the above derivation inherently assumes that the orbital angular momentum, $\vec L_1$, of the star-BH system
(after the first SN) is the dominant angular momentum of the system, so that the change of the direction of the orbital
angular momentum due to the tidal realignment of the (WR) star is negligible.
Eqns.~\ref{eq:th1} and \ref{eq:th1a} can more generally be expressed in terms of the independent azimuthal angels (say,
$\Omega_1$ and $\Omega_2$) of $\aonecap$ and $\atwocap$ - the current form is due to the choice of the
coordinate system as described above. Since we are primarily interested in the distribution of $\xeff$, which quantity
remains invariant during any GR inspiral until the merger, and not in the distributions of the azimuths
themselves in the LVK band, we use the current simplified forms of Eqns.~\ref{eq:th1} and \ref{eq:th1a}.

Assuming small SN tilt angles (as is typical), $\Omega$ practically represents the relative azimuthal angle of the BH
spin vectors on the orbital plane, at the formation of the BBH, \ie, just after the second SN. As the BBH evolves
towards merger and precesses due to GR effects, the azimuth evolves which may be detected in future, more sensitively
detected GW events \citep[\eg,][]{Varma_2022}. However, $\xeff$ will remain constant throughout the inspiral (Sec.~\ref{intro}
and references therein).
Notably, if $\nu_1,\nu_2 \rightarrow 0$ then $\theta_1,\theta_2 \rightarrow 0$, but $\theta_1$ and $\theta_2$ would generally
remain unequal (Eqns.~\ref{eq:th2} and \ref{eq:th1}). In such a case,
the precessing binary can get into a spin-orbit resonance when it is very close
to its merger (usually, when the binary is well inside the LVK frequency band;
see, \eg, \citealt{Gerosa_2013} and references therein).  
Also, for small $\theta_1$ and $\theta_2$, $\xeff$ is determined primarily by the BHs' masses and their spin magnitudes,
$\abhone$ and $\abhtwo$. For significantly mass-ratio-reversed and spun-up BBHs
with $\abhtwo>>\abhzero$, or for spun-up BBHs that are outcomes of double-cored CE (Sec.~\ref{bhspin}),
$\xeff$ is primarily determined by the spun-up spin magnitude.


\end{document}